\documentclass[11pt,a4paper]{article}

\usepackage{amsmath}
\usepackage{amssymb,amsfonts,bm}
\usepackage{dsfont}
\usepackage{color,graphicx,caption,subcaption}
\usepackage[hidelinks]{hyperref}
\usepackage{xspace}
\usepackage{fullpage}
\usepackage{placeins}
\usepackage{enumitem}
\usepackage{url}
\usepackage{verbatim}
\usepackage{amsthm}
\usepackage{commath}
\newcommand{\lela}{\left \langle}
  \newcommand{\rira}{\right \rangle}
\newtheorem{thm}{Theorem}

\newtheorem{remark}[thm]{Remark}

\newtheorem{problem}[thm]{Problem}
\DeclareMathOperator{\Ran}{Ran}
\renewcommand{\Re}{\operatorname{Re}}
\renewcommand{\Im}{\operatorname{Im}}
\renewcommand{\norm}[1]{\left\| #1 \right\|}
\newcommand{\Z}{\mathbb Z}
\newcommand{\C}{\mathbb C}

\newcommand{\argmin}{\operatorname {argmin}}

\def\<{\langle}
\def\>{\rangle}

\def\cR{{\mathcal R}}

\def\ba{{\bf a}}

\def\bK{{\bf K}}

\def\onemat{\mathrm{Id}_J}
\def\zeromat{0_J}

\newcommand{\eps}{\varepsilon}
\newcommand{\R}{\mathbb R}
\newcommand{\Uobs}{U_{\mathrm{obs}}}
\newcommand{\BZ}{{\mathcal B}}
\newcommand{\RBZ}{{\mathcal B_{\rm red}}}

\newcommand{\rme}{\mathrm{e}}
\newcommand{\ri}{\mathrm{i}}
\newcommand{\per}{\mathrm{per}}
\renewcommand{\leq}{\leqslant}

\renewcommand{\geq}{\geqslant}

\newcommand{\bk}{\mathbf{k}}
\newcommand{\bb}{\mathbf{b}}
\newcommand{\br}{\mathbf{r}}
\newcommand{\bR}{\mathbf{R}}

\newcommand{\UC}{Y}
\newcommand{\Uheal}{U_{\mathrm{heal}}}
\newcommand{\dps}{\displaystyle}

\newcommand{\ie}{{\sl i.\,e.\ }}   
\newcommand{\eg}{{\sl e.\,g.\ }} 

\newcommand{\virg}[1]{``#1''}

\newcommand{\half}{\mbox{\footnotesize $\frac{1}{2}$}}
\newcommand{\mhalf}{\mbox{\footnotesize $- \frac{1}{2}$}}

\newcommand{\ins}[1]{ \left\{  #1 \right\}} 

\title{Robust determination of maximally-localized Wannier functions}
\author{\'Eric Canc\`es$^1$, Antoine Levitt$^2$, Gianluca Panati$^3$ and Gabriel Stoltz$^1$ \\
\footnotesize
$^1$: Universit\'e Paris-Est, CERMICS (ENPC), Inria, F-77455 Marne-la-Vall\'ee\\
\footnotesize
$^2$: Inria Paris, F-75589 Paris Cedex 12, Universit\'e Paris-Est, CERMICS (ENPC), F-77455 Marne-la-Vall\'ee\\
\footnotesize
$^3$: Dipartimento di Matematica, \virg{\textsc{La Sapienza}} Universit\`a di Roma, Roma, Italy
}

\begin{document}
\maketitle

\begin{abstract}
  We propose an algorithm to determine Maximally Localized Wannier
  Functions (MLWFs). This algorithm, based on recent theoretical
  developments, does not require any physical input such as initial
  guesses for the Wannier functions, unlike popular schemes based on
  the projection method. We discuss how the projection method can fail
  on fine grids when the initial guesses are too far from MLWFs. We
  demonstrate that our algorithm is able to find localized Wannier
  functions through tests on two-dimensional systems, simplified
  models of semiconductors, and realistic DFT systems by interfacing
  with the Wannier90 code. We also test our algorithm on the Haldane
  and Kane-Mele models to examine how it fails in the presence of
  topological obstructions.
\end{abstract}

\section{Introduction}

Wannier functions are a well-established tool in the solid state
physics community. In addition to providing intuition on chemical
bonding, they are the theoretical and computational underpinning of
many developments such as tight-binding approximations \cite{MMYSV12},
interpolation of band structures \cite{yates2007spectral}, and the
so-called modern theory of polarization or orbital magnetization
\cite{resta2007theory,spaldin2012beginner}.

Wannier functions are however not uniquely defined, and the choice of
the phase of the (quasi-)Bloch functions can have a dramatic impact on
their spatial localization. {As for a single isolated Bloch band,} it
was realized already in the sixties that in specific situations,
appropriate choices of gauge lead to exponentially localized Wannier
functions~\cite{Kohn1959, Cloizeaux1964a, Cloizeaux1964b}.  Progress
was made in the eighties by Nenciu~\cite{Nenciu1983} as well as
Helffer and Sj\"ostrand~\cite{HelfferSjostrand1989}, {who proved in
  the general case the existence of a Bloch gauge yielding
  exponentially localized Wannier functions. }

Whenever Bloch bands intersect each other, as it generically happens
in 3D crystals, Bloch functions cannot be smooth -- and not even
continuous -- at the crossing points (excluding the exceptional case
of 1D systems).  In this situation, a multi-band approach is
mandatory. As early realized \cite{Blount1962,Cloizeaux1964a,Cloizeaux1964b}, in
insulators and semiconductors it is convenient to consider all the
Bloch bands below the Fermi level as a whole, and to replace Bloch
functions with {quasi-Bloch functions}, namely eigenfunctions of the
projector $P(\bk)$ on the occupied bands.  Their preimages under the
Bloch transform are called {composite Wannier functions}, since they
refer to a composite family of energy bands. {For the sake of a
  simpler terminology, we will skip the adjective \virg{composite} in
  the next sections.}

The existence of an orthonormal basis of well-localized
Wannier functions is equivalent to the existence of an orthonormal
frame of quasi-Bloch functions, spanning $\Ran P(\bk)$, which is both
smooth and periodic as a function of $\bk$.  Such an existence problem has been a
long-standing problem in theoretical solid-state physics. The solution
for 1D systems was provided in \cite{Nenciu1982, Nenciu1991},
while a solution for 2D and 3D systems required the crucial use of
geometric ideas and methods \cite{Brouder2007, Panati2007}.

From a practical viewpoint, Marzari and Vanderbilt introduced an
optimization procedure to minimize the spread of Wannier functions
\cite{marzari1997maximally} and compute Maximally-Localized Wannier
Functions (MLWFs). The corresponding localization functional is now
known as the Marzari-Vanderbilt (MV) functional. The approach based on
the minimization of the MV functional was developed before a full
understanding of the theoretical criteria for localization was
gained. It however gives very satisfactory results in many situations,
and has become one of the standard tools of computational solid state
physics. It was conjectured in \cite{marzari1997maximally}
that global minimizers of the MV functional are exponentially localized, a fact
which was proved recently in \cite{panati2013bloch}, thereby providing
a firm and mathematically consistent ground for the MV optimization.

It has however been observed that, in some situations, the
minimization of the MV functional could fail because the algorithm
remained trapped in ``false local minima'' presenting unphysical
oscillations \cite{marzari1997maximally,mostofi2008wannier90}. This
issue is to be distinguished from physically relevant ``real'' local
minima \cite{corsetti2012properties} and was associated with bad
initial guesses obtained by the projection method and fine $\bk$-point
meshes. This problem has been considered recently in
\cite{mustafa2015automated}, where an alternative, more robust method
is presented. This method is however still based on a projection and
requires a physical input in the choice of basis functions, while the
algorithm we present here does not require any input or parameter
tuning. In that sense it is similar to the recently proposed SCDM
algorithm \cite{damle2015compressed,damle2015scdm}, which uses columns
of the density matrix to provide localized Wannier functions. Compared
to this method, our approach works directly in Bloch space, avoiding any
representation of the density matrix in real space, and is readily
implementable using the same input as standard MLWFs computation. It
also does not depend on a potentially ill-conditioned matrix
inversion.

We will show in this work that the issue of ``false local minima''
occurs when the initial guess corresponds to a Bloch gauge with
vortex-like discontinuities, which prevent the convergence of the MV
optimization algorithm on fine samplings of the Brillouin zone. To provide an admissible initial guess
for the MV algorithm, we should really find a continuous Bloch gauge,
a mathematically non-trivial task. This issue has been studied
recently in \cite{cornean2015construction} and
\cite{fiorenza2016construction}, where the authors develop
constructive algorithms {which complement the abstract existence
  results of 
\cite{Brouder2007,Panati2007}}. 
Both methods are however not trivial to implement in practice, that of
\cite{cornean2015construction} requiring a perturbation argument that
might yield to Bloch gauge with very sharp transitions, and that of
\cite{fiorenza2016construction} relying on cumbersome interpolation
procedures.

In this work, we present an algorithm inspired by
the theoretical works of \cite{cornean2015construction} and \cite{fiorenza2016construction} but
more suited for actual implementation. This method fixes the gauge at
the $\Gamma$ point in the Brillouin zone and then incrementally on its
neighboring points, until the whole Brillouin zone is covered. Certain
pseudo-periodicity conditions have then to be satisfied, which we
enforce by the method of obstruction matrices developed in
\cite{fiorenza2016construction}. This requires finding continuous and
periodic logarithms of families of unitaries, a problem discussed in
\cite{cornean2015construction}. We propose an algorithm to solve this
problem. Although we are unable to prove that this step {always} produces a
continuous Bloch gauge, we conjecture it to be the case, and support
this claim by numerical tests.

Our algorithms only provides a continuous Bloch gauge, which yields
algebraically-decaying Wannier functions. Therefore, we see this
algorithm not as an end in itself but as a way to obtain good initial
guesses for the MV localization procedure. We show by numerical
examples on two- and three-dimensional systems that this two-step
process always yields localized Wannier functions, which is not the
case for state-of-the-art methods based on the projection method. We
therefore advocate the use of our method when good initial guesses are
not available. We however emphasize that our method only finds
``real'' local minima of the Marzari-Vanderbilt functional, which
might or might not possess optimal spread (\ie be a global minimum). Since they are devoid of
any physical input, they might also not correspond to physically
intuitive Wannier functions. Therefore, we do not see our method as a
replacement of the traditional construction of Wannier functions for
systems where physical insight are available, but rather as an
automated procedure that will always produce exponentially-localized
Wannier functions in the case when intuition is lacking.

\medskip

This article is organized as follows. Section~\ref{sec:statement}
introduces the notation and recalls some well-known facts on Wannier
functions. We next present our algorithm in a one-dimensional case in
Section~\ref{sec:1D}, then its extensions to the two- and
three-dimensional settings in Sections~\ref{sec:2D}
and~\ref{sec:3D}. The subtle issue of the appropriate choice of
continuous and periodic logarithms is discussed in
Section~\ref{sec:logpb}. We finally illustrate the approach with
numerical results in Section~\ref{sec:num}, and present some
perspectives in Section~\ref{sec:conclusion}.

\section{Notation and statement of the problem}
\label{sec:statement}

We consider a $d$-dimensional crystal. We denote by
$\cR=\sum_{j=1}^d \Z \ba_j$ the corresponding Bravais lattice, by $\UC$ a unit
cell centered on the origin, and by $\cR^\ast =\sum_{j=1}^d \Z \ba_j^\ast$ the reciprocal
lattice, with $\ba_i \cdot \ba_j^\ast = 2\pi\delta_{ij}$. We choose
the vectors $\{\ba_j^\ast\}$ as a basis for the reciprocal lattice. In
these coordinates, the Brillouin zone is {
\[
\BZ =  \ins{ \bk \in \R^d : \bk =  \sum_{j =1}^d k_j \, \ba_j^\ast  
\text{ for } k_j \in \left[-\frac 1 2, \frac 1 2\right] }    
\simeq \left[-\frac12, \frac12\right]^{d},
\]
where the symbol $\simeq$ refers to the fact that hereafter we will identify
the vector  $\bk =  \sum_{j=1}^d k_j \, \ba_j^\ast $  with its components $(k_1, \ldots, k_d)$ with respect to the reciprocal 
basis. 
}

\subsection{Smooth families of projectors and their symmetries}
The key input of our algorithm is a family $P(\bk)$ of orthogonal
projectors of rank $J$, which is smooth with respect to the wave
vector  $\bk =  \sum_{j} k_j \, \ba_j^\ast \in \R^d$. Such projectors are obtained,
in standard solid-state physics computations, by spectral projections of the
effective one-body Schr\"odinger operator
\begin{equation}
  \label{eq:Schrodinger}
  H = -\frac12 \Delta + V_\per,
\end{equation}
where $V_\per$ is a real-valued $\cR$-periodic potential. More
precisely, introduce the Bloch orbitals $\psi_{n,\bk}$, with periodic
parts $u_{n,\bk}$, so that
\begin{align*}
  \psi_{n,\bk}(\br) = u_{n,\bk}(\br) \, \rme^{\ri \bk \cdot \br}.
\end{align*}
The integer $n$ labels the band index, while $\bk \in \BZ$ denotes the
quasi-momentum. For a fixed $\bk \in \BZ$, the periodic functions
$(u_{n,\bk})_{n \geq 1}$ form an orthogonal
basis of $L^{2}_{\per}(Y)$ consisting of solutions of the following eigenvalue problem:
\[
H(\bk) u_{n,\bk} = \varepsilon_{n,\bk} u_{n,\bk}, \qquad H(\bk) =
\frac12 \left(-\ri \nabla + \bk \right)^2 + V_\per, \qquad \int_Y
\overline {u_{n,\bk}(\br)} u_{m,\bk}(\br) \, d\br = \delta_{nm},
\]
where the $\varepsilon_{n,\bk}$ are labeled by $n$ in increasing order. Our
convention for the Bloch transform is
\[
f_\bk(\br) = \sum_{\bR\in\cR} f(\br+\bR) \rme^{-\ri \bk \cdot (\br+\bR)}. 
\]
In order to construct Wannier functions from the Bloch orbitals, we
identify a set $\mathcal I$ of $J$ {Bloch} bands in the energy spectrum.
We assume that they are isolated in the sense that
\begin{align*}
  \inf_{\bk \in \BZ, n \in \mathcal I, m \not \in \mathcal I} |\eps_{n,\bk} -
  \eps_{m,\bk}| > 0.
\end{align*}
This gap condition is satisfied for instance by the occupied bands of an
insulator or a semiconductor. It ensures that the spectral
projector
\begin{equation}
  \label{eq:P_as_sum}
  P(\bk) = \sum_{n \in \mathcal I} |u_{n,\bk}\rangle\langle u_{n,\bk}|
\end{equation}
is a smooth (and even analytic) function of $\bk$ \cite{Nenciu1991,panati2013bloch}.


The underlying symmetries of the Hamiltonian $H(\bk)$ translate into
corresponding symmetries of the family of projectors $P(\bk)$. We
first note that, because the potential $V_{\per}$ is real, $P(\bk)$
satisfies the time-reversal property
\begin{align}
  \label{eq:P_C}
  P(-\bk) = C P(\bk) C,
\end{align}
where $C$ is the anti-unitary operator corresponding to complex conjugation, namely
\[
C u_{n,\bk} = \overline{u_{n,\bk}}.
\]
We also note that, for any $\bK \in \cR^{\ast}$,
\begin{align}
  \label{eq:P_tau}
  P(\bk+\bK) = \tau_\bK P(\bk) \tau_{-\bK},
\end{align}
with the following unitary translation operators
\[
(\tau_{\bK} f)(\br) = \rme^{-\ri \bK \cdot \br} f(\br).
\]
The convention we use here is slightly different from the one
in~\cite{fiorenza2016construction}: our definition of $\tau_\bK$
ensures that $\tau_\bK u_{n,\bk}$ is an eigenvector of $H(\bk+\bK)$ (so that $\tau_\bK$ is a translation by $-\bK$). The family of operators $\tau_{\bK}$ is a {unitary} group representation of $\cR^{\ast}$, in the sense that
\begin{align}
  \label{eq:tautau}
  \tau_{\bK+\bK'} = \tau_{\bK} \tau_{\bK'}  {\qquad \text{ and } \qquad    \tau_{-\bK} = \tau_{\bK}^{*}.}
\end{align}

In view of the time-reversal symmetry~\eqref{eq:P_C} and the
translation property~\eqref{eq:P_tau}, we can restrict ourselves to studying the reduced Brillouin zone
\begin{align*}
  \RBZ &=   
\ins{ \bk =  \sum_{j =1}^d k_j \, \ba_j^\ast  
\text{ with  }   
\begin{array}{ll}
k_1 \in [0, \half]  &    \\
k_j \in [\mhalf, \half]  & \text{ for } j \geq 2 
\end{array}
} 
\simeq  \left[0,\frac12\right] \times \left[-\frac12, \frac12\right]^{d-1},
\end{align*}
and map functions defined for $\bk \in \RBZ$ to functions defined for any $\bk \in \R^d$ by reflection and translation. 
\begin{remark}
  Similarly to \cite{fiorenza2016construction}, we do not use the
  specific structure of the model \eqref{eq:Schrodinger}, so that the
  approach presented here can be applied to various other periodic
  models of quantum mechanics, such as tight-binding models or
  relativistic models described by a Dirac operator. We only require a
  continuous family of projectors $P(\bk)$ satisfying \eqref{eq:P_C}
  and \eqref{eq:P_tau}, for some unitary operators $\tau_{\bK}$ and an
  anti-unitary operator $C$ satisfying
  \begin{align*}
    \tau_{\bK+\bK'} = \tau_{\bK} \tau_{\bK'},{\quad    \tau_{-\bK} =
    \tau_{\bK}^{*}},\quad 
  C^{2} = \mathrm{Id},  \quad \text{and}\quad    \tau_{\bK} C  = C \tau_{-\bK}.
\end{align*}
\end{remark}
\begin{remark}
\label{rmk:more_symmetries}
It is possible to further restrict the Brillouin zone under
consideration when $P(\bk)$ has more symmetries than the time-reversal
and translation symmetries \eqref{eq:P_C} and \eqref{eq:P_tau}. The
construction we present here does not take into account these possible
additional symmetries, and therefore may produce Wannier functions
that do not respect these symmetries. Our algorithm could be extended to
work only on the irreducible Brillouin zone, but this is outside the
scope of this paper.
\end{remark}

\subsection{Orthonormal frames}

The fundamental element to construct well localized Wannier functions
and the output of our algorithm is a Bloch frame depending smoothly on
the wave vector $\bk$. By definition, a {{Bloch} frame} (or,
shortly, a frame) is a mapping from $\bk \in \R^{d}$ to an orthonormal
basis $u_\bk = (u_{1,\bk}, \dots, u_{J,\bk})$ of $\Ran(P(\bk))$, such
that each component satisfies the pseudo-periodicity property
\begin{align}
u_{n, \bk+\bK} = \tau_{\bK} u_{n, \bk}  \qquad 
\forall (\bk,\bK) \in \R^{d} \times \cR^{\ast}.  
  \label{frame_tau}
\end{align}
The latter condition expresses the fact that such functions are
compatible with the symmetries of the family of projectors given by
\eqref{eq:P_tau}.

We emphasize that, despite the similarity in the notation, we do {not}
assume that the functions $u_{n, \bk}$ are (the periodic part of)
Bloch functions.  It is only assumed that
$$
P(\bk) u_{n, \bk} = u_{n, \bk},
$$
while $H(\bk) u_{n,\bk} = \varepsilon_{n,\bk} u_{n,\bk}$ does not hold
true in general.  Such functions are sometimes called {quasi-Bloch
  functions} in the literature.

\medskip

{As already noticed \cite{Nenciu1991}, the existence of a smooth Bloch frame is not trivial in view of the competition between the smoothness of $u$ and the property \eqref{frame_tau}, which encodes a global topological constraint. For instance, it is known that in some models with broken time-reversal symmetry (\eg in the presence of
a magnetic field or in a Chern insulator) there cannot exist any such continuous
frame, due to a topological obstruction  \cite{Dubrovin1980, Haldane1988, Brouder2007, MPPT2016}.
}

For two given frames $u$ and $v$, we write $u_\bk^{*} v_\bk$ for the $J \times J$ matrix with entries
\[
(u_\bk^{*}v_\bk)_{nm} = \int_Y \overline{u_{n,\bk}(x)} v_{m,\bk}(x) \, dx.
\]
For a $J \times J$ matrix $U(\bk)$, we write
\begin{align*}
(u_\bk U(\bk))_{n} = \sum_{m \in \mathcal I} u_{m,\bk} U(\bk)_{mn}.
\end{align*}
Note that, if $u$ is a frame, then so is $u U$ when the matrices
$U(\bk)$ are unitary for all values of $\bk \in \R^d$.  Similarly, for
an operator $A$, $Au_{\bk}$ is obtained by applying $A$ to all
components $u_{j,\bk}$ of $u_\bk$ independently. These conventions
allow for a uniform notation whether $u_{j,\bk}$ are functions or
coefficients in a basis set of size $N$ (in which case $u_\bk$ is a
$N \times J$ matrix, and the previous definitions are simply
restatements of matrix multiplication rules).

\subsection{Well localized Wannier functions}

(Composite) Wannier functions are defined, for a given frame $u$, by
\begin{equation}
  \label{eq:def_Wannier_unit_cell}
  \forall \br \in \R^d, \qquad w_n(\br) = \frac{1}{|\BZ|}\int_\BZ u_{n,\bk}(\br) \, \rme^{\ri \bk \cdot \br}\, d\bk,
\end{equation}
and the translations of these functions:
\[
\forall (\br,\bR) \in \R^d \times \R, \qquad w_{n,\bR}(\br) := w_n(\br - \bR) = \frac{1}{|\BZ|} \int_\BZ u_{n,\bk}(\br) \, \, \rme^{\ri \bk \cdot (\br-\bR)} \, d\bk.
\]
The Wannier functions
$\{ w_{n,\bR} \}_{n \in \mathcal I, \, \bR \in \cR}$ form a complete
orthonormal basis of the subspace of $L^{2}(\R^{d})$
associated with the chosen $J$ bands.

The localization properties of the Wannier functions are determined by
the regularity of the frame $u$. This can be seen from the fact that,
because $u$ satisfies the pseudo-periodicity property
\eqref{frame_tau}, an integration by parts gives
\begin{align*}
  \frac{1}{|\BZ|}\int_\BZ \left(\frac{\partial}{\partial k_{i}}u_{n,\bk}(\br)\right) \,
  \rme^{\ri \bk \cdot \br}\, d\bk &= -\ri r_{i} w_{n}(\br),
\end{align*}
Therefore, a $C^{\infty}$ frame yields Wannier functions that decay
faster than any polynomial, and an analytic frame yields exponentially
localized Wannier functions (see Section 2 of \cite{panati2013bloch}
for a more precise statement involving the Sobolev regularity of
$u_{n,\bk}$). In practice, the frame straightforwardly coming out of
the simulations, computed on the reduced Brillouin zone, is usually
not smooth because of the arbitrary phases of the $u_{n,\bk}$ and of
the possible crossings between the energy levels
$\eps_{n,\bk}$. Therefore, some correction through a family of unitary
matrices $U(\bk)$ has to be applied to the frame in order to transform
it into a smooth frame $uU$. Although it is easy to fix phases to
impose local smoothness, global smoothness is a more complicated issue
because the frame, extended by reflection and translation to the full
space $\R^d$, might not be continuous at the boundary of the reduced
Brillouin zone. Certain compatibility or gluing conditions should be
satisfied in order for this extension to be smooth.

Our problem is to compute real-valued localized Wannier functions,
which, {after Bloch transform, is equivalent to the following}.
\begin{problem}
\label{pbm}
Find a smooth frame $u$ such that
\begin{align}
  \label{frame_C}
  \forall \bk \in \R^d, \qquad u_{-\bk} = C u_\bk. 
\end{align}
\end{problem}
The additional property \eqref{frame_C} ensures that the Wannier
functions are real-valued.

We discuss in the next section an algorithm for constructing
continuous frames. We do not attempt to impose a higher degree of
regularity, as done in
\cite{fiorenza2016construction,cornean2015construction} via abstract
procedures, but consider a more pragmatic approach where a subsequent
smoothing of the continuous frame is obtained with the MV
algorithm. Our hope is that the continuous frame we obtain at the end
of our procedure is a sufficiently good initial guess for the
minimization of the MV functional in order to actually converge
towards a minimizer.

\section{Algorithm in the one-dimensional case}
\label{sec:1D}

We present here a practical algorithm to solve Problem~\ref{pbm} in
dimension~1. It can be seen as a discretization of well-known
procedures to construct smooth frames
\cite{fiorenza2016construction}, dating back to at least
\cite{Nenciu1991}, and discussed in \cite{marzari1997maximally}
section IV.C.1. We however carefully present this
simple case since, as in~\cite{cornean2015construction}, it is the
building block of the algorithm in higher dimensions.

We start from a given frame $u$ on the reduced Brillouin zone
$\RBZ = [0, 1/2]$, which we then extend to the whole space $\R$ by the
relations~\eqref{frame_tau} and~\eqref{frame_C}. More precisely, we
first extend $u$ to $k \in (-1/2, 0)$ by $u_k = C u_{-k}$, and then to
any $k \in \R$ by $u_{k} = \tau_{-K} u_{k+K}$ where $K \in \Z$ is such
that $k + K \in (-1/2, 1/2]$. This procedure yields a globally
continuous frame if and only if the frame on $\RBZ$ is continuous and
satisfies the following compatibility conditions at $0$ and $1/2$:
\begin{align}
  \tag{P1}
  \label{cond_1D_0}
  u_0 &= C u_0, \\
  \tag{P2}
  \label{cond_1D_1/2}
  u_{1/2} &= \tau_{1} C u_{1/2}.
\end{align}
The algorithm we propose consists of four steps: choosing a starting
{frame} at $0$ satisfying \eqref{cond_1D_0}, propagating this to
$[0, 1/2]$, enforcing \eqref{cond_1D_1/2} at $1/2$, and propagating
this fix back to $[0, 1/2]$.

\paragraph{Step 1: Fix $0$.} The first step is to choose a frame
$u_{0}$ satisfying~\eqref{cond_1D_0}. Since $H(0)$ is real-valued,
this can be done by choosing an orthonormal set of real-valued
eigenfunctions of $H(0)$.

\paragraph{Step 2: Propagate from $k=0$ to $k = 1/2$.}
Evolving the eigenvector at $k=0$ to $k=1/2$ can be done in various
ways, for instance by the Sz.-Nagy intertwining unitary as done
in~\cite{cornean2015construction}. Here we describe a natural way of
performing this operation on a mesh of $k$-points, using a
L\"owdin orthogonalization procedure. Assume that a mesh
$0 = k_{0} < k_{1} < \dots < k_{N} = 1/2$ of $\RBZ = [0, 1/2]$ is
given. We construct iteratively
\begin{align}
  \label{eq:propagation}
  \begin{cases}
    \widetilde{u}_{k_i}\hspace{-3mm} &= P(k_{i}) u_{k_{i-1}}, \\ 
    u_{k_i}\hspace{-3mm} &= \widetilde{u}_{k_i} \left(\widetilde{u}_{k_i}^{*} \widetilde{u}_{k_i}\right)^{-1/2}.
  \end{cases}
\end{align}
Since $P$ is continuous, provided the mesh spacing is sufficiently
small, the overlap matrix $\widetilde{u}_{k_i}^{*}
\widetilde{u}_{k_i}$ has its eigenvalues bounded away from 0, so that ${u}_{k_i}$ is a well-defined orthonormal basis of $\Ran(P(k_{i}))$. 

\begin{remark}
When the mesh spacing tends to zero, it can be shown that $u$ converges to the solution of the following ODE, with initial condition $u_{0}$:
\begin{equation}
\label{eq:continuous_dyn_u}
\frac{du_k}{dk} = \frac{dP(k)}{dk} u_k - \frac12 u_k \left( \left[\frac{dP(k)}{dk} u_k\right]^{*} u_k + u_k^{*} \left[\frac{dP(k)}{dk} u_k\right] \right).  
\end{equation}
In particular, the function $k \mapsto u_k$ is smooth.
\end{remark}

\paragraph{Step 3: Enforcing \eqref{cond_1D_1/2} at $k = 1/2$.}
The previous step yields a frame that is continuous, satisfies the
compatibility condition \eqref{cond_1D_0}, but
not~\eqref{cond_1D_1/2}. Note however that
$\Ran(P(1/2)) = \Ran(\tau_{1} P(-1/2)) = \Ran(\tau_{1} C P(1/2))$, so
that $u_{1/2}$ and $\tau_{1} C u_{1/2}$ are both orthonormal bases of
the same space. There exists therefore a unitary matrix $\Uobs$, which
we call ``obstruction matrix'' following
\cite{fiorenza2016construction}, such that
\begin{align}
  \label{eq:def_uobs}
  u_{1/2}  \Uobs & = \tau_{1} C u_{1/2}.
\end{align}
This matrix can be explicitly computed as $\Uobs = u_{1/2}^{*} (\tau_{1} C u_{1/2})$. A simple computation also shows that $\Uobs^{T} = \Uobs$:
\[
\begin{aligned}
\phantom{} [\Uobs]_{nm} & = \langle u_{n,1/2}, \tau_{1} C u_{m,1/2} \rangle = \langle C \tau_{1} C u_{m,1/2}, C u_{n,1/2}\rangle = \langle \tau_{-1} u_{m,1/2}, C u_{n,1/2}\rangle \\
& = \langle u_{m,1/2}, \tau_{1} C u_{n,1/2}\rangle = [\Uobs]_{mn}.
\end{aligned}
\]
Therefore $\overline{\Uobs} = \Uobs^{*} = \Uobs^{-1}$.

We now look for a matrix $\Uheal$ such that $u_{1/2} \Uheal$ satisfies
the compatibility condition \eqref{cond_1D_1/2}. Applying $\tau_{1}
C$ and using \eqref{eq:def_uobs}, we get
\begin{align*}
  \tau_{1} C \left(u_{1/2} \Uheal\right) &= \left(\tau_{1} C u_{1/2}\right)\overline\Uheal\\
  &= u_{1/2} \left(\Uobs \overline\Uheal\right)\\
  &= \left(u_{1/2} \Uheal\right) \left(\Uheal^{*} \Uobs \overline\Uheal\right).
\end{align*}
We see that $u_{1/2} \Uheal$ satisfies the compatibility condition
\eqref{cond_1D_1/2} if and only if
$\Uheal^{*} \Uobs \overline\Uheal = \onemat$, \ie
$\Uheal \Uheal^{T} = \Uobs$. Since
$\left(\Uobs^{1/2}\right)^{T} =\left(\Uobs^{T}\right)^{1/2} =
\Uobs^{1/2}$, we can choose $$\Uheal = \Uobs^{1/2}.$$ With this
choice, $u_{1/2} \Uheal$ satisfies \eqref{cond_1D_1/2}. We define the
fractional power of a normal matrix in the usual way, by applying the
exponent to its eigenvalues. For matrices with complex eigenvalues $z$
such as $\Uobs$, defining $z^{\alpha} = \rme^{\alpha \log z}$ requires
choosing a branch cut in the complex plane for the logarithm, which we
fix by using the principal determination, \ie for all
$z \in \C^{*}, \Im (\log z) \in (-\pi, \pi]$.



\paragraph{Step 4: Correct the frame on $\RBZ$.}
The last step is to globally define a new frame $u'$ on $\RBZ$ by interpolation: 
\begin{align*}
  u'_k &= u_k\, \Uheal^{2k} = u_{k}\, \Uobs^{k}.
\end{align*}
Note that $u'_0 = u_0 = C u_0 = C u'_0$, and so $u'$ satisfies
\eqref{cond_1D_0}. In the limit of infinitely small mesh spacing, the
mapping $k \mapsto u_{k}$ is continuous, and so is
$k \mapsto \Uobs^{k}$, which implies that $k \mapsto u'_k$ is
continuous on $\RBZ$.

The compatibility conditions finally ensure that $u_k$ can be extended
to a continuous frame for all $k \in \R$.

\begin{remark}
  We have only imposed continuity in our construction, but we actually
  get a smooth frame. This is because the frame we constructed is the same
  as the one we would obtain by defining $u$ on $\R$ via the
  propagation \eqref{eq:propagation}, then ${u'_{k}} = u_{k}
  \Uobs^{k}$. The latter frame is smooth because {both $k \mapsto u_{k}$
  and $k \mapsto \Uobs^{k}$ are}.
\end{remark}

\section{Algorithm in the two-dimensional case}
\label{sec:2D}

As in the one-dimensional case, we first construct a frame on the reduced Brillouin zone
\begin{align*}
  \RBZ \simeq \left[0, \frac12\right] \times \left[-\frac12, \frac12\right],
\end{align*}
and then extend it by symmetry to the whole space using~\eqref{frame_tau} and~\eqref{frame_C}. This yields a globally continuous frame if and only if the frame on the reduced Brillouin zone is continuous and the following compatibility conditions are satisfied:   
\begin{align}
  \label{edge_1}
  \tag{E1}
  u_{(0, k_{2})} &= C u_{(0,-k_{2})},\\
  \label{edge_2}
  \tag{E2}
  u_{(k_{1}, 1/2)} &= \tau_{(0,1)} u_{(k_{1}, - 1/2)},\\
  \label{edge_3}
  \tag{E3}
  u_{(1/2, k_{2})} &= \tau_{(1,0)} C u_{(1/2, -k_{2})}.
\end{align}

\begin{figure}[h]
\centering
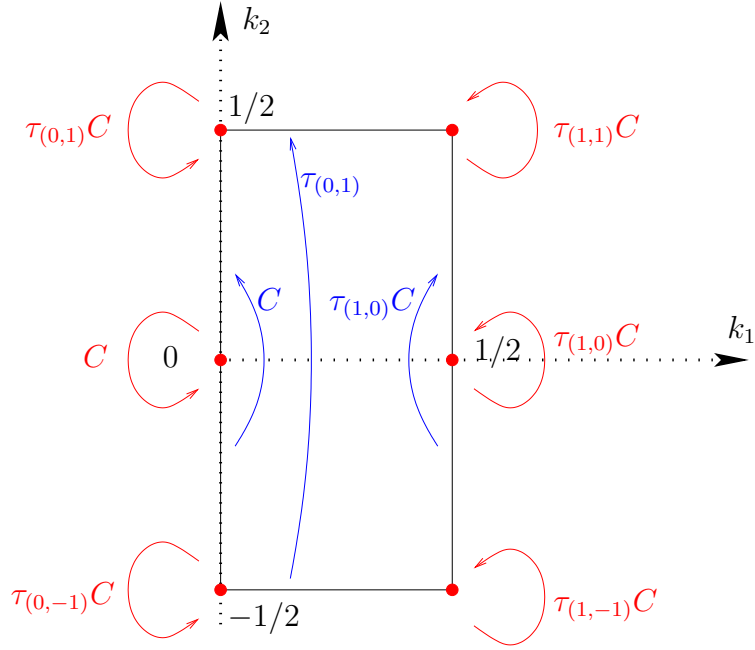
\caption{Symmetries in the two-dimensional reduced Brillouin zone. The
  blue arrows correspond to the edge conditions \eqref{edge_1},
  \eqref{edge_2} and \eqref{edge_3}, and the red arrows to the point
  conditions implied by the edge conditions.}
\label{fig:BZ2D}
\end{figure}

Note that these conditions are now edge conditions, relating two
points of the same edge (for \eqref{edge_1} and \eqref{edge_3}) or two
points on opposite edges (for \eqref{edge_2}), in contrast to the
point conditions of the one-dimensional case. These edge conditions
imply compatibility conditions at the special points
$(0,0), (0, \pm 1/2), (1/2, \pm 1/2)$ and $(1/2, 0)$, which are the
fixed points of the symmetry group. All these conditions can be
visualized on Figure \ref{fig:BZ2D}.

We take care of these conditions in a sequential manner: we first find
a frame on $\{0\} \times [-1/2, 1/2]$ that satisfies \eqref{edge_1},
extend this to a frame satisfying \eqref{edge_2}, then enforce the
condition \eqref{edge_3} on the right edge.

\paragraph{Step 1: Constructing a frame satisfying~\eqref{edge_1} and~\eqref{edge_2}.}
We first use the one-dimensional construction to obtain a frame $u_{(0,k_{2})}$ with $-1/2 \leq k_2 \leq 1/2$, satisfying \eqref{edge_1} as well as 
\begin{align}
  \label{prev_equation}
  u_{(0, 1/2)} &= \tau_{(0,1)} u_{(0, -1/2)}.
\end{align}
Then, for every $k_{2} \in [-1/2, 1/2]$, we propagate from $(0, k_{2})$ to $(1/2, k_{2})$ using the propagation procedure described in~\eqref{eq:propagation} in the one-dimensional case. This defines a frame on $\RBZ$. 

The compatibility condition~\eqref{prev_equation} at $k_1 = 0$ is propagated for any $k_{1} \in [0, 1/2]$, which implies that \eqref{edge_2} is satisfied. Let us prove this when starting from $k_1=0$ and going to $k_1 = \Delta k$. We define $\widetilde{u}_{(\Delta k,k_2)} = P(\Delta k,k_2) u_{(0,k_2)}$. Then, using~\eqref{prev_equation} and then~\eqref{eq:P_tau},
\[
\widetilde{u}_{(\Delta k,1/2)} = P(\Delta k,1/2) \tau_{(0,1)} u_{(0,-1/2)} = \tau_{(0,1)}P(\Delta k,-1/2)u_{(0,-1/2)} = \tau_{(0,1)} \widetilde{u}_{(\Delta k,-1/2)},
\]
so that \eqref{edge_2} follows upon normalizing the frames as $u_{\bk} = \widetilde{u}_{\bk} \left(\widetilde{u}_{\bk}^{*} \widetilde{u}_{\bk}\right)^{-1/2}$. However, \eqref{edge_3} is not satisfied in general.

\paragraph{Step 2: Enforcing the compatibility conditions~\eqref{edge_3}
  at the corners $(1/2, \pm 1/2)$.}
Before satisfying \eqref{edge_3} on the right edge, we enforce the condition at the corner $(1/2, 1/2)$, where the condition \eqref{edge_3} implies
\begin{align}
  \label{corner}
  u_{(1/2, 1/2)} &= \tau_{(1,1)} C u_{(1/2, 1/2)}.
\end{align}
This is a point compatibility condition analogous to the one encountered in the one-dimensional case. To enforce it, we consider the unitary obstruction matrix
\begin{align*}
  \Uobs &= u_{(1/2, 1/2)}^{*}\tau_{(1,1)} C u_{(1/2, 1/2)},
\end{align*}
and define the modified frame
\begin{equation}
  \label{eq:healing_E3_3D}
  u'_{\bk} = u_\bk \Uobs^{k_{1}}.
\end{equation}
The modified frame $u'$ satisfies \eqref{corner} by construction. Note also that the transformation~\eqref{eq:healing_E3_3D} preserves the compatibility conditions~\eqref{edge_2} and~\eqref{edge_1}. Moreover, \eqref{edge_2} and \eqref{corner} imply that $u'$ also satisfies the corner condition at $(1/2, -1/2)$:
\[
u'_{(1/2, -1/2)} = \tau_{(1,-1)} C u'_{(1/2, -1/2)}.
\]
In order to simplify the notation, we drop the prime in the remainder of the algorithm, and simply denote by $u$ the modified frame obtained at the end of this step. 

\paragraph{Step 3: Enforcing~\eqref{edge_3} on the right edge.}
We still need to modify $u$ to satisfy \eqref{edge_3} for any value of
$k_2 \in [-1/2, 1/2]$. We define to this end the following family of obstruction matrices:
\begin{align*}
  \Uobs(k_{2}) &= u_{(1/2, k_{2})}^{*} \tau_{(1,0)} C u_{(1/2, -k_{2})}.
\end{align*}
These matrices satisfy by construction
\[
\Uobs(-k_{2}) =  \Uobs(k_{2})^{T}, 
\qquad 
\Uobs(1/2) = \Uobs(-1/2) = \onemat.
\]
At this point, we would like to define a new frame by
\begin{align*}
  u'_{(k_{1}, k_{2})} &= u_{(k_{1}, k_{2})} \Uobs(k_{2})^{k_{1}}.
\end{align*}
The frame $u'$ satisfies all the compatibility
conditions~\eqref{edge_1}-\eqref{edge_2}-\eqref{edge_3}, is continuous
with respect to $k_{1}$, but may fail to be continuous with respect to
$k_{2}$, because the eigenvalues of $\Uobs$ may pass through the
branch cut of the logarithm in the negative real axis. We are
therefore faced with the problem of the continuous periodic logarithm:
find a matrix valued function $k_2 \mapsto L(k_{2})$, with $L(k_2)$
hermitian, which is continuous and satisfies the following conditions
for all $k_{2} \in [-1/2, 1/2]$:
\begin{align*}
  \exp(\ri L(k_{2})) &= \Uobs(k_{2}),\\
  L(-k_{2}) &= L(k_{2})^{T},\\
  L(-1/2) &= L(1/2) = \zeromat.
\end{align*}
If such a function exists, the frame
\begin{align*}
  u'_\bk &= u_\bk \exp\Big(\ri k_1 L(k_{2})\Big)
\end{align*}
is continuous and satisfies the compatibility conditions~\eqref{edge_1}-\eqref{edge_2}-\eqref{edge_3}. We delay the discussion of this problem to Section~\ref{sec:logpb}.

\medskip

\begin{figure}[h!]
  \centering
    \begin{subfigure}[b]{0.33\textwidth}
    \centering
    \includegraphics[width=\textwidth]{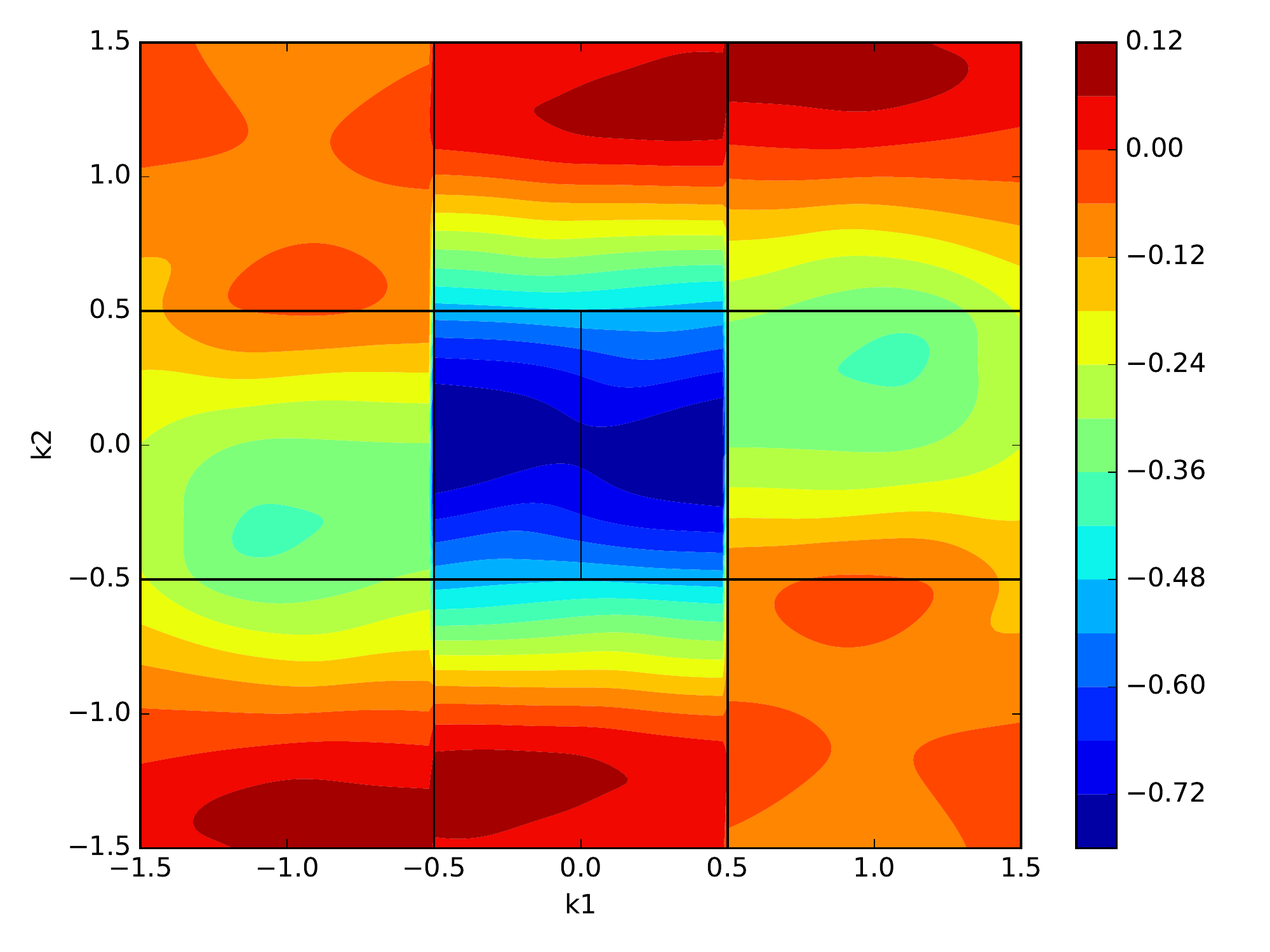}
  \end{subfigure}%
  \begin{subfigure}[b]{0.33\textwidth}
    \centering
    \includegraphics[width=\textwidth]{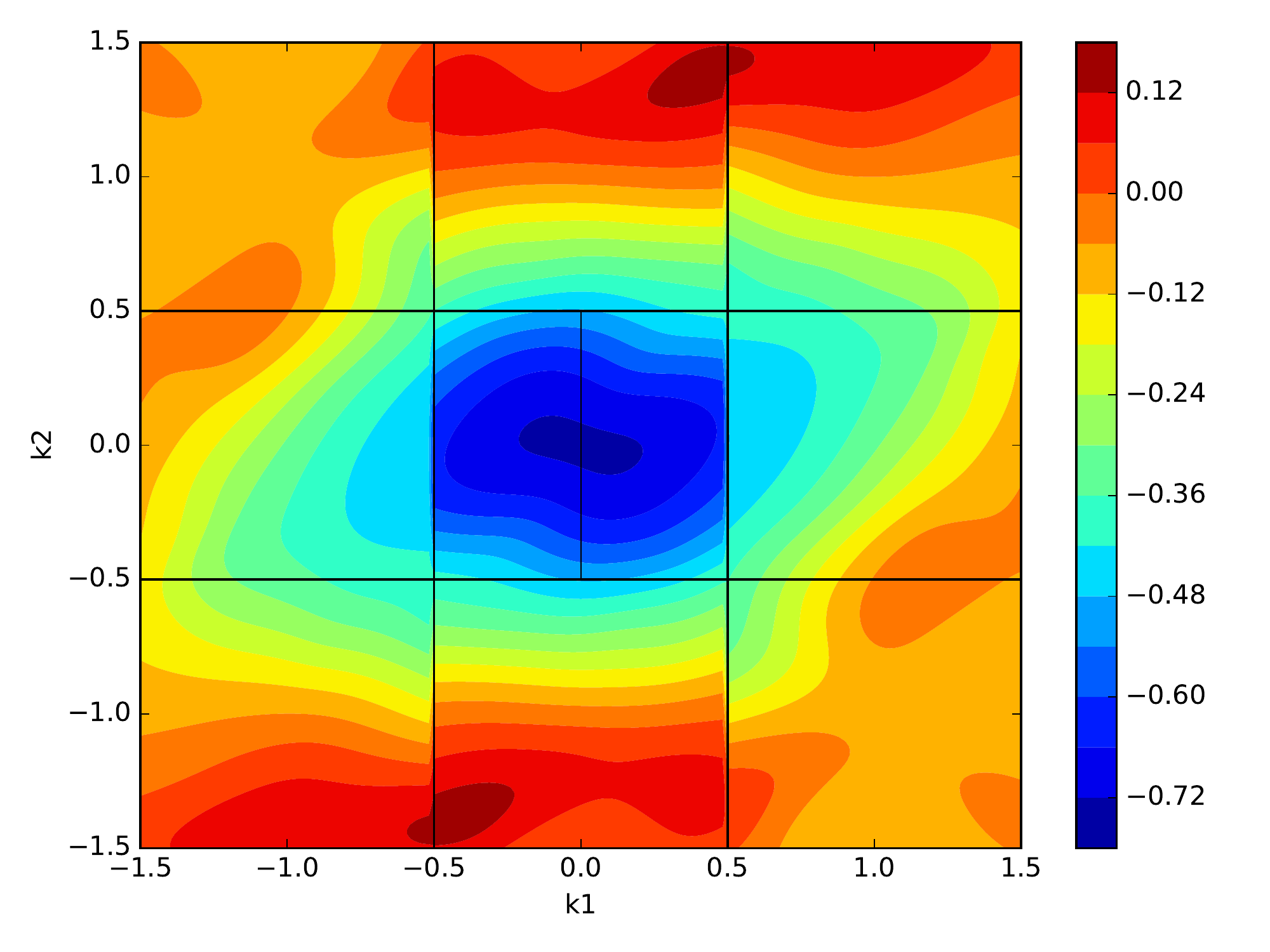}
  \end{subfigure}%
  \begin{subfigure}[b]{0.33\textwidth}
    \centering
    \includegraphics[width=\textwidth]{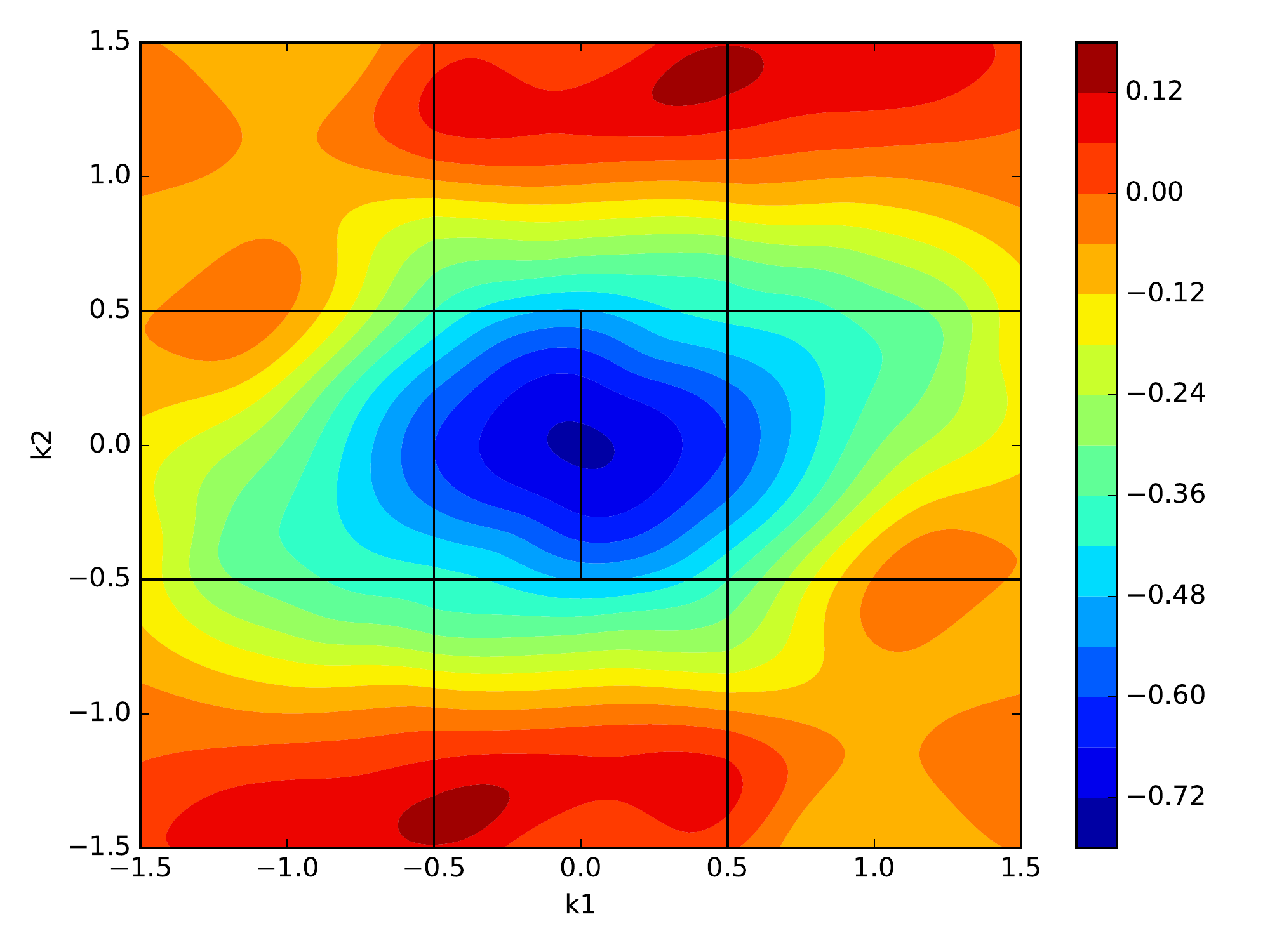}
  \end{subfigure}%

  \begin{subfigure}[b]{0.33\textwidth}
    \centering
    \includegraphics[width=\textwidth]{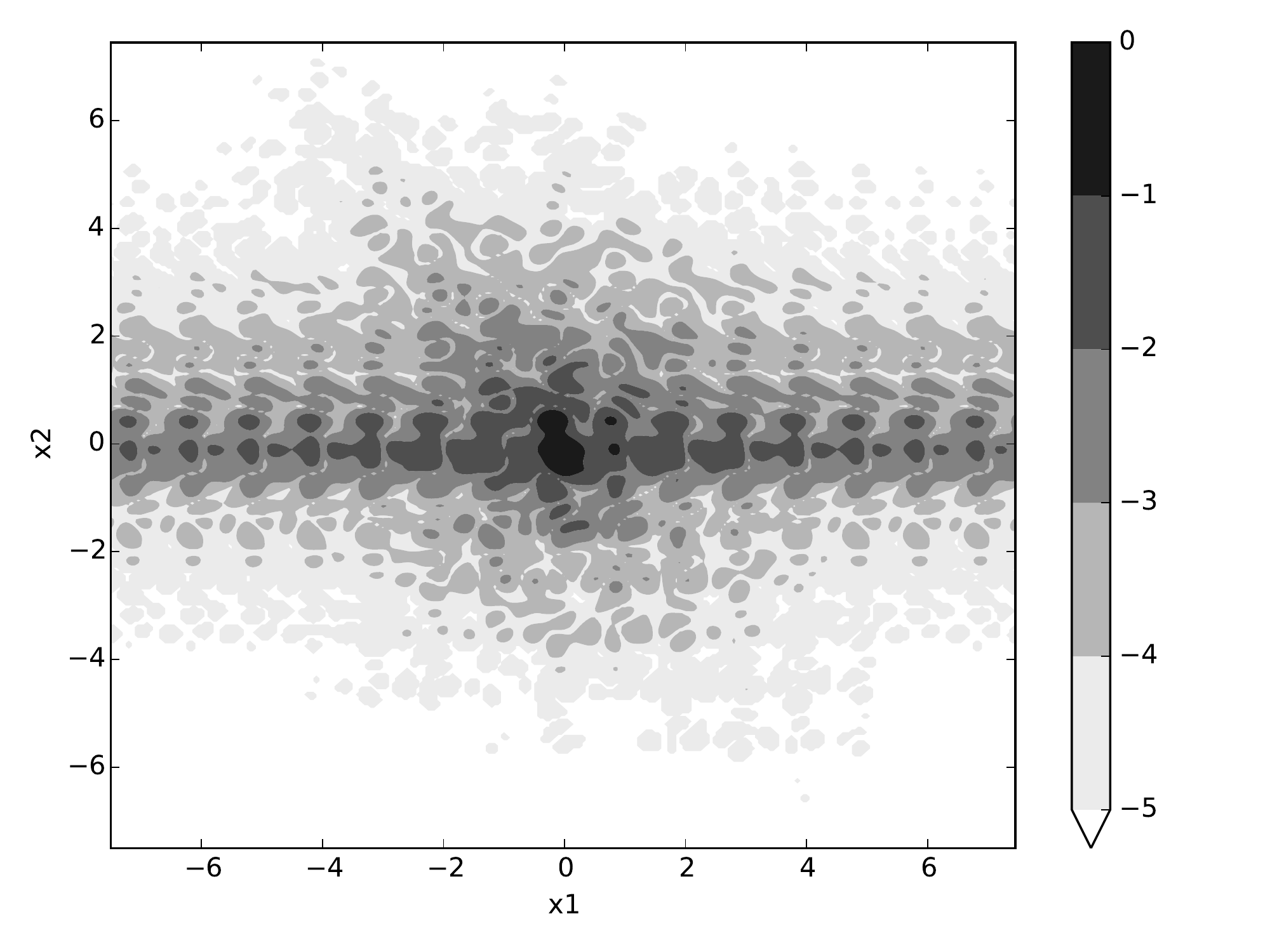}
    \caption{Step 1.}
  \end{subfigure}%
  \begin{subfigure}[b]{0.33\textwidth}
    \centering
    \includegraphics[width=\textwidth]{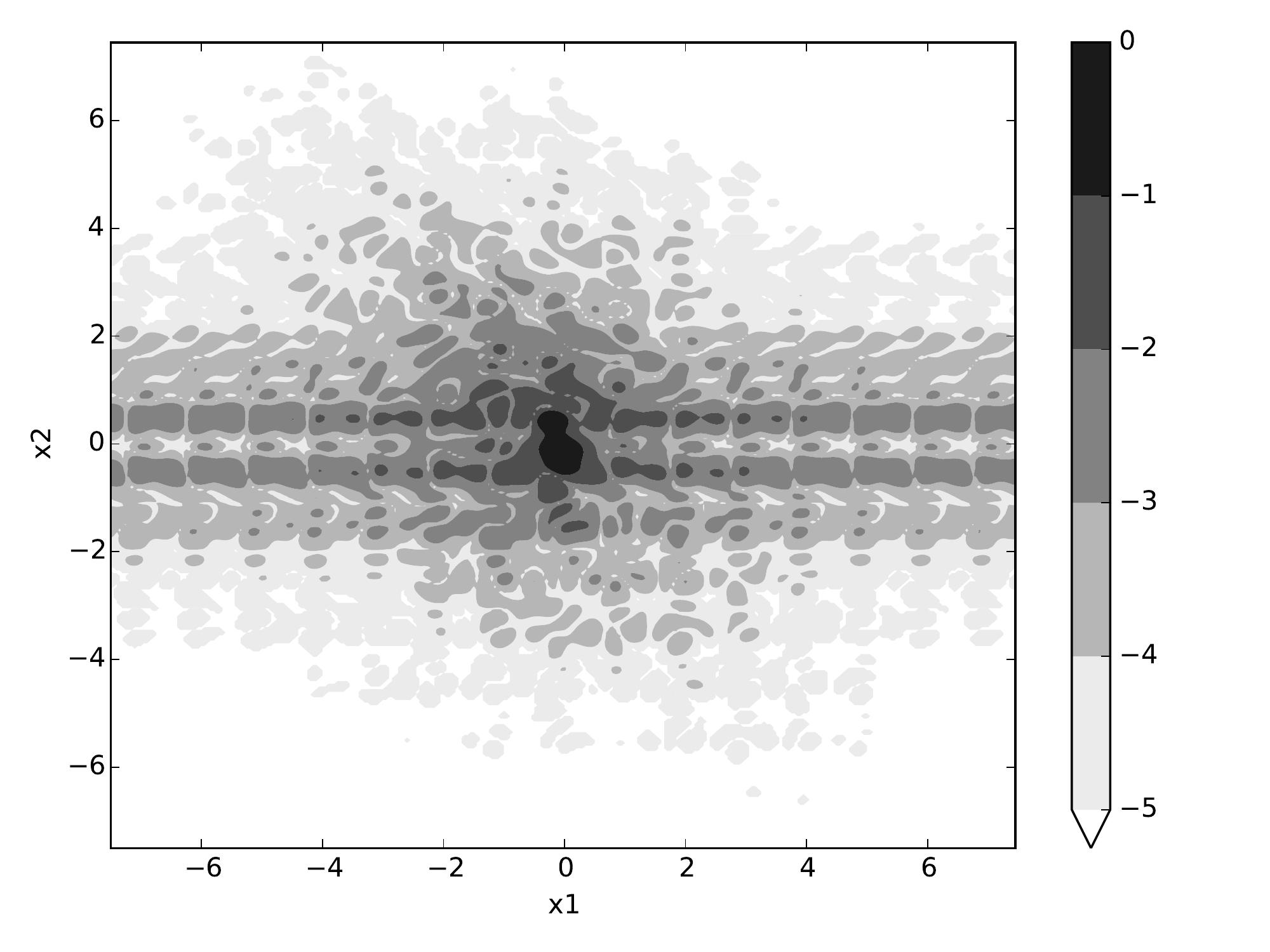}
    \caption{Step 2.}
  \end{subfigure}%
  \begin{subfigure}[b]{0.33\textwidth}
    \centering
    \includegraphics[width=\textwidth]{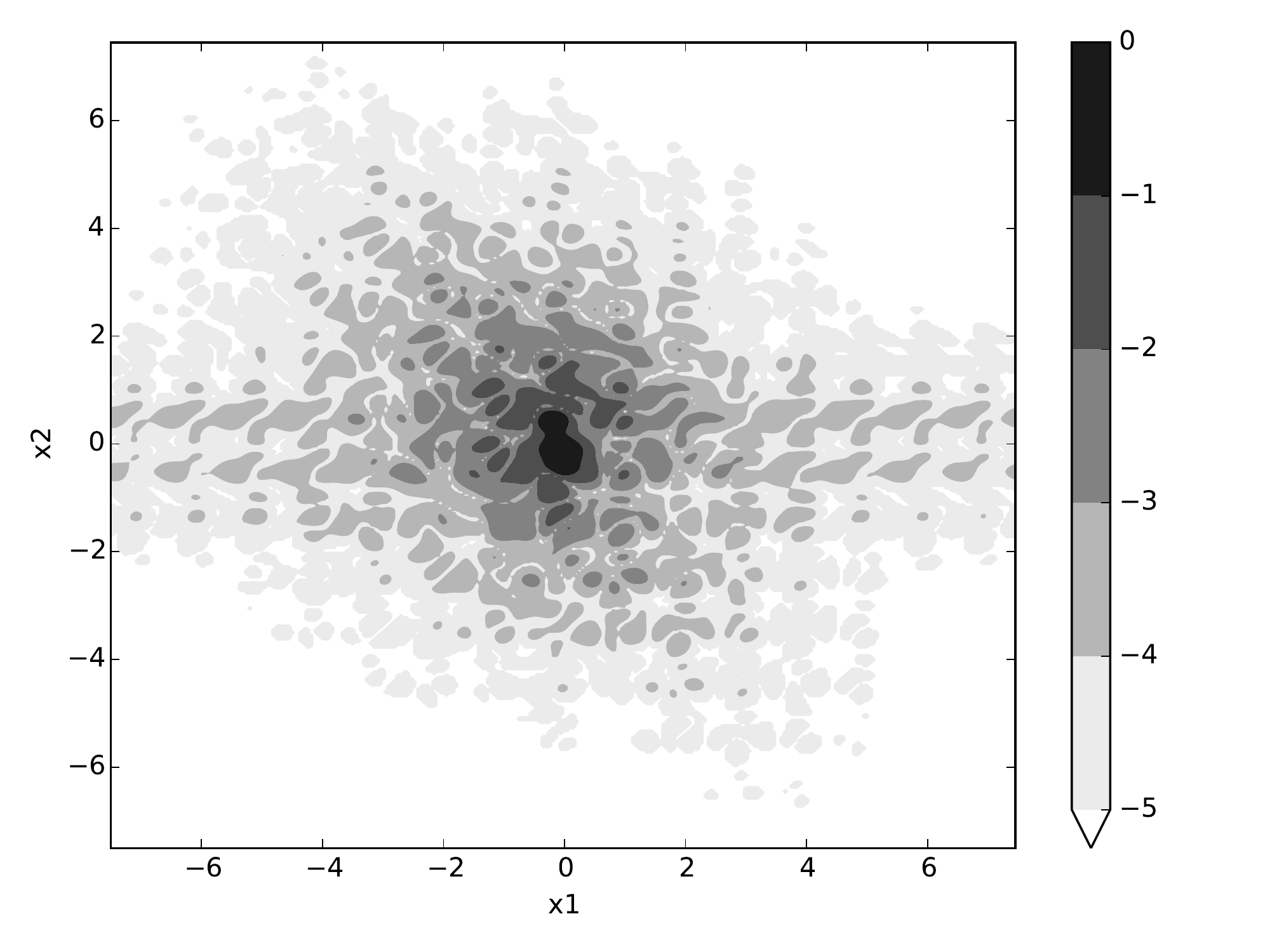}
    \caption{Step 3.}
  \end{subfigure}%
  \caption{\label{fig:ex_alg} Real part of the average of the first
    component, $\dps \int_{Y} \Re\left( u_{1,\bk}(r) \right) d\br$, of a
    Bloch frame corresponding to $J=5$ Bloch bands, as a
    function of $\bk \in [-3/2, 3/2]^{2}$ (top) and associated Wannier
    function $w_{1}$ in logarithmic scale (bottom). The real part of
    the average of $u_{1,\bk}$ is used as a proxy to visualize the regularity of the
    frame
    $u_{\bk}$.\\
    After step 1, the frame satisfies the compatibility conditions
    related to vertical translations, and the associated Wannier
    function is localized vertically, but not horizontally (left
    panel). After step 2, the frame satisfies all the compatibility
    conditions at the corners $(1/2 + n_1, 1/2 + n_2)$ (with
    $n_1,n_2 \in \mathbb{Z}$), but not on the vertical edges between
    these points. The associated Wannier function is still delocalized
    horizontally (middle panel). After step 3, the frame satisfies all
    the compatibility conditions and is globally continuous, although
    not smooth. The associated Wannier function is localized, but
    exhibits a slow algebraic horizontal decay (right panel). See
    Section \ref{sec:num} for the details of the model used.}
\end{figure}

The behavior of the algorithm on a simple example can be visualized in Figure~\ref{fig:ex_alg}. 
\begin{remark}
  The algorithm produces a continuous frame. However, in contrast to
  the 1D case, it does not provide any additional regularity. Because
  the 1D construction guarantees a smooth frame, there are no
  discontinuities of derivatives in the vertical direction, \ie at
  the points $(k_{1}, 1/2 + \Z)$. However, the condition for the first
  derivative to be continuous at a point $(1/2, k_{2})$ is
  \begin{align*}
    \left.\frac{\partial u'}{\partial k_{1}}\right|_{(1/2, k_{2})}
    &=-\tau_{(1,0)} C\left.\frac{\partial u'}{\partial k_{1}} \right|_{(1/2, -k_{2})}
  \end{align*}
  which is in general incompatible with~\eqref{edge_3}. This yields
  discontinuous derivatives at $(1/2 + \Z, k_{2})$.

  This means that the Wannier function will only decay algebraically
  in the $\ba_{1}$ direction, as can be seen in the right panel of
  Figure \ref{fig:ex_alg}. This is however easily fixed by the
  application of the MV minimization algorithm, as will be shown on
  Figure \ref{fig:ours} in Section \ref{sec:num}. We note that the MV
  algorithm is essentially equivalent to other forms of smoothing,
  such as for instance the one employed in
  \cite{cornean2015construction}: the MV algorithm can be seen as the
  gradient flow of a Dirichlet-like energy functional, \ie a heat
  flow, whose solution can be written as a convolution similar to the
  one used in \cite{cornean2015construction}.
\end{remark}

\section{Algorithm in the three-dimensional case}
\label{sec:3D}

We use the same induction to pass from the two-dimensional to the
three-dimensional case as the one to pass from the one-dimensional to the two-dimensional case. We define a frame on the reduced Brillouin zone 
\[
\RBZ \simeq \left[0, \frac12\right] \times \left[-\frac12, \frac12\right]^{2},
\]
and extend it to the full reciprocal space by symmetry. The
compatibility conditions to be satisfied by the frame $u$ are now
\begin{align}
  \label{face_1}
  \tag{F1}
  u_{(0, k_{2}, k_{3})} &= C u_{(0,-k_{2}, -k_{3})}, \\
  \label{face_2}
  \tag{F2}
  u_{(k_{1}, \pm 1/2, k_{3})} &= \tau_{(0,\pm 1,0)} u_{(k_{1}, \mp 1/2,k_{3})}, \\
  \label{face_3}
  \tag{F3}
  u_{(k_{1}, k_{2}, \pm 1/2)} &= \tau_{(0,0,\pm 1)} u_{(k_{1}, k_{2},\mp 1/2)}, \\
  \label{face_4}
  \tag{F4}
  u_{(1/2, k_{2}, k_{3})} &= \tau_{(1,0,0)} C u_{(1/2, -k_{2}, -k_{3})}.
\end{align}

\paragraph{Step 1: Constructing a frame satisfying \eqref{face_1}-\eqref{face_2}-\eqref{face_3}.}
Our first step is to use the two-dimensional algorithm to build a frame $u_{(0, k_{2}, k_{3})}$ on the face $k_{1} = 0$. This frame satisfies~\eqref{face_1}, and additionally 
\begin{align*}
  u_{(0, \pm 1/2, k_{3})} &= \tau_{(0,\pm 1,0)} u_{(0, \mp 1/2, k_{3})}, \\
  u_{(0, k_{2}, \pm 1/2)} &= \tau_{(0,0,\pm 1)} u_{(0, k_{2},\mp 1/2)}.
\end{align*}
For any $(k_{2}, k_{3}) \in [-1/2, 1/2]^{2}$, we propagate from
$(0, k_{2}, k_{3})$ to $(1/2, k_{2}, k_{3})$
using~\eqref{eq:propagation}. This defines a continuous frame on
$\RBZ$ which satisfies \eqref{face_1}-\eqref{face_2}-\eqref{face_3},
but not \eqref{face_4}.

\paragraph{Step 2: Enforcing \eqref{face_4} on the edges of the face $k_{1} = 1/2$.}
As in the two-dimensional case, before enforcing~\eqref{face_4} on the
whole face $k_{1} = 1/2$, we enforce it on the four edges of its boundary. In order to do this, we first fix the corner $(1/2, 1/2, 1/2)$, for which the compatibility condition is
\begin{align}
  \label{corner_3D}
  u_{(1/2, 1/2, 1/2)} &= \tau_{(1,1,1)} C u_{(1/2,1/2, 1/2)}.
\end{align}
We define the obstruction matrix
\begin{align*}
  \Uobs &= u_{(1/2, 1/2, 1/2)}^{*} \tau_{(1,1,1)} C u_{(1/2,1/2, 1/2)},
\end{align*}
and introduce 
\begin{align*}
  u'_{\bk} &= u_\bk \Uobs^{k_{1}}.
\end{align*}
By construction, $u'$ satisfies~\eqref{corner_3D}. The compatibility conditions \eqref{face_1}-\eqref{face_2}-\eqref{face_3} are also still valid for the modified frame $u'$. In addition, in view of~\eqref{corner_3D} and~\eqref{face_2}-\eqref{face_3}, the frame $u'$ satisfies the compatibility conditions at the other corners $(1/2, 1/2, -1/2), (1/2, -1/2, 1/2)$ and $(1/2, -1/2, -1/2)$.

We next enforce the condition \eqref{face_4} on the edge $(k_{1},k_2) = (1/2,1/2)$, namely:
\begin{align*}
  u_{(1/2, 1/2, k_{3})} &= \tau_{(1,1,0)} C u_{(1/2, 1/2, -k_{3})}.
\end{align*}
The corresponding unitary obstruction matrix is
\begin{align*}
  \Uobs(k_{3}) &= u_{(1/2, 1/2, k_{3})}^* \tau_{(1,1,0)} C u_{(1/2, 1/2, -k_{3})}.
\end{align*}
Note that $\Uobs(-k_{3}) = \Uobs(k_{3})^{T}$ and
$\Uobs(\pm 1/2) = \onemat$. As in the two-dimensional case, provided
we can solve the logarithm problem, we find a hermitian logarithm
$L(k_{3})$ satisfying $L(-k_{3}) = L(k_{3})^{T}$,
$L(\pm 1/2) = \zeromat$ and $\rme^{\ri L(k_{3})} = \Uobs(k_{3})$, and
modify the frame as
\begin{align*}
  u''_\bk &= u'_\bk \exp\Big(\ri k_{1} L(k_{3}) \Big).
\end{align*}
The modified frame $u''$ then satisfies \eqref{face_4} on the two edges $(k_{1},k_2) = (1/2, \pm 1/2)$. 

We repeat this construction on the edge $(k_{1},k_3) = (1/2,1/2)$, and introduce appropriate hermitian matrices $L(k_{2})$ such that
\begin{align*}
  u'''_\bk &= u''_\bk \exp\Big(\ri k_{1} L(k_{2}) \Big)
\end{align*}
satisfies \eqref{face_4} on the two edges $(k_{1},k_3) = (1/2,\pm 1/2)$. 

To simplify the notation, we drop the primes in the remainder of the algorithm, and simply denote by $u$ the modified frame obtained at the end of this step. 

\paragraph{Step 3: Enforcing \eqref{face_4} on the whole face $k_{1} = 1/2$.}
The compatibility condition~\eqref{face_4} is, on the whole face $k_1 = 1/2$:
\begin{align*}
  u_{(1/2, k_{2}, k_{3})} &= \tau_{(1,0,0)} C u_{(1/2,-k_{2},-k_{3})}.
\end{align*}
We therefore introduce the family $\Uobs(k_{2}, k_{3})$ of unitary obstruction matrices 
\begin{align*}
  \Uobs(k_{2}, k_{3}) &= u_{(1/2, k_{2}, k_{3})}^* \tau_{(1,0,0)} C u_{(1/2,-k_{2},-k_{3})}.
\end{align*}
Because of the previous step, 
\begin{align*}
  \Uobs(1/2, k_{3}) = \Uobs(-1/2, k_{3}) = \Uobs(k_{2}, 1/2) = \Uobs(k_{2}, -1/2) = \onemat.
\end{align*}
As in the two-dimensional case, we would like to define a new frame satisfying all the compatibility conditions as 
\begin{align*}
  u'_\bk &= u_\bk \Uobs(k_{2},k_{3})^{k_{1}},
\end{align*}
We again face with the logarithm problem, which we discuss in Section~\ref{sec:logpb}.

\begin{remark}
  As a mathematical curiosity, this construction immediately extends
  to higher dimensions, again provided the logarithm problem can be
  solved. By induction, we first build a frame on
  $\{0\} \times [-1/2, 1/2]^{d-1}$, then propagate it to
  $[0, 1/2] \times [-1/2, 1/2]^{d-1}$. We next fix the obstruction on the
  boundary of $\{1/2\} \times [-1/2, 1/2]^{d-1}$, and conclude by finding a
  continuous logarithm of the obstruction matrix on
  $\{1/2\} \times [-1/2, 1/2]^{d-1}$ to correct the frame.
\end{remark}

\section{The logarithm problem}
\label{sec:logpb}

We come back in this section to the so-called logarithm problem,
summarized in Problem~\ref{pbm:log} below.


An element $\bk \in \R^d$ is denoted $(k_1,\bk')$ in this section. We
also introduce the set of edges in dimension 3 (vertices in dimension
2)
\[
\mathcal{E} = \left\{ \bk' = (k_2,\dots,k_d) \in \left[-\frac12, \frac12\right]^{d-1} \, \left| \, k_i \in \left\{ -\frac12,\frac12 \right\} \textrm{ for some } 2 \leq i \leq d \right. \right\}.
\]
With this notation, we can summarize the logarithm problem as follows.

\begin{problem}
\label{pbm:log}
Given a continuous mapping $U$ from $[-1/2, 1/2]^{d-1}$ to the set of unitary matrices, such that
\begin{align}
  \label{Uedgeeq1}
\forall \bk' \in \mathcal{E}, \qquad U(\bk') = \onemat, 
\end{align}
and
\begin{align}
  \label{UeqUT}
\forall \bk' \in [-1/2, 1/2]^{d-1}, \qquad U(-\bk') = U(\bk')^{T},
\end{align}
find a continuous mapping $L$ from $[-1/2, 1/2]^{d-1}$ to the set of hermitian matrices, such that
\[
\forall \bk' \in \mathcal{E}, \qquad L(\bk') = \zeromat, 
\]
and, for all $\bk' \in [-1/2, 1/2]^{d-1}$,
\begin{align}
  \label{logpb}
  \rme^{\ri L(\bk')} &= U(\bk'),\\
  \notag L(-\bk') &= L(\bk')^{T}.
\end{align}
\end{problem}
This is the problem as used by our construction in the previous two
sections: \eqref{Uedgeeq1} is satisfied because we fixed the
obstruction matrix on the edges, while \eqref{UeqUT} stems from the
structure of the conditions \eqref{edge_3} and \eqref{face_4}. Let us
already note that this problem cannot be solved in general (see the
counter-example in~\cite{cornean2015construction}, which we recall in
Section~\ref{sec:eig_collisions}). We describe an algorithm that
solves the problem provided that the type of eigenvalue collisions
present in this counter-example does not occur. In our numerical
tests, we did not encounter such difficulties and were always able to
solve the logarithm problem. We refer to Section
\ref{sec:eig_collisions} for a discussion of this issue.

\subsection{The two-dimensional case}
\label{sec:log_pbm_2D}

A logarithm of $U$ can be found by starting at $k_2 = -1/2$, and by following the eigenvalues of $U$ to ensure the continuity of $L$. This procedure can be seen as a discrete counterpart of the phase following procedure used in~\cite[Lemma~2.13]{cornean2015construction} for a single band.

Assume that $[-1/2, 0]$ is partitioned as $-1/2 =
k_{2,0} < k_{2,1} < \dots < k_{2,N} = 0$. We set $L(k_{2,0}) = 0$, and
iteratively determine $L(k_{2,j})$ as a logarithm of $U(k_{2,j})$, taking into
account the phase information in $L(k_{2,j-1})$. To this end, we
diagonalize $U(k_{2,j})$ as $U(k_{2,j}) = W(k_{2,j}) D(k_{2,j})
W(k_{2,j})^{*}$, where $W(k_{2,j})$ is unitary and $D(k_{2,j})$ a
diagonal matrix whose entries have modulus~1. We then set $L(k_{2,j})
= W(k_{2,j}) E(k_{2,j}) W(k_{2,j})^{*}$, where $E(k_{2,j})$ is the
diagonal matrix given by
\begin{equation}
  \label{phase_fixing}
  \begin{aligned}
    \ri E(k_{2,j})_{n} &= \log\left( D(k_{2,j})_{n} \right) + 2\pi\ri \, p_{j,n},\\
    p_{j,n} &= \mathop{\argmin}_{p \in \Z}\; \mathrm{dist}\Big( \log(D(k_{2,j})_{n}) + 2\ri\pi \, p,  \,E(k_{2,j-1}) \Big),
  \end{aligned}
\end{equation}
where $\mathrm{dist}(x, E(k_{2,j-1}))$ is the distance of the imaginary number $x$ to the set
of diagonal elements of $E(k_{2,j-1})$. The matrix-valued function $L$
is finally extended from $[-1/2,0]$ to $[-1/2, 1/2]$ by the relation
$L(-k_2) = L(k_2)^{T}$. The continuity at $k_2 = 0$ is ensured because
$L(0)^{T} = L(0)$ (see~\eqref{UeqUT}).

\begin{figure}[h!]
  \centering
  \includegraphics[width=0.45\textwidth]{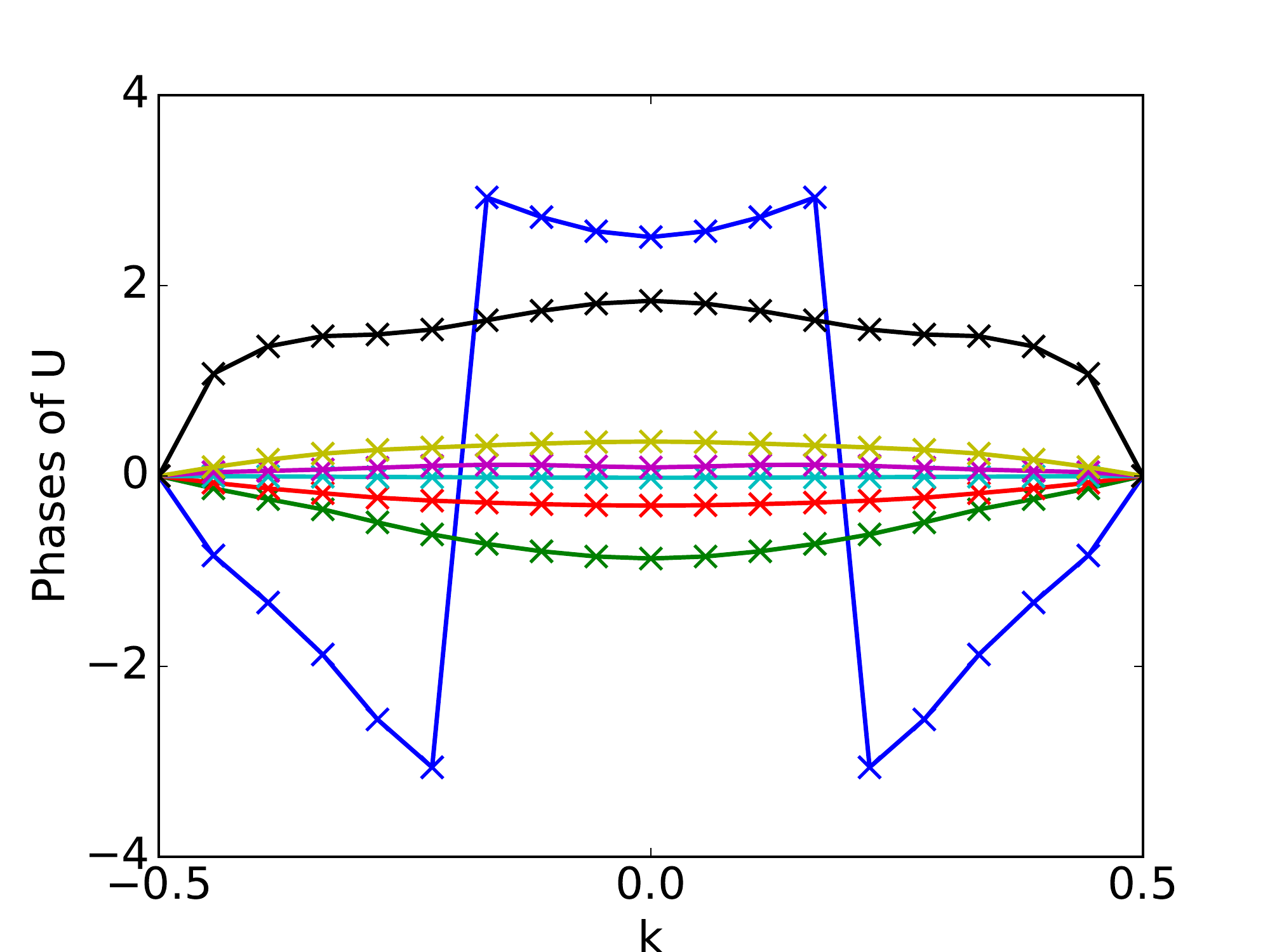}
  \includegraphics[width=0.45\textwidth]{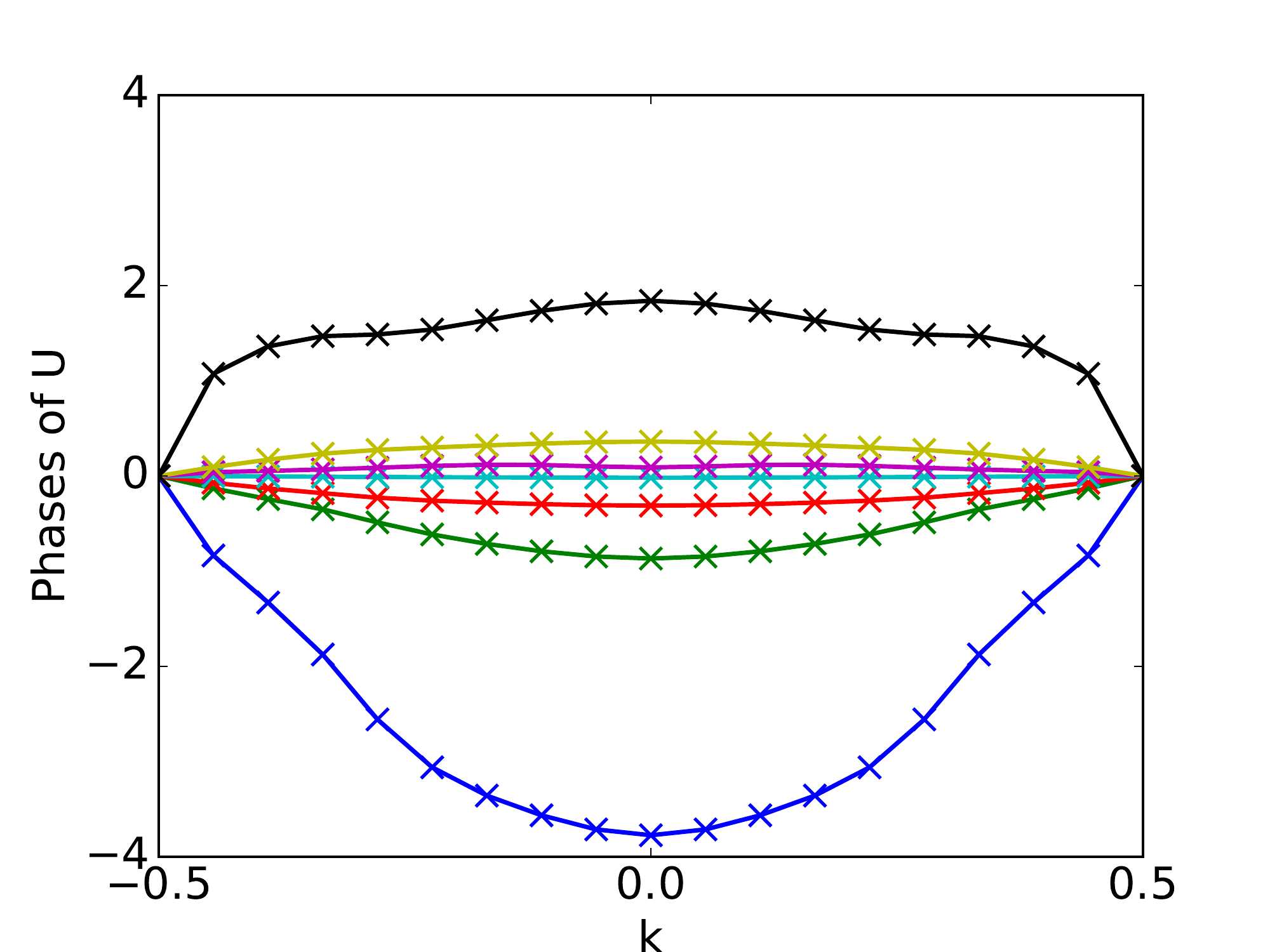}
  \caption{Eigenvalues of $L$, with a naive determination (left) and
    with our algorithm (right). This example was obtained by a
    variation of the potential presented in Section \ref{sec:num} with
    $J = 7$.}
  \label{fig:2D_log}
\end{figure}

Figure~\ref{fig:2D_log} shows the algorithm in action on one
example. With a naive phase determination, \ie
$E(k_{2,j})_{n} = \log(D(k_{2,j})_{n})$, the matrix-valued function
$L$ is not continuous when the eigenvalues of $U$ cross $-1$, and
therefore the eigenvalues of $L$ cross $\pm \pi$ (left panel). This is
because the complex logarithm is only continuous outside its branch
cut, chosen here on the negative real axis. When that happens in our
procedure, the integer $p$ jumps to accommodate this, and the
eigenvalues of $L$ evolve smoothly (right panel).

If the eigenvalues of $U$ never collide for values of $k_2$ in the
open interval $(-1/2,1/2)$, then $L$ is continuous in the limit of
infinitesimal mesh spacing. This is a consequence of the fact that the
minimizer in~\eqref{phase_fixing} is always uniquely determined, and
the phase following procedure ensures that the eigenvalues are
determined continuously when they cross the value $-1$ (see also the
continuity result stated
in~\cite[Lemma~2.13]{cornean2015construction}). The continuity of the
frame is preserved even if a crossing happens, as long as the integer
$p_n$ is the same for all the colliding eigenvalues. On the other
hand, when two eigenvalues with different values of $p_{n}$ collide,
$L$ is discontinuous. In the example of Figure~\ref{fig:2D_log}, this
would correspond to a collision of the top and bottom eigenvalues. In
this case, our algorithm fails. We however never observed such a
situation in all the tests we have performed. We discuss this in more
details in Section \ref{sec:eig_collisions}.

\subsection{The three-dimensional case}
\label{log_pbm_3D}

The three-dimensional case is handled similarly to the two-dimensional
case, in slices: for every~$k_{2}$, we find a continuous logarithm of
$k_{3} \mapsto U(k_{2}, k_{3})$, which we call $L(k_{2}, k_{3})$. As long
as there are no crossings between eigenvalues with different values of
$p_{n}$, this $L$ is continuous with respect to both $k_{2}$ and $k_{3}$,
which solves the problem.
\subsection{Eigenvalue collisions}
\label{sec:eig_collisions}

\begin{figure}[h!]
  \centering
  \includegraphics[width=.5\textwidth]{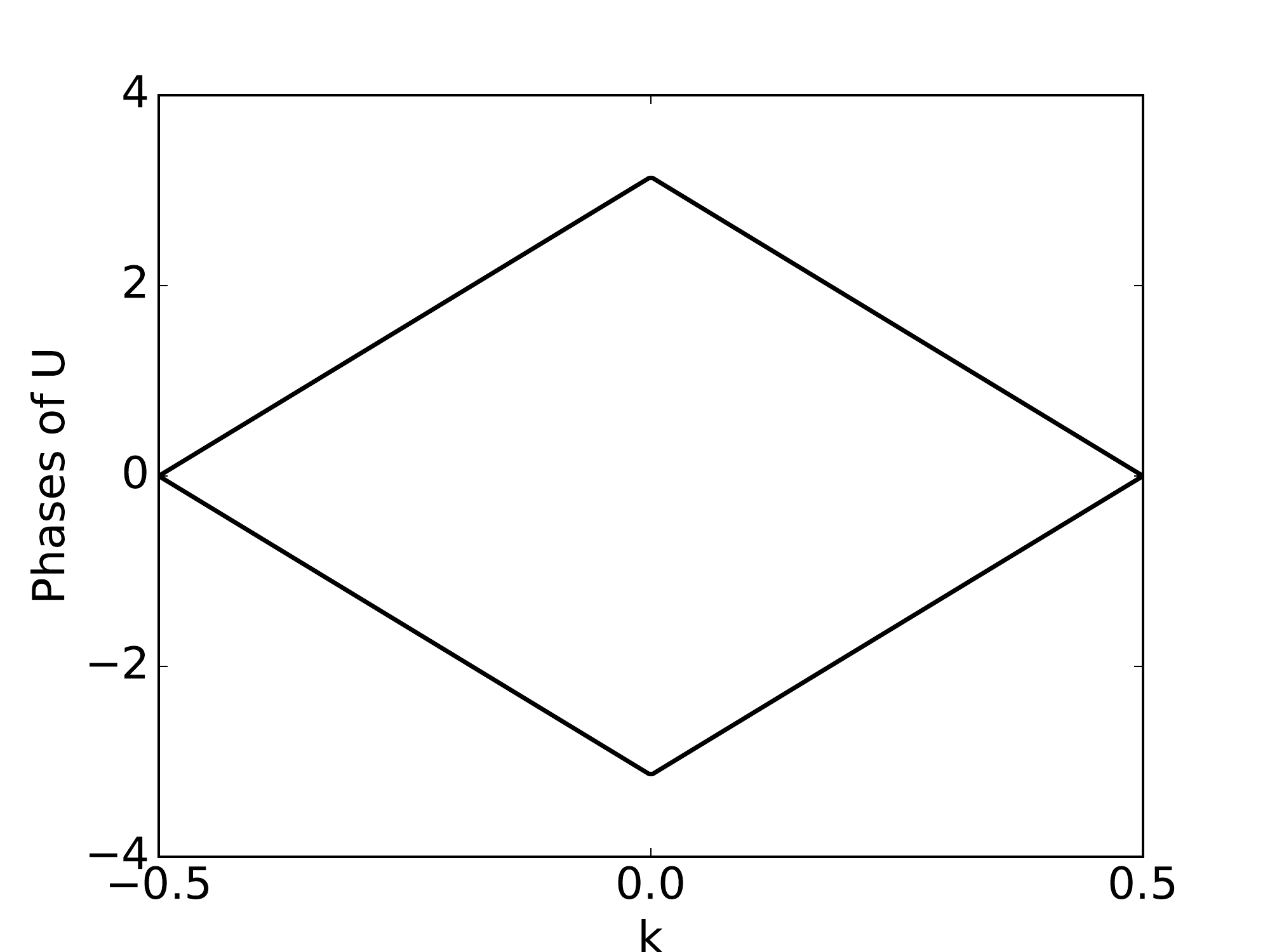}
  \caption{Imaginary part of $\log\left(\lambda^{\pm}(k)\right)$, the principal logarithm of the eigenvalues of~\eqref{bad_log_example}.}
  \label{fig:badlog}
\end{figure}

The prototypical case of a problematic eigenvalue collision is exemplified by the following situation (see~\cite[Example~2.25]{cornean2015construction}): {let} $J=2$ and, for $k \in \R$, {define}
\begin{align}
  \label{bad_log_example}
  U(k) = -
  \begin{pmatrix}
    \cos(2\pi k)& -\sin(2\pi k)\\
    \sin(2\pi k) & \cos(2\pi k)
  \end{pmatrix}.
\end{align}
The eigenvalues of this matrix are (see Figure~\ref{fig:badlog})
\begin{align*}
  \lambda^{\pm}(k) &= \rme^{\pm 2\ri \pi (k + 1/2)}.
\end{align*}
If the matrix-valued function $L$ is continuous at $k=0$, the phases of its eigenvalues must respectively vary by amounts of $2\pi$ and $-2\pi$ when evolving from $k=-1/2$ to $k=1/2$. Therefore, $L(1/2)$ must have eigenvalues $\pm 2\pi$ and cannot be equal to $0$. This shows that no matrix logarithm can simultaneously be continuous and satisfy $L(-1/2) = L(1/2)$.

In \cite{cornean2015construction}, the authors avoid this problem by
relaxing Problem~\ref{pbm:log} to the following weaker problem: Find a
sequence of hermitian, continuous matrices $L_{1}, \dots, L_{\mathcal N}$ for
$\bk' \in [-1/2, 1/2]^{d-1}$, equal to $\zeromat$ on $\mathcal E$, and such that
\begin{align}
  \label{multilogpb}
  &\rme^{- \ri L_{\mathcal N}(\bk')/2}\dots \rme^{- \ri L_{1}(\bk')/2} U(\bk')
    \rme^{-\ri L_{1}(\bk')/2} \dots \rme^{-\ri L_{\mathcal N}(\bk')/2} = \onemat,\\
  &\forall i \in \{ 1, \dots, \mathcal N\}, \qquad L_{i}(-\bk') = L_{i}(\bk')^{T}.
\end{align}
This is in fact sufficient for our purposes. Coming back to the last step of the algorithm in the three-dimensional case, assume that, instead of finding $L$ satisfying \eqref{logpb} and setting
\begin{align*}
  u'_\bk &= u_\bk \, \rme^{\ri k_{1} L(k_{2}, k_{3})},
\end{align*}
we find a family of functions $(L_{i})_{i = 1, \dots, \mathcal N}$ with values
in the space of hermitian matrices, which in addition
satisfy~\eqref{multilogpb}. Then we can fix the frame in $\mathcal N$ steps by defining 
\begin{align*}
  u_\bk' &= u_\bk \, \rme^{\ri k_{1} L_{1}(k_{2}, k_{3})} \dots
           \rme^{\ri k_{1} L_{\mathcal N}(k_{2}, k_{3})}.
\end{align*}
The corrected frame $u'$ is continuous and satisfies all the required
compatibility conditions.

In \cite{cornean2015construction}, the authors use the relaxed
problem~\eqref{multilogpb} in two steps. Their proof relies crucially
on the analyticity properties of the eigenvalues and eigenvectors, and
is tied to the one-dimensional case $d-1 = 1$, \ie $d=2$. Moreover, it
does not provide a practical way of computing the hermitian matrices
$L_i$. A natural extension to their method {to} dimension $d=3$ is to
slightly perturb $U(\bk)$ to eliminate any problematic eigenvalue
collision, and replace eigenvalue crossings with avoided crossings,
which the authors allude to in \cite[Remark 1.10]{cornean2015construction}. For the two-dimensional
and three-dimensional cases, where it is necessary to avoid crossings
in one-dimensional and two-dimensional settings respectively, generic
perturbations will turn the crossings into avoided crossings, because
the space of unitary matrices with repeated eigenvalues is of
codimension 3. This argument can be made precise using the
transversality theorem \cite{guillemin2010differential}, but the
construction is technical and we refrain from doing so here. This
allows us to find a continuous logarithm ${L_{1}}(\bk)$ of a
perturbation $\widetilde{U}(\bk)$ of $U(\bk)$ by the procedure
above. Then,
\begin{align*}
  \widehat{U}(\bk) = \rme^{-iL_{1}(\bk)/2} U(\bk) \rme^{-iL_{1}(\bk)/2}
\end{align*}
is very close to $\onemat$, and therefore its eigenvalues never cross $-1$. There is therefore no obstruction in finding a logarithm $L_{2}(\bk)$ of $\widehat{U}(\bk)$. Finally,
\begin{align*}
  \rme^{-iL_{2}(\bk)/2} \rme^{-iL_{1}(\bk)/2} U(\bk) \rme^{-iL_{1}(\bk)/2} \rme^{-iL_{2}(\bk)/2} = \onemat.
\end{align*}
The problem with this approach is that it is not clear how to best
implement in practice the perturbation argument, because small gaps in
$\widetilde{U}(\bk)$ result in eigenvectors that are continuous but
have very large variations, yielding large derivatives for
$L_{1}(\bk)$. Therefore, although we could in principle design a
scheme to treat such eigenvalue collisions, it is likely to be
unstable on coarse meshes, and would require some parameters to be
fine-tuned.

We do not discuss such extensions here since, as explained in
Section~\ref{sec:num}, we have never encountered any problematic
eigenvalue collisions in our numerical tests on two- and
three-dimensional systems. We do not know whether there is a
topological reason forbidding such crossings, or whether crossings
only occur on subspaces of codimension~3 and are therefore generically
absent in two- and three-dimensional situations, where the obstruction
matrices depend on 1 and 2 parameters respectively.


\section{Numerical results}
\label{sec:num}
This section presents numerical tests of the method proposed here, as
well as a comparison with the projection method. We first detail our
methodology. We then test the algorithms on two-dimensional toy
models, three-dimensional semiconductors with effective potentials,
and DFT computations. Results on topological insulators are presented
in an appendix.

\subsection{Spatial discretization}

We present in this section numerical experiments illustrating our
algorithm. We solve the periodic Schr\"odinger
equation~\eqref{eq:Schrodinger} using a Galerkin basis consisting of
all plane waves
\begin{align*}
  e_{\bK}(\br) &= \rme^{\ri \bK \cdot \br},
\end{align*}
for $\bK \in \cR^{\ast}$ such that $|\bK|^{2} \leq E_{\rm c}$, for
some fixed energy cut-off $E_{\rm c}$. Note that the basis we consider
does not depend on the value of $\bk \in \RBZ$, in contrast to most
plane-wave codes, which use the $\bk$-dependent cut-off condition
$|\bk + \bK|^{2} \leq E_{\rm c}$. This is done for convenience in our
rudimentary implementation. This has a number of undesirable
properties. For instance, the equality
\begin{align*}
    H(\bk + \bK) &= \tau_{\bK} H(\bk) \tau_{-\bK}
\end{align*}
is only approximately valid, in the regime where $\bk$ and $\bK$ are
not too large. In addition, $\tau_{\bK}$ is not unitary when
restricted to the vector space spanned by the Galerkin basis. These
properties are however recovered in the limit when
$E_{\rm c} \to +\infty$. In our tests, we take $E_{\rm c}$ large
enough to avoid these problems.

\subsection{Two-dimensional case}

As a first test, we use an artificial potential with randomly chosen coefficients, with unit cell $Y = [-1/2,1/2]^2$. More precisely,
\begin{align} \label{Model definition}
  V(\br) &= \sum_{(j,k) \in \mathbb{Z}^2} \widehat{V}_{j,k} \, \rme^{-2\ri\pi (j,k) \cdot \br} + \text{c.c.},
\end{align}
where the only nonzero coefficients $\widehat{V}_{j,k}$ were selected randomly. The results we present here correspond to the following set of coefficients: 
\begin{align*}
  \widehat{V}_{(0,0)} &= 0,\enskip
  \widehat{V}_{(0,1)} = 5.50+10.37 \, \ri,\enskip
  \widehat{V}_{(0,2)} = -12.50+11.52 \, \ri,\enskip\widehat{V}_{(1,-1)} = 11.70+10.21 \, \ri,\\
  \widehat{V}_{(1,-1)} &= 11.70+10.21 \, \ri,\enskip
  \widehat{V}_{(1,0)} = -10.19+10.04 \, \ri,\enskip
  \widehat{V}_{(1,1)} = -7.85+1.97 \, \ri,\enskip
  \widehat{V}_{(2,0)} = -2.02+4.61 \, \ri.
\end{align*}
With this potential, the first five bands are isolated from the higher
energy ones, so we choose $J = 5$.

We note that this system is entirely unphysical, and was chosen to be
a challenging test for our algorithm. On this system, it is hard to
get good initialization for the projection method, because the centers
and shapes of the Wannier functions are a priori unknown.

\subsubsection{Results obtained with the proposed algorithm}
\begin{figure}[h!]
  \centering
  \begin{subfigure}[b]{\textwidth}
    \centering
    \includegraphics[width=0.49\textwidth]{ours_noMV_frame}
    \includegraphics[width=0.49\textwidth]{ours_noMV_wannier}
    \caption{{With our algorithm, before} MV minimization.}
    \label{fig:ours_noMV}
  \end{subfigure}%

  \begin{subfigure}[b]{\textwidth}
    \centering
    \includegraphics[width=0.49\textwidth]{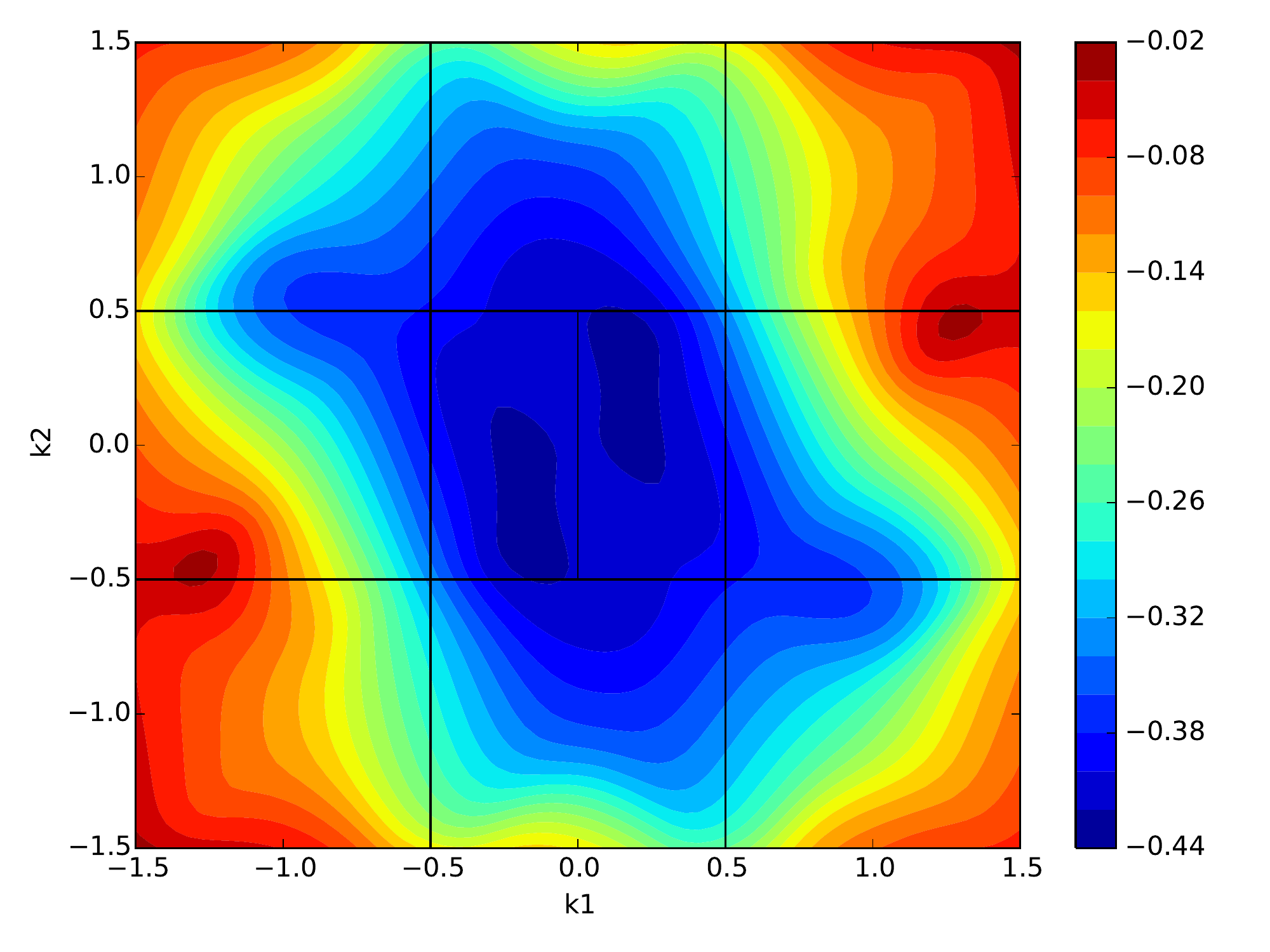}
    \includegraphics[width=0.49\textwidth]{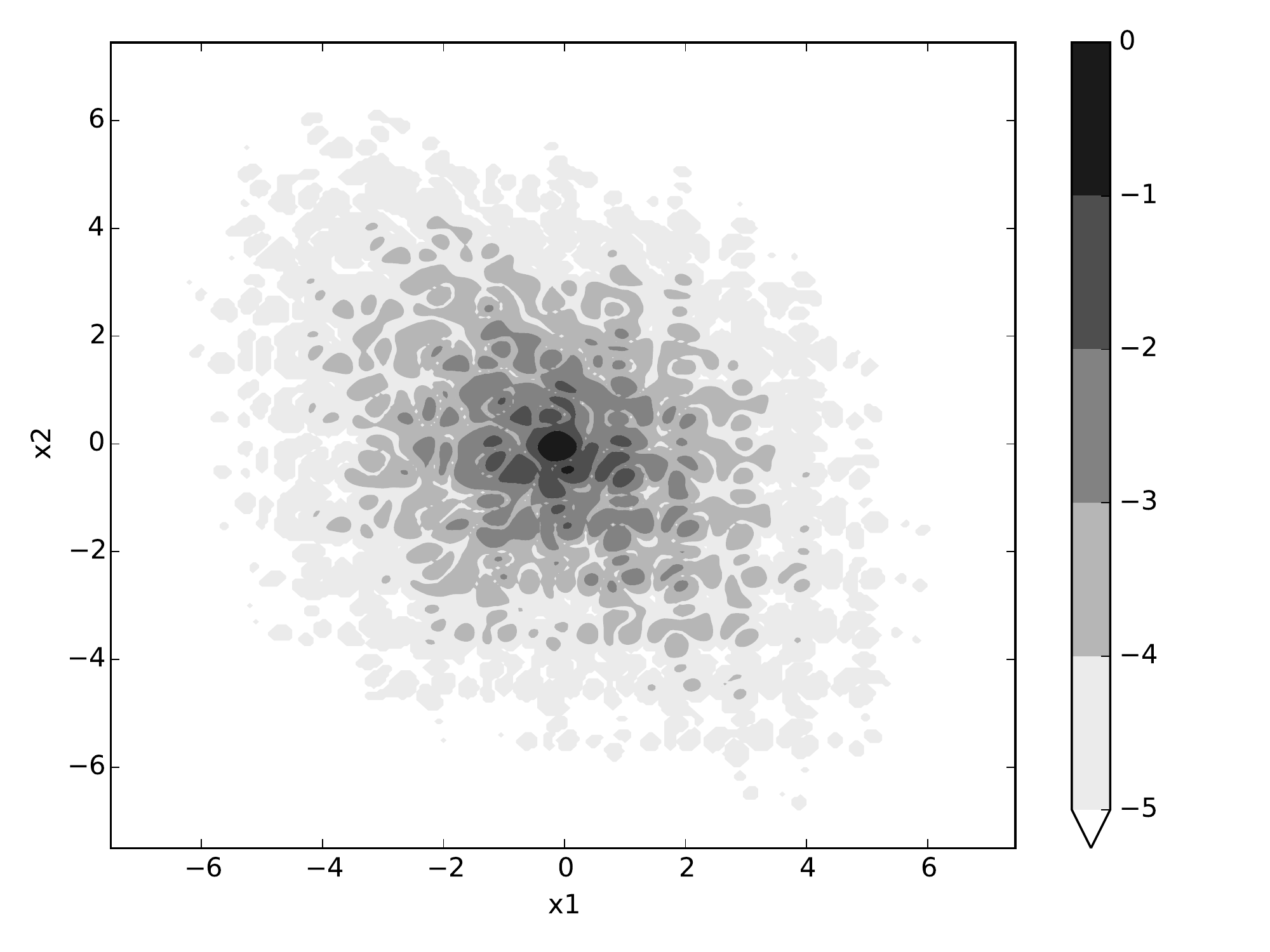}
    \caption{{With our algorithm, after} MV minimization.}
    \label{fig:ours_MV}
  \end{subfigure}%
  \caption{Result of our algorithm for the model specified by
    \eqref{Model definition}. As in Figure~\ref{fig:ex_alg}, we plot
    the average real part of the first component $u_{1,\bk}$ of the
    Bloch frame with $J=5$ as a function of $\bk$ (left),
    and corresponding Wannier function (right).}
  \label{fig:ours}
\end{figure}

\begin{figure}[h!]
  \centering
  \includegraphics[width=0.5\textwidth]{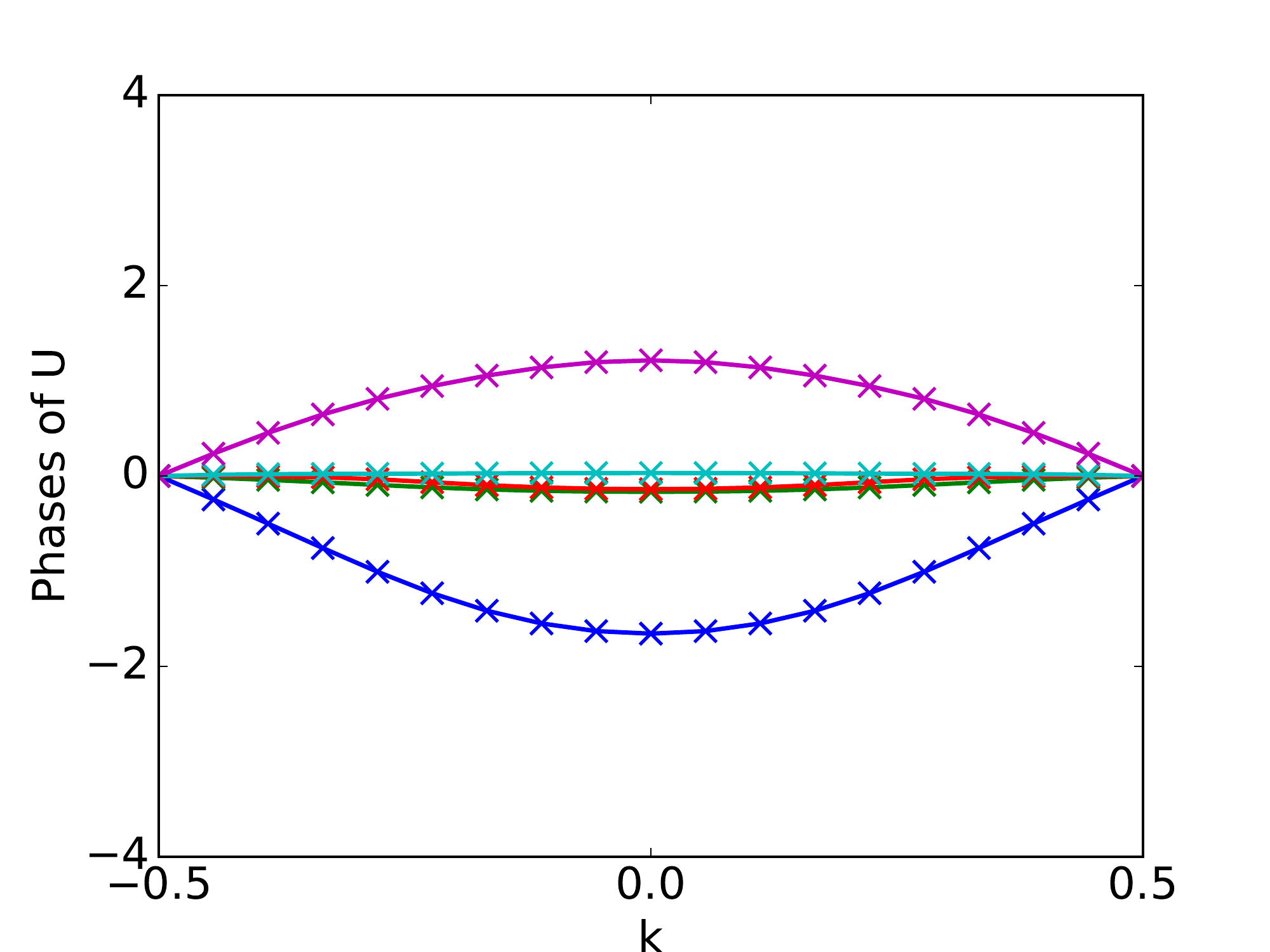}
  \caption{Phase of the obstruction matrix.}
  \label{fig:phase_2D}
\end{figure}

The result of our algorithm can be seen in Figure~\ref{fig:ours},
where we represent the first component $u_{1,\bk}$ of our Bloch frame
as a function of $\bk$, using the same methodology as in
Figure~\ref{fig:ex_alg}. We also plot the associated Wannier function
$w_{1}$ to check its spatial decay. On this very simple example, there
is no ambiguity in finding the logarithm on the right edge, because
the eigenvalues of the obstruction matrix never cross the value $-1$,
as can be seen in Figure~\ref{fig:phase_2D}. We therefore obtain a
continuous frame, which leads to localized Wannier functions. However,
there are slight discontinuities in the first derivative at the edges
$k_{1} = 1/2 + p$ for $p\in \Z$. This translates into a slow spatial
decay of the Wannier function in the $x_{1}$ direction.

To smooth the discontinuities of $\partial_{k_1} u_{n,k}$, we use the
MV procedure. The output of this minimization is given in
Figure~\ref{fig:ours_MV}. On this simple example, the main features of
the frame do not change, but some extra regularization occurs. This
removes the spatial tail of the Wannier function in the $x_1$
direction. We checked that the Wannier function we obtain presents an
exponential decay, provided the cut-off energy $E_{\rm c}$ and the
number of iterations of the MV algorithm are sufficiently large.

\subsubsection{Comparison with the projection method}

\begin{figure}[h!]
  \centering
  \begin{subfigure}[b]{\textwidth}
    \centering
    \includegraphics[width=0.49\textwidth]{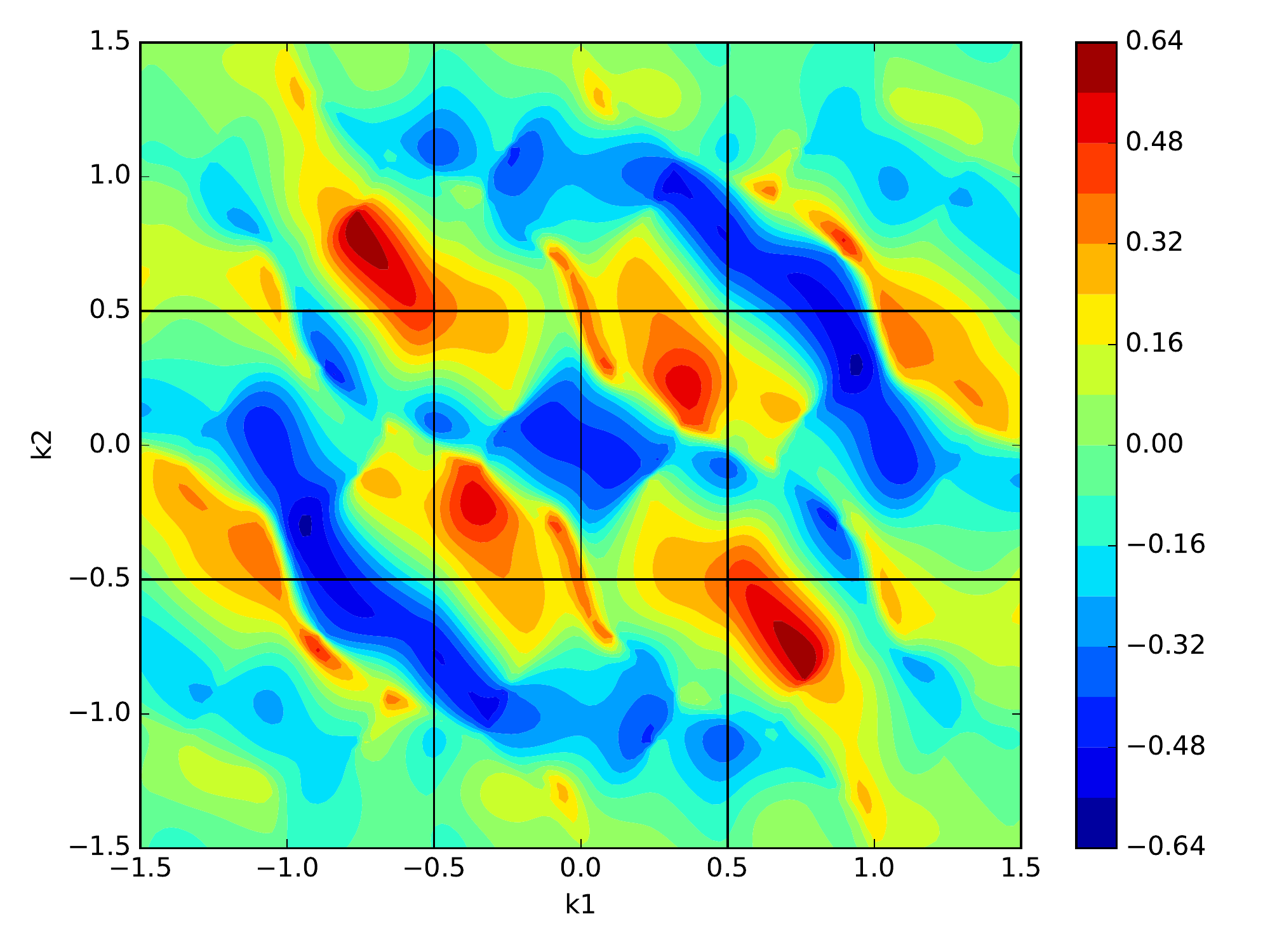}
    \includegraphics[width=0.49\textwidth]{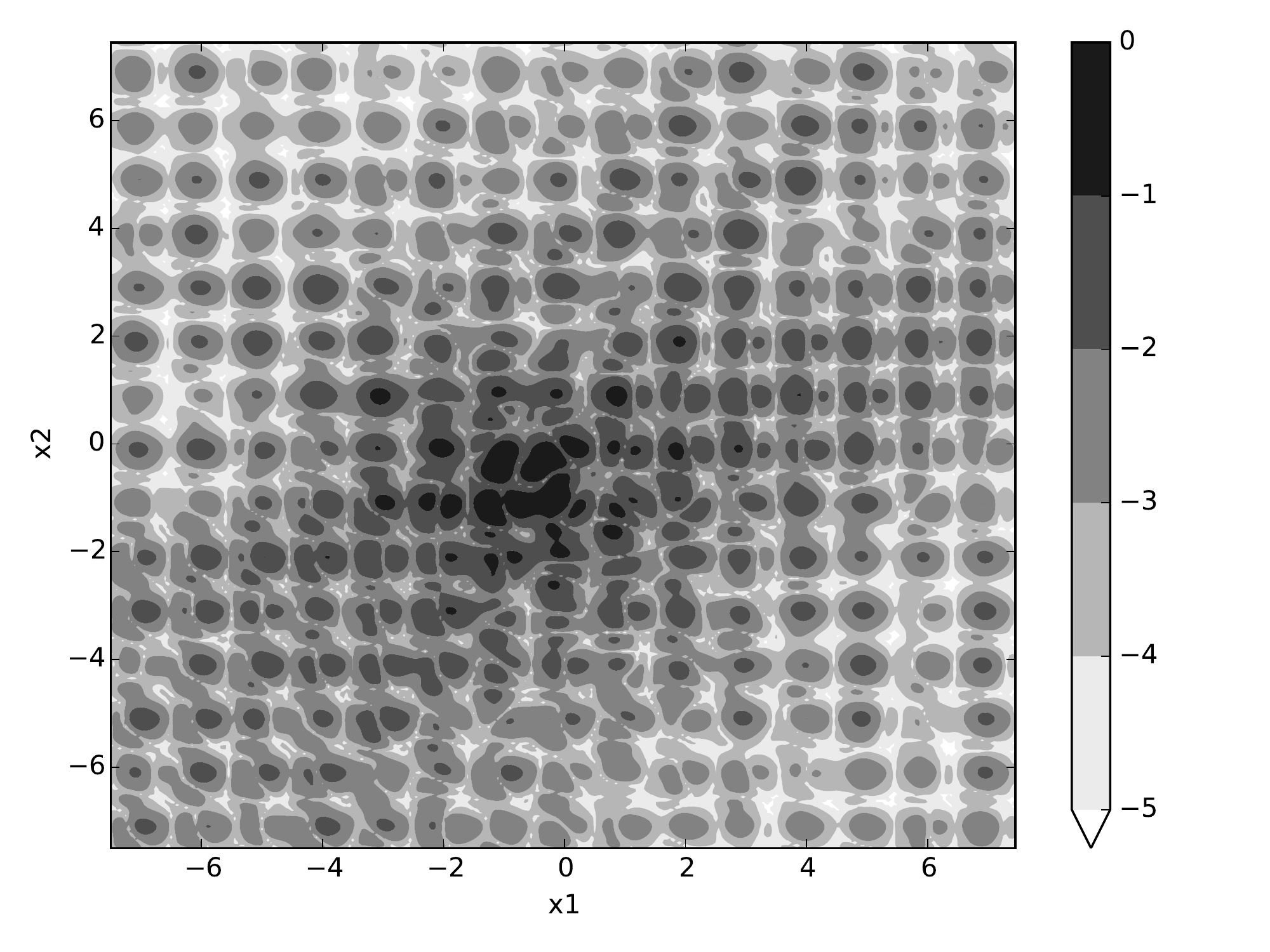}
    \caption{{With the projection method, before} MV minimization.}
    \label{fig:proj_noMV}
  \end{subfigure}%

  \begin{subfigure}[b]{\textwidth}
    \centering
    \includegraphics[width=0.49\textwidth]{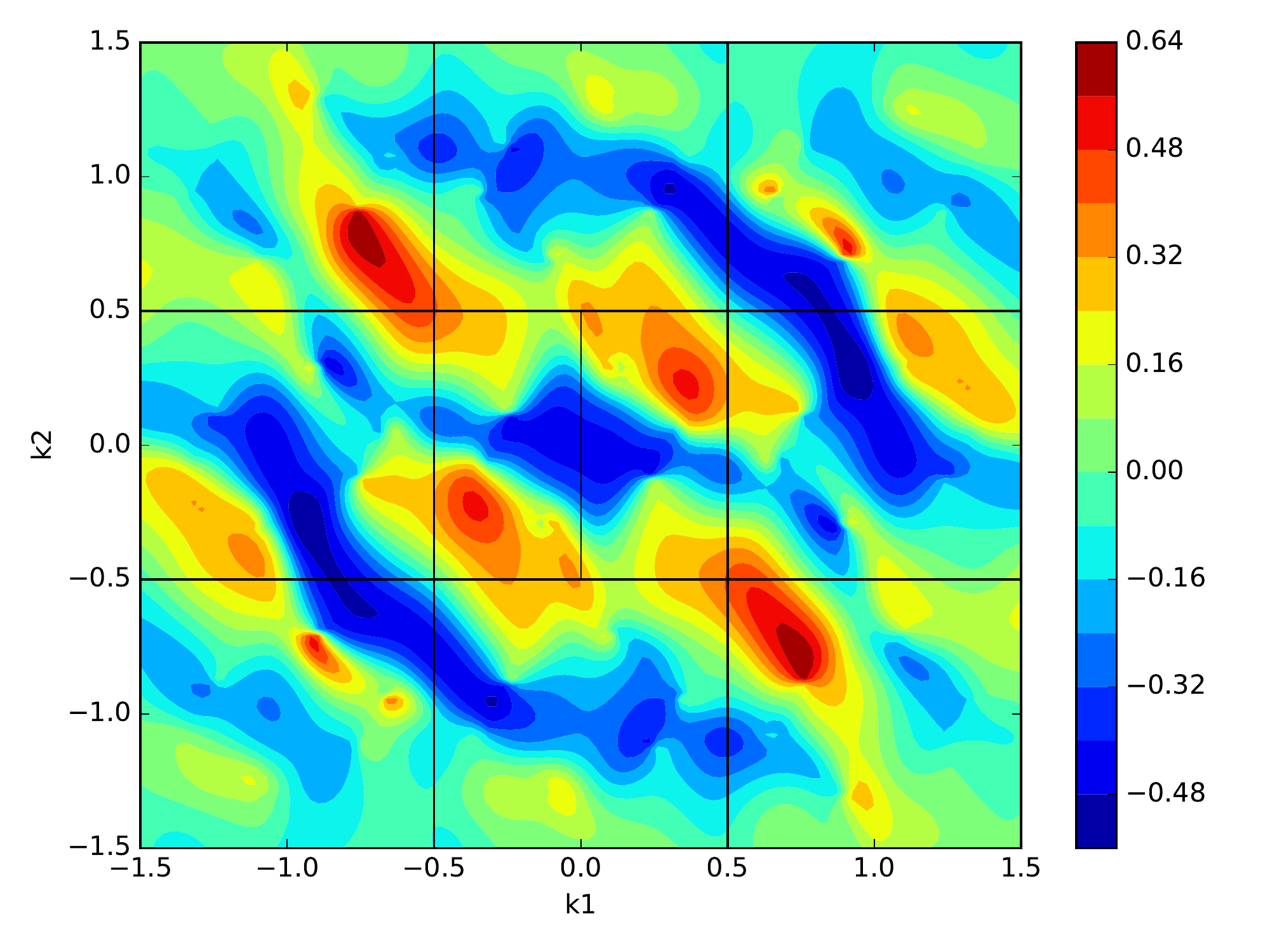}
    \includegraphics[width=0.49\textwidth]{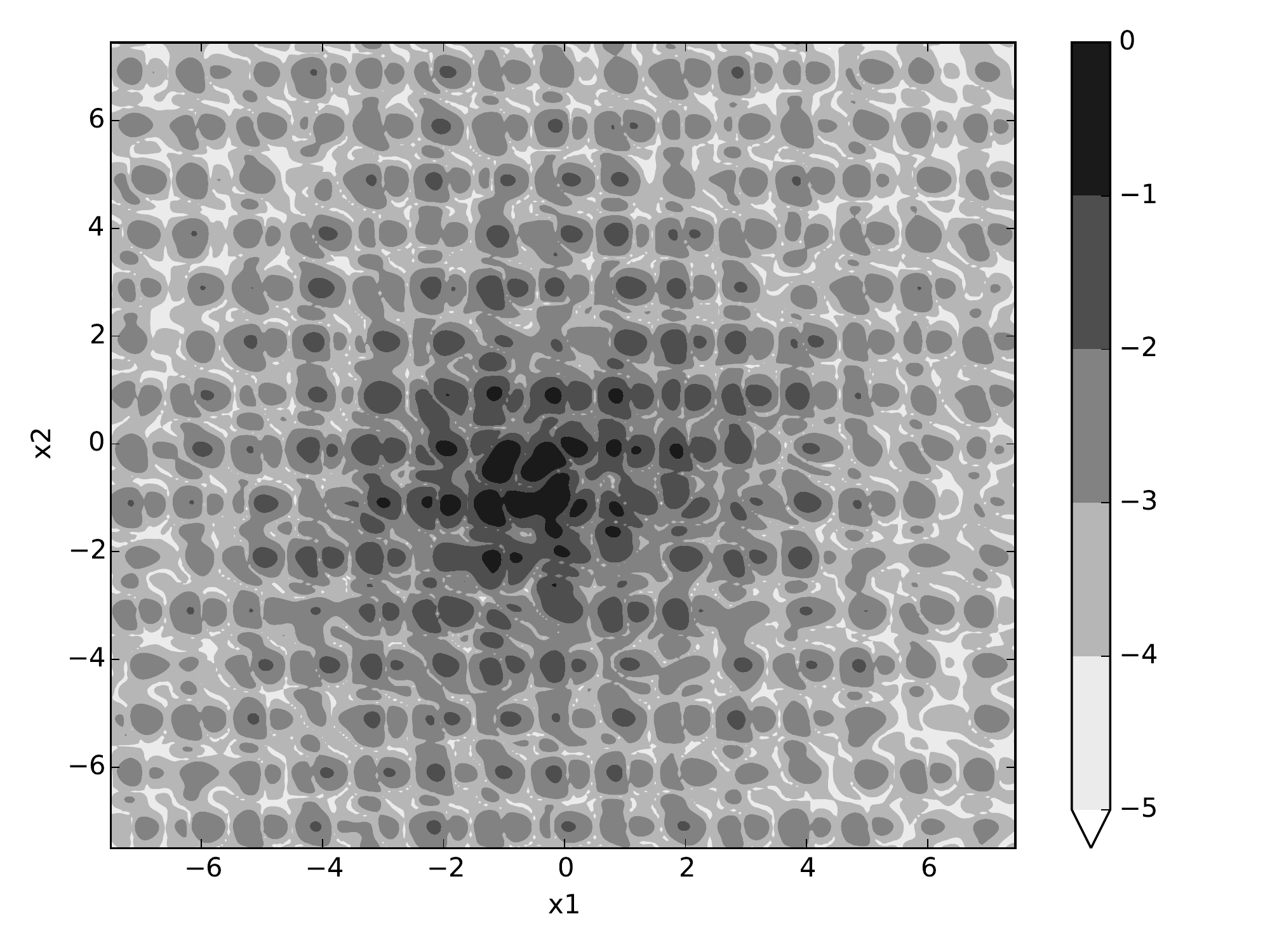}
    \caption{ {With the projection method, after} MV minimization.}
    \label{fig:proj_MV}
  \end{subfigure}%
  \caption{{Result of the projection method with an initial guess
      corresponding to Gaussian functions centered at random
      positions, for the same model and methodology as in
      Figure~\ref{fig:ours}.}}
  \label{fig:proj}
\end{figure}

We next compare the results of our algorithm to the ones obtained from
the projection method commonly used to initialize the MV algorithm
\cite{marzari1997maximally}. The initial guesses for the projection
method are $J$ Gaussian functions centered at random positions. The
width of these Gaussian functions is chosen to match approximately the
size of the localized Wannier functions we obtained with our
method. Figure~\ref{fig:proj} represents the Bloch frame obtained by
one example of such a projection, before and after running the MV
minimization.

Although the frame is smooth in most regions, and obeys by
construction the symmetry properties, we can see isolated points of
discontinuity, which visually manifest themselves as rapid changes of
color. Moreover, the MV {minimization procedure} fails to further
localize the initial guess, a situation the authors of
\cite{marzari1997maximally} refer to as ``false local minima''.

The behavior around the isolated points of singularity can be understood as follows. We denote by $v_\bk$ the periodic parts of the Bloch transform of the trial Wannier functions used as an initial guess. When $J=1$, {according to the projection method}, the corresponding frame is given by
\begin{align*}
  u_\bk &= \frac{P(\bk) v_\bk}{\norm{P(\bk) v_\bk}} = \frac{\lela w_\bk, v_\bk \rira w_\bk}{\abs{\lela w_\bk, v_\bk \rira}} = \rme^{\ri\theta(\bk)} w_\bk,
\end{align*}
where $w_\bk$ is any vector in the one-dimensional vector space
$\Ran P(\bk)$, which can be chosen locally continuous, and
$\theta(\bk)$ is the phase of $\lela w_\bk, v_\bk \rira$.  This phase
is well-defined and continuous as long as $\lela w_\bk, v_\bk\rira$ is
non-zero. The condition $\lela w_\bk, v_\bk\rira = 0$ is a set of two
real equations, which generically admits point solutions in
dimension~2, and solutions on a one-dimensional manifold in
dimension~3. {The frame $u$ is singular at these values.}
The analysis when $J>1$ is more involved but yields the same conclusions:
the frame is singular when the overlap matrix
$(P(\bk) v_\bk)^{*}(P(\bk) v_\bk)$ is not invertible, which generically is a
set of two equations. By formal analogy with similar phenomena in
other application fields, we call such singular sets
\emph{vortices}. Vortices are points in dimension~2, and lines in
dimension~3. A more quantitative information about the behavior of
frames around vortices can be obtained by introducing the notion of
{eigenspace vorticity} \cite{MonacoPanati2014}, which generalizes the
{pseudospin winding number} appearing in the literature on graphene
\cite{Park2011}.

As can be seen in Figure~\ref{fig:proj_MV}, the MV minimization fails
to resolve these vortices: on a finite mesh, these pathological
singularities appear as local minima of the functional. This is
because the MV algorithm was designed with a continuous input in
mind. The continuity ensures that, when the mesh is fine enough, the
matrix elements $M_{nn, \bk, \bb}$, defined in \eqref{eq:Mnmkb} and
used crucially in the algorithm, are close to $1$. This allows to
define their logarithm unambiguously. In the presence of vortices,
this quantity is not close to $1$, even on fine meshes, and the
algorithm is not well-defined.

In order to confirm that the undesirable behavior of the MV algorithm
arises from the existence of initial vortices, we consider the case of
a single band. The dominant contribution to the MV functional is given
by \cite{panati2013bloch}
\begin{align}
  \label{eq:omega_tilde}
  \tilde \Omega[u] &= \int_{\BZ}\int_{Y}|\nabla_{\bk}
                  u_{\bk}(\br)|^{2} d\br\, d\bk.
\end{align}
Consider a model vortex at $\bk = 0$ (without loss of generality), and let
$\bk = (\rho \cos \theta, \rho \sin \theta)$. We consider the model vortex
$u_\bk = \rme^{\ri \theta} u_0$ in a neighborhood of $0$. Then
$\nabla_{\bk} u_{\bk}(\br)$ has a $1/\rho$ singularity at
$0$. Therefore, on a $N \times N$ grid, we can expect that the
MV functional \eqref{eq:omega_tilde} diverges like
$\log(N)$. In dimension~3, where vortices are lines,
$\nabla_\bk u_\bk$ also presents a $1/\rho$ divergence, where $\rho$
is the distance from $\bk$ to the line of singularities. This also
yields a logarithmic divergence of the functional. The reasoning is
similar for $J > 1$.

Such a divergence is observed in our simulations: for the simple case
presented here, the values of the MV functional at convergence of the
algorithm on a $20 \times 20$, $40 \times 40$, $80 \times 80$ and
$160 \times 160$ grid respectively, are 3.7, 5.7, 7.9 and 9.6. This is
consistent with a logarithmic divergence. However, since the
divergence is relatively mild (logarithmic only), when the mesh is
sufficiently coarse, these vortices are found not to impact the
algorithm, which converges to non-singular minima. For instance, the
MV algorithm on the example considered here converges to a smooth
minimum on a $7 \times 7$ grid in $\bk$-space, but stalls without
removing the vortices on $10 \times 10$ grids and finer.

This explains the observation in
\cite[Section~IV.D.2]{marzari1997maximally}, where the MV minimization
is reported to sometimes fail for random initial guesses. The authors
state that ``this problem [false local minima] is not associated with
the presence of a large number of bands, but instead with the use of
fine $\bk$-point meshes'', consistent with our observation that the
energy contribution of a vortex diverges logarithmically with the grid
spacing. They also ``never observed the system to become trapped in a
false local minimum when starting from reasonable trial projection
functions''. In the situation we consider here (a random potential),
``reasonable trial projection functions'' are hard to devise in
advance. We have found vortices to occur for a large class of initial
inputs, although of course not for ones close to real
maximally-localized Wannier functions, where the overlap matrix
$(P(\bk) v_\bk)^{*}(P(\bk) v_\bk)$ has its eigenvalues bounded away
from zero.

By contrast, our algorithm always succeeded in constructing a good
frame, even on more complicated systems where the eigenvalues of the
obstruction matrix cross the value $-1$, as in
Figure~\ref{fig:2D_log}.  In order to investigate whether the type of
problematic eigenvalue collisions of the obstruction matrix described
in Section \ref{sec:logpb} was possible, we played with the Fourier
coefficients~$\widehat{V}$, but were not able to see any. In fact, the
only way we could make the eigenvalues of the obstruction matrix
collide is by closing the gap
$\inf_{\bk \in \BZ}\left(\eps_{J+1}(\bk) - \eps_{J}(\bk)\right)$, which
violates the hypotheses of our problem (we only consider isolated
bands). We do not know if it is fundamentally impossible for
eigenvalues of the obstruction matrix to cross at a finite gap (for
topological reasons, for instance), or if it is simply an exceptional
situation which we failed to encounter in our tests.

\subsection{Three-dimensional case}

The previous two-dimensional example with a random potential was artificial and not
representative of real systems. To test our algorithm on more
realistic cases, we use the simple effective one-body potentials of \cite{cohen1966band} in a zincblende structure, discretized on a plane-wave basis. These pseudopotentials were chosen to obtain representative band structures at minimal cost. In this framework, the potential $V$ reads
\begin{align*}
  V(\br) &= \sum_{\bK \in \mathcal R^{\ast}} \widehat{V}(K) \, \rme^{-\ri \bK \cdot \br},\\
  \widehat{V}(\bK) &= V_{\rm S}(|\bK|) \cos(\bK \cdot \tau) + \ri V_{\rm A}(|\bK|) \sin(\bK \cdot \tau),
\end{align*}
where $\tau = a(1,1,1)/8$ and $a$ is the lattice constant of the zincblende structure. The form factors $V_{\rm S}$ and $V_{\rm A}$, as well as the parameters $a$ for various compounds, are tabulated in~\cite{cohen1966band}.

\begin{figure}[h!]
  \centering
  \begin{subfigure}[b]{0.49\textwidth}
    \centering
    \includegraphics[width=\textwidth]{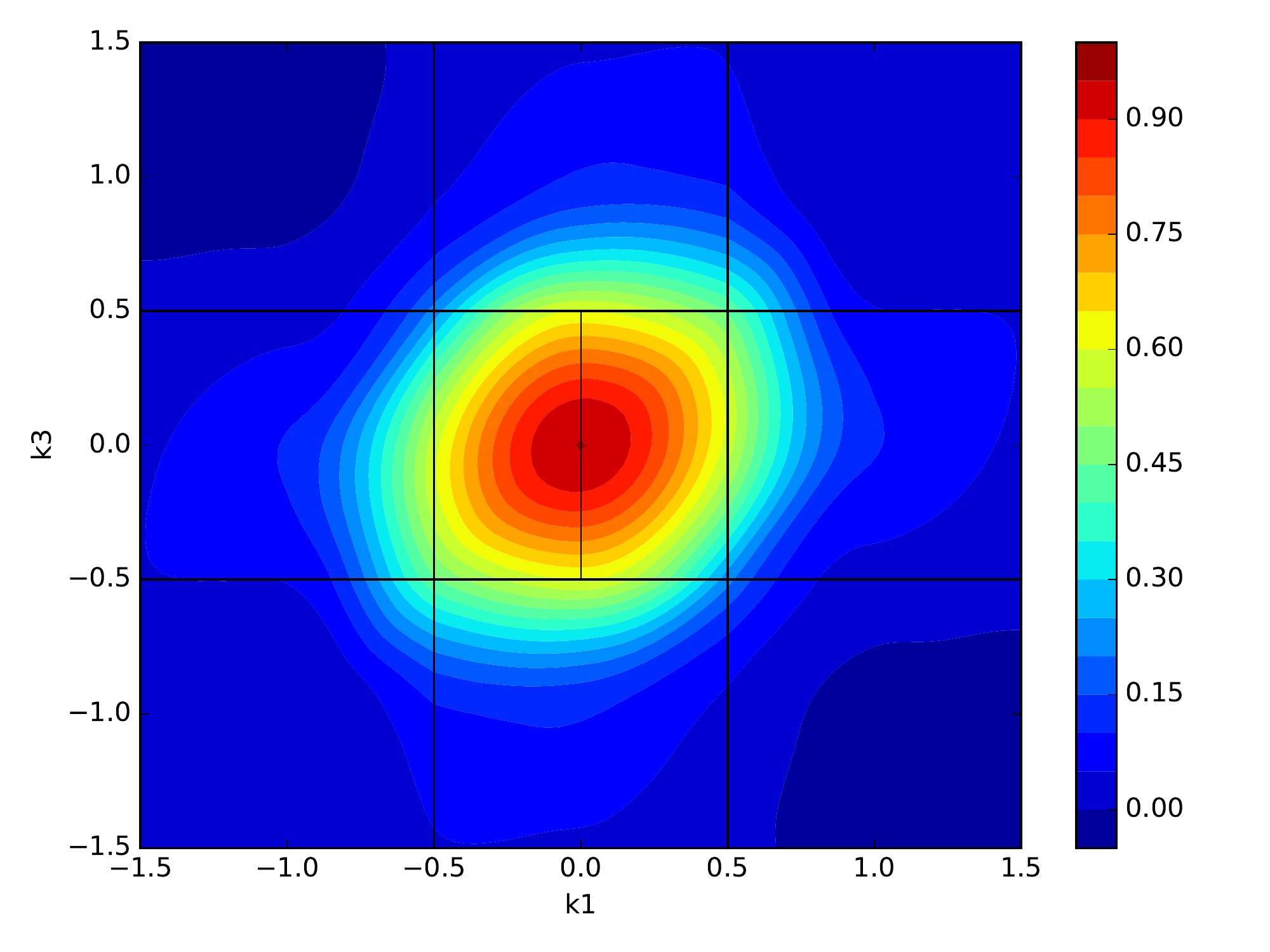}
    \caption{Before MV {minimization}}
  \end{subfigure}%
  \begin{subfigure}[b]{0.49\textwidth}
    \centering
    \includegraphics[width=\textwidth]{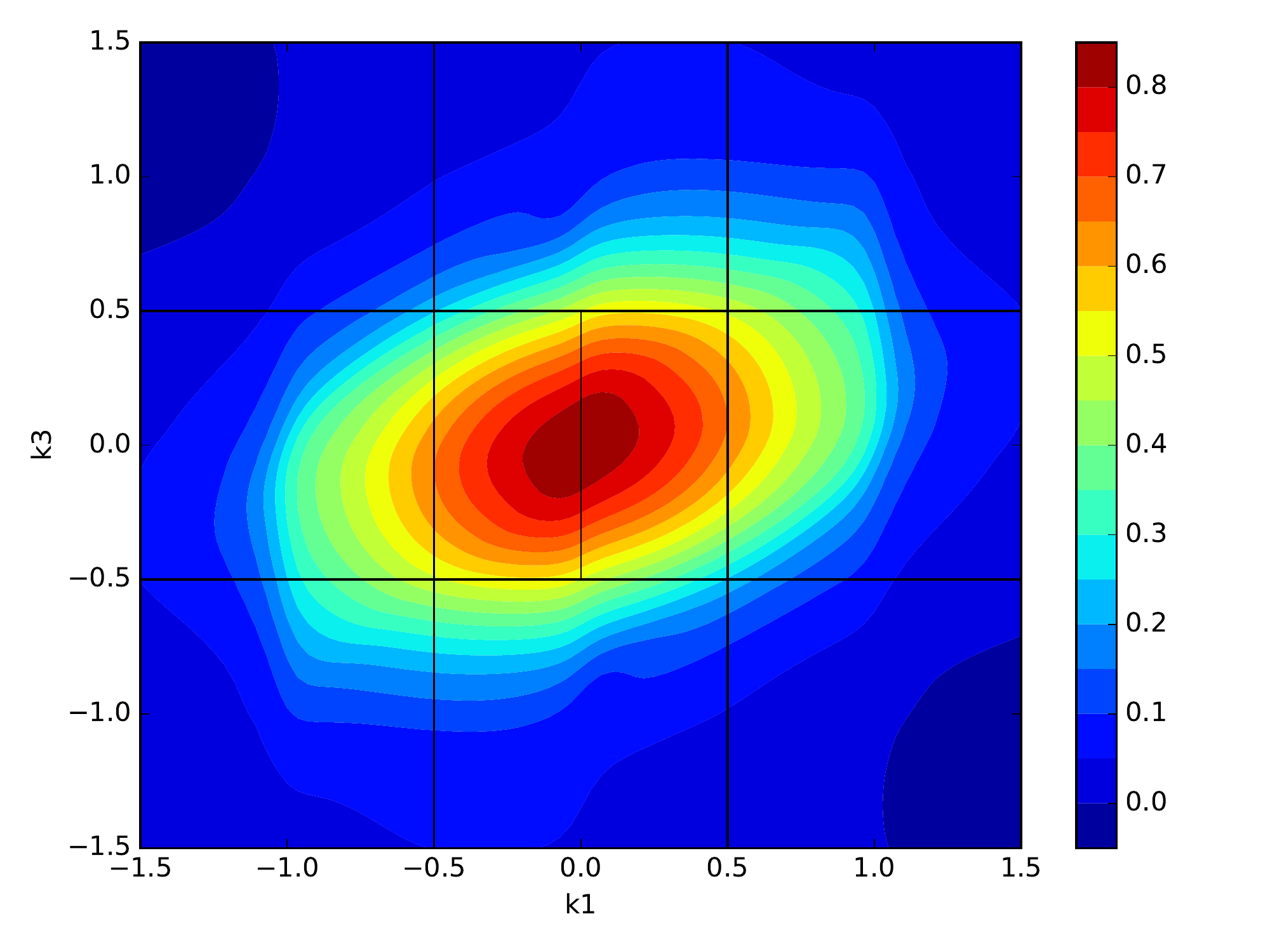}
    \caption{After MV {minimization}}
  \end{subfigure}%
  \caption{Average real part of the {first component of the} Bloch frame
    obtained by our method on the cut plane $k_{2} = 0$, before (left)
    and after (right) the MV algorithm, on a $24 \times 24 \times 24$
    grid.}
  \label{fig:3D_ours}
\end{figure}

We first test our algorithm on Silicon. We study the first four bands,
which are isolated from the others. As in the 2D case, our algorithm was able to produce a continuous
frame, as seen in Figure~\ref{fig:3D_ours}, where we plot again the
average real part of the first component of the Bloch frame, on the cut plane
$k_{2} = 0$.

\begin{figure}[h!]
  \centering
  \begin{subfigure}[b]{0.6\textwidth}
    \centering
    \includegraphics[width=\textwidth]{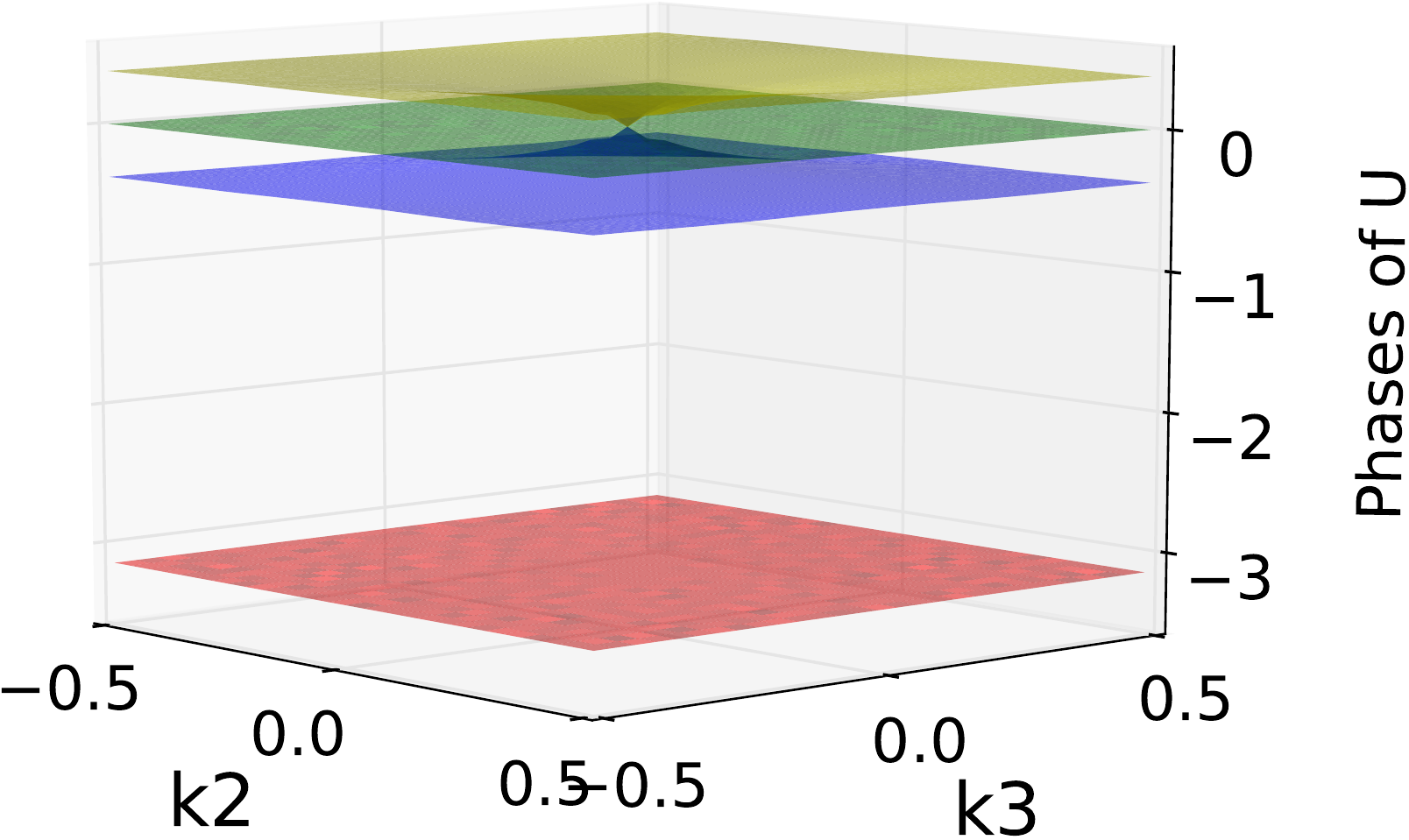}
    \caption{Before fixing the corners.}
    \label{fig:phase_3D_before_corners}
  \end{subfigure}%
  \vspace{1cm}
  \begin{subfigure}[b]{0.4\textwidth}
    \centering
    \includegraphics[width=\textwidth]{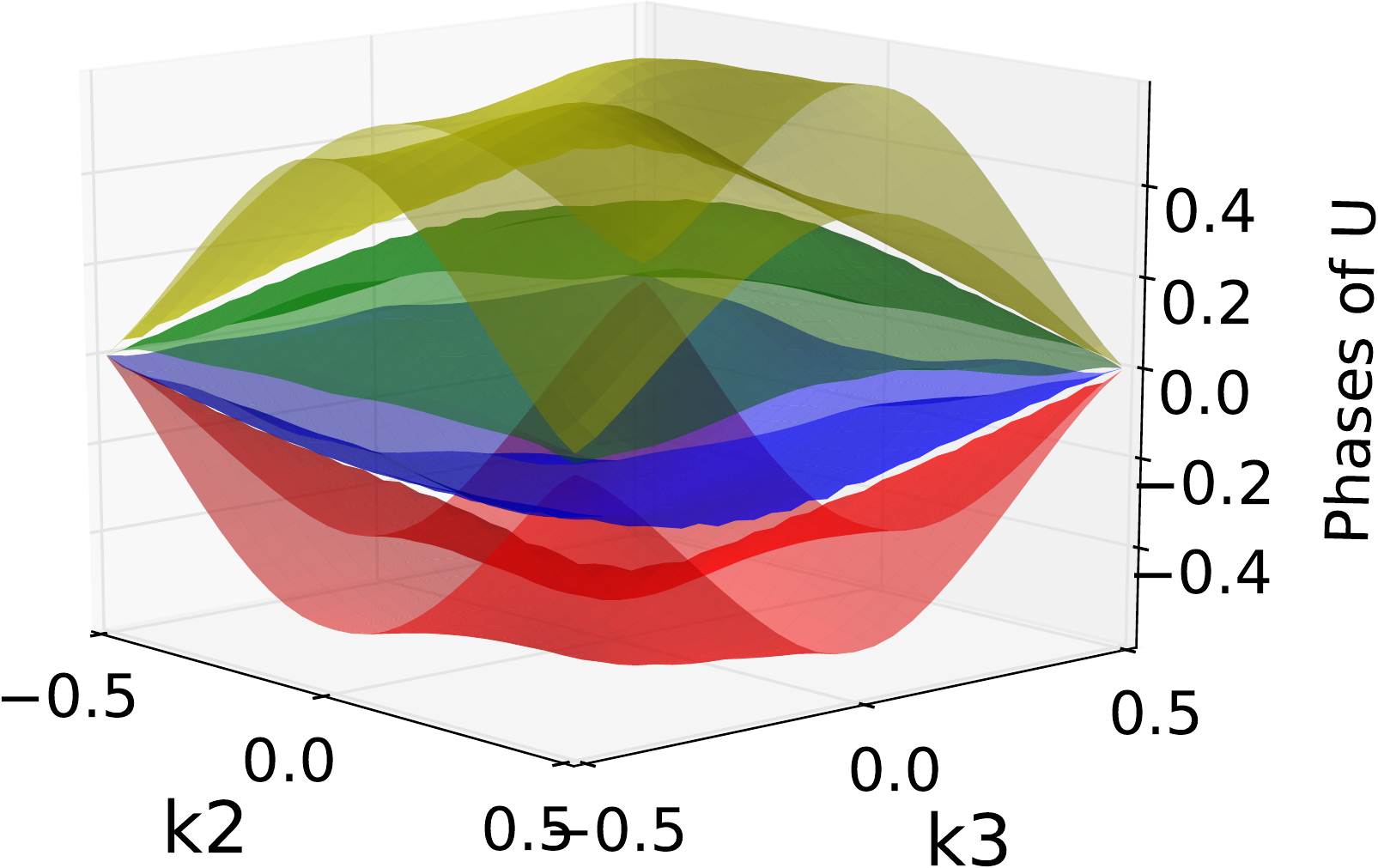}
    \caption{Before fixing the edges.}
    \label{fig:phase_3D_before_edges}
  \end{subfigure}\hspace{1cm}
  \begin{subfigure}[b]{0.4\textwidth}
    \centering
    \includegraphics[width=\textwidth]{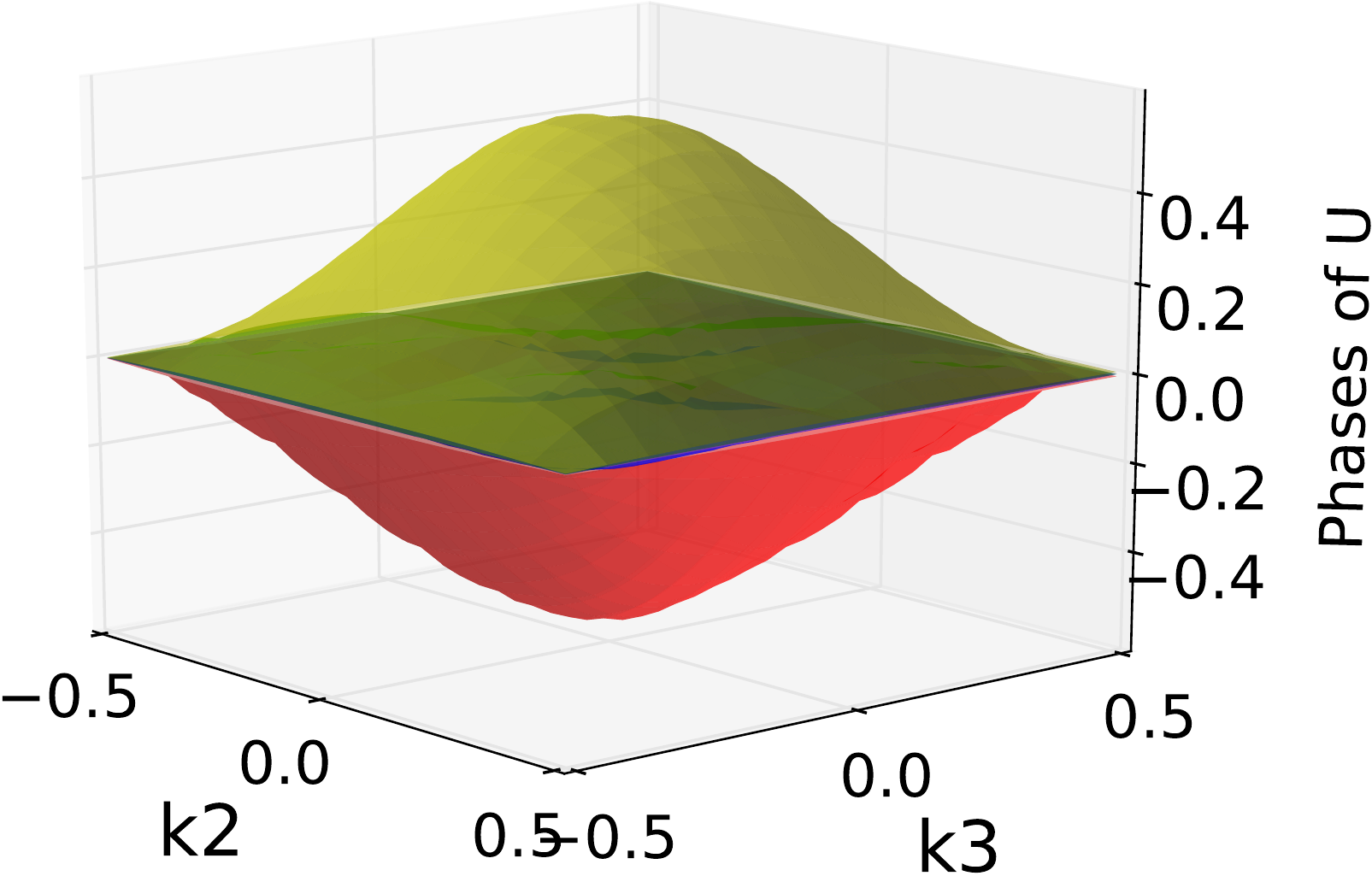}
    \caption{After fixing the edges.}
    \label{fig:phase_3D}
  \end{subfigure}%
  \caption{Phase of the obstruction matrix for the Silicon example.}
  \label{fig:phase_si}
\end{figure}

Figure \ref{fig:phase_si} represents the $J=4$ different phases
of the {eigenvalues of the} obstruction matrix $\Uobs(k_{2}, k_{3})$ in the three stages of
the algorithm: fixing the corners, the edges and the face.

Before fixing the corners, there is a conical intersection in the
eigenvalues at $(0,0)$. This is a non-generic case, presumably due to
eigenvalue degeneracy at $\Gamma$ in the model, itself related to a
particular symmetry of the potential. However, this intersection is
harmless, as it connects eigenvalues with the same phase
determination. After fixing the corners, the conical intersection moves to the
corners, and the phase of all the eigenvalues has a maximum of about
$0.6$, well below $\pi$. Fixing the edges is not necessary in this
case.

\begin{figure}[h!]
  \centering
  \includegraphics[width=\textwidth]{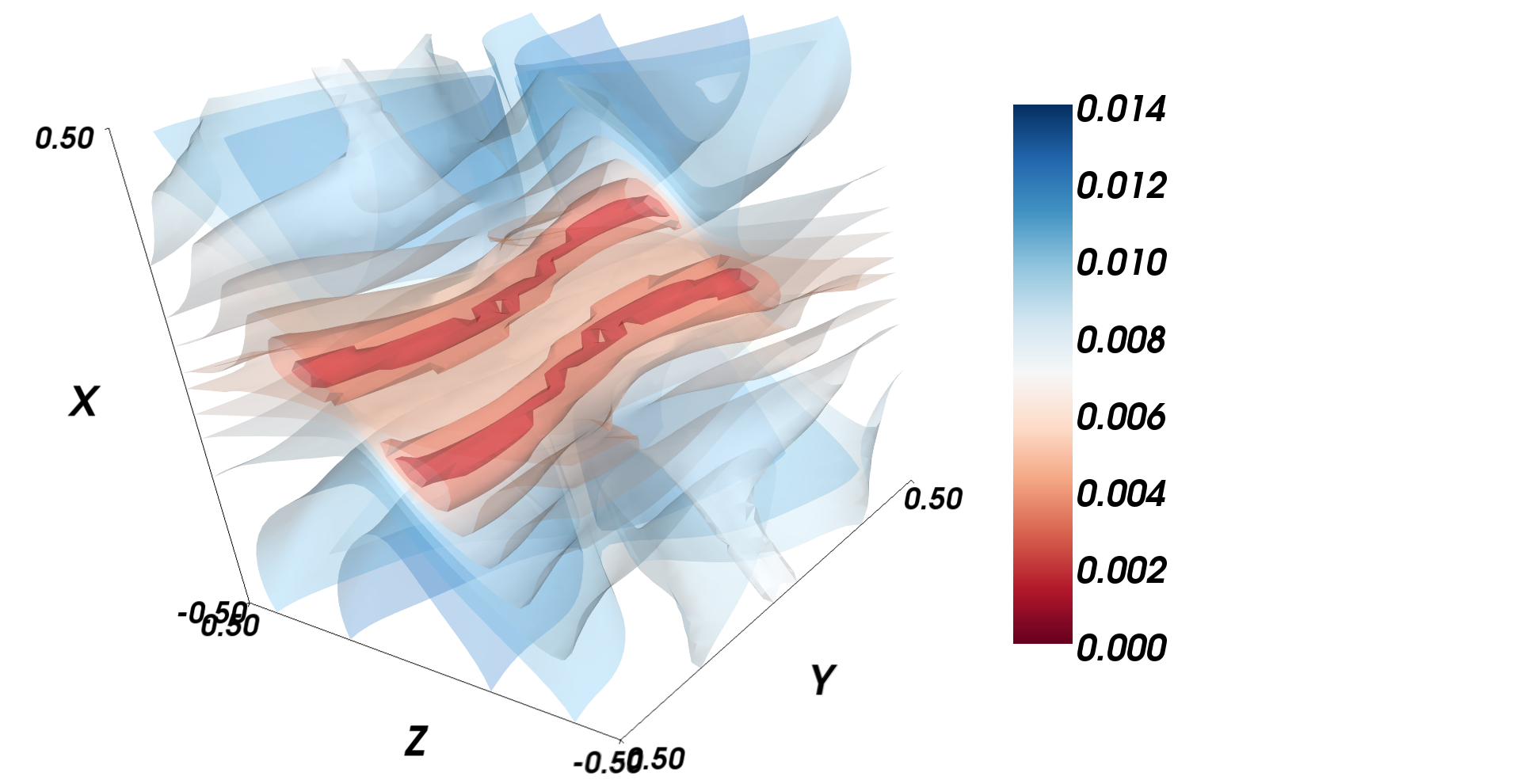}
  \caption{Smallest eigenvalue of the overlap matrix
    \eqref{proj_grammatrix}. The lowest isosurface corresponds to the
    region where this matrix is nearly indefinite, and forms a vortex
    line.}
  \label{fig:mineig}
\end{figure}

\begin{figure}[h!]
  \centering
  \begin{subfigure}[b]{0.49\textwidth}
    \centering
    \includegraphics[width=\textwidth]{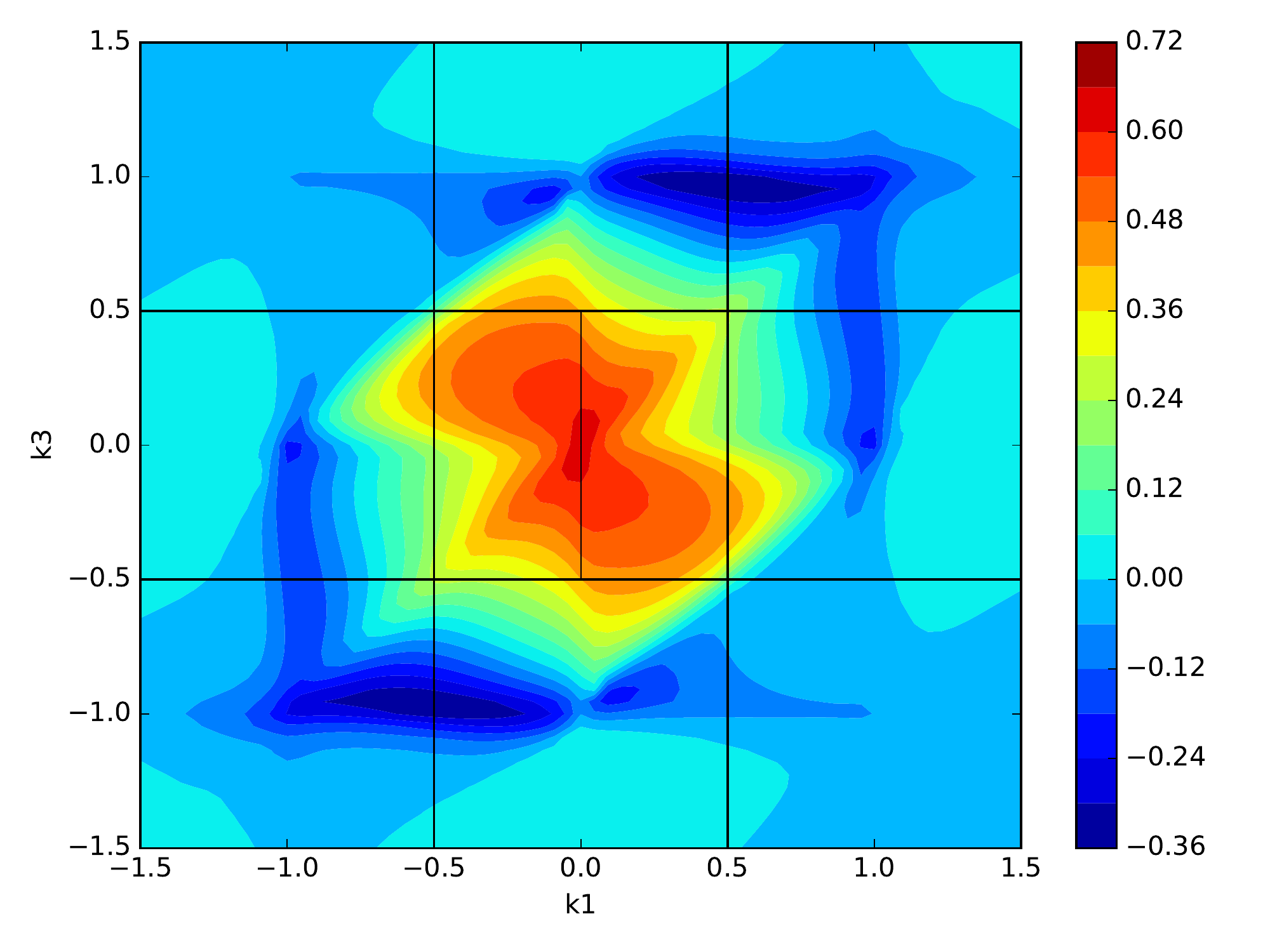}
    \caption{Before MV {minimization}}
    \label{fig:3D_before}
  \end{subfigure}%
  \begin{subfigure}[b]{0.49\textwidth}
    \centering
    \includegraphics[width=\textwidth]{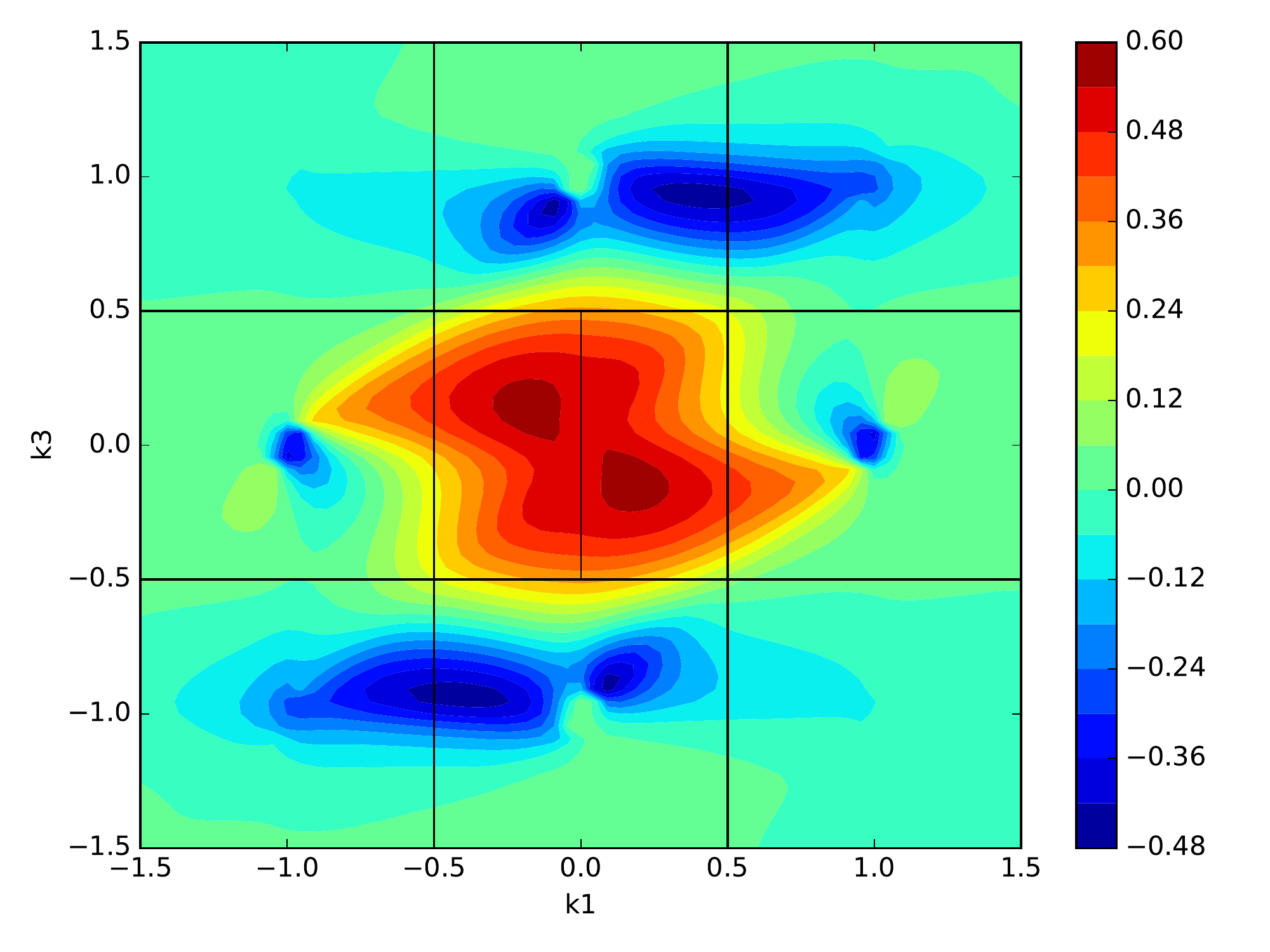}
    \caption{After MV {minimization}}
    \label{fig:3D_after}
  \end{subfigure}%
  \caption{Average real part of the {first component of the} Bloch frame obtained by the
    projection method on the cut plane $k_{2} = 0$, before (left) and
    after (right) the MV algorithm, on a
    $24 \times 24 \times 24$ grid. The MV algorithm is able to smooth out
    the general features of the frame, but not to remove the
    vortices.}
  \label{fig:3D}
\end{figure}

To compare our method to the projection method, as in the 2D case, we
use as initial Wannier functions Gaussians centered on random
points. The projection method yields a continuous frame as long as the
overlap matrix
\begin{align}
  \label{proj_grammatrix}
  O(\bk) &= \left(P(\bk)v_{\bk}\right)^{*} \left(P(\bk)v_{\bk}\right)
\end{align}
is positive-definite for all $\bk \in \BZ$. As can be seen in Figure
\ref{fig:mineig}, this matrix becomes indefinite on vortex lines. This
yields discontinuous frames, as illustrated in Figure
\ref{fig:3D_before}. In our tests with random positions of the
Gaussians, this occurred about half the time. As in the 2D case, the
MV algorithm is unable to remove these vortices when the sampling of
the Brillouin zone is fine enough. In this example, the MV algorithm
was able to remove the vortices on grids of size up to $20 \times 20
\times 20$, but not on finer grids. As expected, when the Wannier centers are
selected appropriately, the MV algorithm converges to localized Wannier functions.

Silicon represents an easy test case because it possesses a relatively
large direct gap between the valence and conduction bands. This makes
the projector on the valence bands a smooth function of $\bk$, which
in turn produces well-localized Wannier functions. We also tried our
algorithm on more complicated semiconductors such as indium arsenide,
still using the pseudopotentials of \cite{cohen1966band}. Indium
arsenide has a very small direct gap, resulting in sharp variations in
the Brillouin zone near the band edges which require a fine sampling
of the Brillouin zone to resolve accurately the band structure. Although the obstruction
phases of Figure \ref{fig:phase_si} are more rugged in this case,
their amplitude at the last step is still less than about $0.6$, well
below $\pi$, and our algorithm has no problem distinguishing the
bands to fix the phases.

\subsection{Interface with Wannier90 and tests on DFT systems}
We have implemented our algorithm in a way that is compatible with the
standard code Wannier90 \cite{mostofi2008wannier90,
  mostofi2014updated}. To that end, we note that our method, while
presented here with frames, can also be implemented in the Wannier90
paradigm where one computes a fixed set of electronic orbitals
$u_{n \bk}$ for each $\bk$-point, and then finds a set $U_{\bk}$ of
unitaries from which the final frame is constructed as
$u'_{\bk} = u_{\bk} U_{\bk}$. The advantage of this approach for the
MV algorithm is that only the low-dimensional unknowns $U_{\bk}$ have
to be optimized, and that the only inputs from electronic structure
codes are the overlap matrices
\begin{align}
  \label{eq:Mnmkb}
  M_{nm, \bk, \bb} &= \lela u_{n \bk+\bb}, u_{m \bk}\rira,
\end{align}
where $\bb$ runs over nearest neighbors of $0$ on the $\bk$-space
mesh. This separates the computation of the $u_{n,\bk}$ from the
computation of Wannier functions, and facilitates the creation of
independent libraries. Similarly to the traditional MV minimization,
in our algorithm we only need as input the overlap matrices
$M_{nm, \bk, \bb}$ (\texttt{.mmn} file) to compute the final frame
$u'_{\bk} = u_{\bk} U_{\bk}$ and output the unitary matrix $U_{\bk}$
at each $\bk$-point (\texttt{.amn} file). Because this format does not
explicitly account for the time-reversal symmetry, it is more
convenient to use a modified version of our algorithm (see Appendix),
which does not use the time-reversal symmetry. The resulting code is
available at \url{https://github.com/antoine-levitt/wannier} and can
readily be inserted in any workflow using Wannier90.

\begin{figure}[h!]
  \centering
  \includegraphics[width=0.6\textwidth]{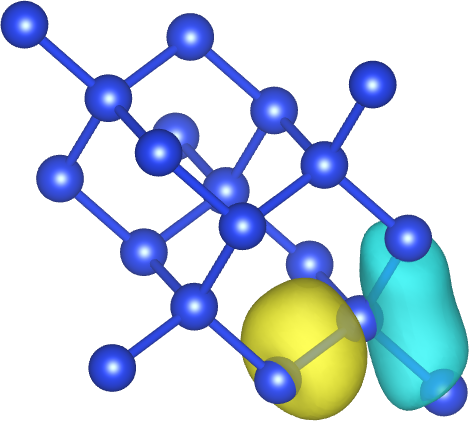}
  \caption{One of the four Wannier functions produced by our algorithm
    for Silicon. The positive and negative isovalues of the Wannier
    function are plotted, at a level equal to 20\% of the maximal
    absolute value of the function. The Wannier functions obtained by
    our algorithm are localized, and appear to be close to
    combinations of the four bond-centered maximally-localized Wannier
    functions.}
  \label{fig:wannier_visu}
\end{figure}

We have tested our algorithm on bulk Silicon computed by the code
Quantum Espresso \cite{giannozzi2009quantum} using the PBE
exchange-correlation functional. Our conclusions are the same as those
of the previous section using an effective potential approach. The
phases of the obstruction matrix are similar to those of Figure
\ref{fig:phase_si}, and our algorithm produces a continuous frame. The
corresponding Wannier functions are localized, as can be seen on
Figure \ref{fig:wannier_visu}.

\begin{figure}[h!]
  \centering
  \begin{subfigure}[b]{0.49\textwidth}
    \centering
    \includegraphics[width=\textwidth]{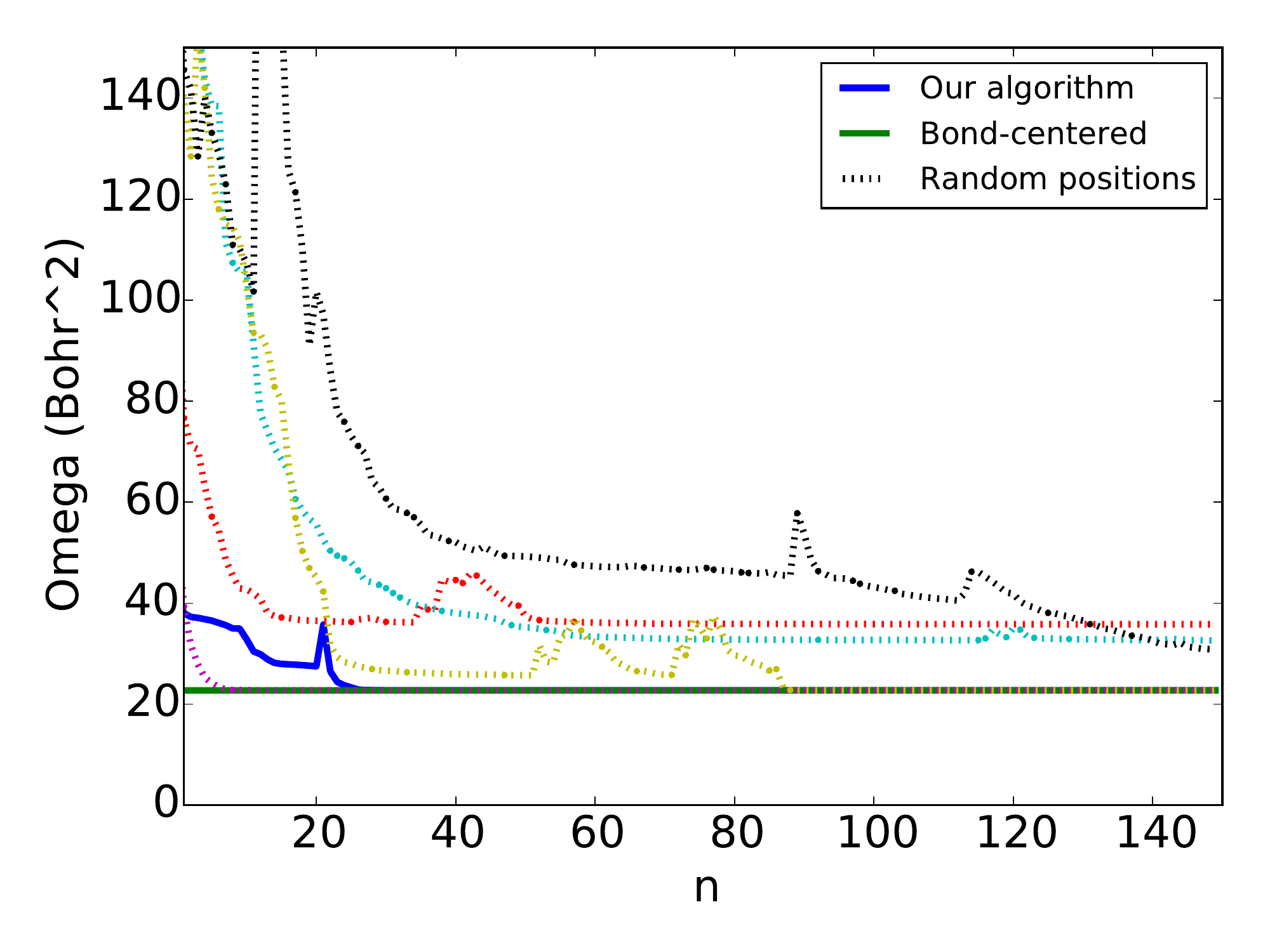}
    \caption{$4 \times 4 \times 4$ mesh}
  \end{subfigure}%
  \begin{subfigure}[b]{0.49\textwidth}
    \centering
    \includegraphics[width=\textwidth]{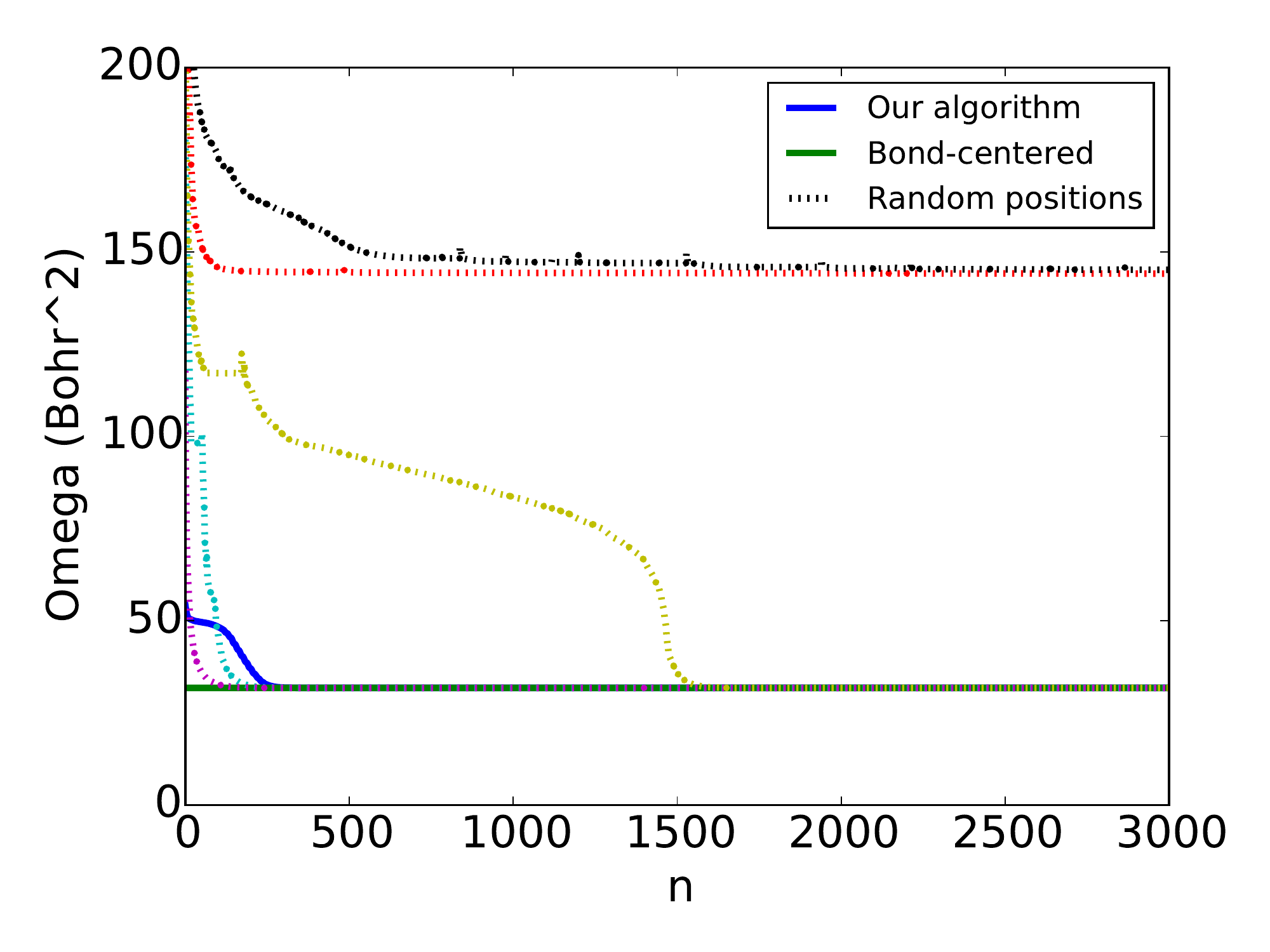}
    \caption{$30 \times 30 \times 30$ mesh}
  \end{subfigure}%
  \caption{Spread as a function of the number of iterations, for a
    coarse and a fine $\bk$-point mesh. We use as initial guess (1)
    the one produced by our algorithm; (2) the projection method with
    bond-centered s-type orbitals; (3) the projection method with
    s-type orbitals with random centers (five different realizations).
    On the left panel, the black and light blue curves eventually
    converged to the global minimum after about 500 iterations, but
    the red one converged to a local minimum to machine precision. The
    red and black curves on the right panel seem to have converged to
    a local minimum, but the gradient is non-zero and the energy keeps
    decreasing, although very slowly; after 100,000 iterations, the
    energy was still above 136 Bohr$^2$.}
  \label{fig:wannier90_conv}
\end{figure}

These Wannier functions are not maximally-localized, but form a good
initial guess for the MV algorithm. By contrast, initializing the
algorithm with randomly-centered s-type orbitals (as is the default in
Wannier90 in the absence of a specific prescription) might yield an
algorithm that can converge to a ``false local minimum'', especially on
fine meshes, as can be seen in Figure
\ref{fig:wannier90_conv}. However, when the initial guess is good
enough, the convergence is satisfactory, and even faster than using
our algorithm to produce an initial guess.

The computational time of our algorithm is dominated by the cost of a
few operations on $J \times J$ matrices at each $\bk$-point, and is
therefore comparable to that of one iteration of the MV algorithm. By
contrast, the projection method requires computations on the
plane-wave grid, and is therefore much more costly. Therefore, the
total time for the computation of MLWFs is lower with our algorithm,
even when it requires more iterations to converge than the projection
method with good initial guesses.
\section{Conclusion and perspectives}
\label{sec:conclusion}
We {proposed} an algorithm to obtain well-localized Wannier functions
without any initial guess or free parameters, and presented numerical
results showing its correctness, as well as its superiority compared
to the projection method on very fine meshes and when good initial guesses are not
available. The flipside to this is that our method, which does not
utilize the physics of the system, might not yield optimal Wannier
functions, but only local minima. We anticipate our method to be
useful for systems where no physical intuition is available, and where
well-localized Wannier functions, even if not optimal or physically
sensible, can be used for example for Wannier interpolation \cite{yates2007spectral}.

Our algorithm can fail when eigenvalues of the obstruction matrix
corresponding to different number of turns $p$ collide. We have never
found this to be the case in practice. However, it is unclear whether
such a phenomenon is impossible because of topological reasons or if
it is simply rare and we have never found it in our tests. A better
understanding of this issue is a worthwhile direction of research. Our
algorithm could be adapted to tackle such collisions, but would be
significantly more complicated, introduce free parameters, and be less
robust for coarse $\bk$-point meshes, which is why we refrain from
doing so in this work.

As we demonstrated, a singular input to the MV algorithm may or
may not yield a physically relevant answer, depending on the size of
the mesh: singularities may not be seen on a coarse mesh, while fine
meshes emphasize the divergent contribution of the vortices to the
functional. In this case, the MV algorithm stalls and is unable to
converge to a ``true'' local minimum. The behavior of the MV algorithm
in this case would be interesting to study from a numerical analysis
point of view. Another important question is whether any continuous
frame will yield exponentially-localized Wannier functions when used
as initial guess to the MV algorithm.

On the numerical side, our algorithm, publicly available at
\url{https://github.com/antoine-levitt/wannier}, can readily be
interfaced in standard workflows using the Wannier90
\cite{mostofi2008wannier90,mostofi2014updated} code. Our construction
preserves the time-reversal symmetry, but not any other symmetries the
crystal might possess, unlike the construction in
\cite{sakuma2013symmetry}. The extension of our method to additional
symmetries is also an interesting direction for future work.

\section*{Acknowledgments}
We wish to thank the anonymous referees for useful suggestions and comments.

\section*{Appendix: systems without time-reversal symmetry and
  topological insulators}
\label{sec:noTRS}

Our algorithm was introduced in the framework of time-reversal
symmetric systems, and more precisely of a symmetry of
\textit{bosonic} type, which is one in which the anti-unitary
time-reversal operator $C$ squares to $1$. This is the case where we
know theoretically that there exist localized Wannier
functions. Although a detailed study is outside the scope of this
paper, it is interesting to explore what happens when this assumption
fails, and in particular investigate the case of topological
insulators. Our algorithm can be adapted very simply to systems
without time-reversal symmetry (or with a \textit{fermionic} one, that
squares to $-1$), and we explain it here for one- and two-dimensional
systems (the three-dimensional case is a simple extension).

In 1D, we first start with a frame at $\Gamma$, which we do not impose
to be real. Then, we propagate it using \eqref{eq:propagation} to the
segments $[0, 1/2]$ and $[-1/2, 0]$. At this point, generically
$u_{1/2} \neq \tau_{1} u_{-1/2}$: we compute the unitary obstruction
matrix $\Uobs = u_{1/2}^{*} (\tau_{1} u_{-1/2})$, and set $u'_{k} =
u_{k} \Uobs^{k}$, which is now a continuous frame.

In 2D, we apply the same construction as in the 1D case to build a
continuous frame on the segment from $(0,-1/2)$ to $(0,1/2)$ that
satisfies $u_{(0,1/2)} = \tau_{2} u_{(0,-1/2)}$. Then, we propagate it
horizontally and obtain a frame $u_{(k_{1}, k_{2})}$ on
$[-1/2, 1/2]^{2}$. Generically,
$u_{(1/2, k_{2})} \neq \tau_{1} u_{(-1/2, k_{2})}$, which we now fix
in two steps. We define the top edge obstruction matrix
$U_{\rm obs,top} = u_{(1/2, 1/2)}^{*} (\tau_{1} u_{(-1/2,1/2)})$, set
$u'_{(k_{1}, k_{2})} = u_{(k_{1}, k_{2})} U_{\rm obs,top}^{k_{1}}$,
and drop the primes for simplicity. We now define the obstruction
matrix $\Uobs(k) = u_{(1/2, k)}^{*} (\tau_{1} u_{(-1/2,k)})$, which,
like in the time-reversal symmetric case, satisfies
$\Uobs(-1/2) = \Uobs(1/2) = \onemat$. Provided we can find a
continuous logarithm $L(k)$ of $\Uobs(k)$, we set
$u'_{(k_{1}, k_{2})} = u_{(k_{1}, k_{2})} \rme^{\ri k_{1} L(k_{2})}$
and obtain a continuous frame.

Note that, as this algorithm always produces a continuous frame if the
logarithm problem is solved, it follows that the logarithm problem
cannot be solved for materials with non-zero Chern numbers, where we
know that there cannot exist a continuous frame \cite{Panati2007}. We
will now show how exactly the existence of a logarithm fails on a
Chern insulator.

\subsection*{Chern insulators: the Haldane model}
\label{sec:haldane}
We test our algorithm on the prototype of Chern insulators, the two
band Haldane model. We use the same notation as in the original paper
\cite{Haldane1988}: the parameters $t_{1}$ and $t_{2}$ are the nearest
and next-nearest neighbors hopping terms, $\phi$ is the phase, which
breaks the time-reversal symmetry, and $M$ is the on-site energy,
which breaks the inversion symmetry. We used as parameters
$t_{1} = t_{2} = 1$ and $M = 0.1$, and consider the first band
($J = 1$). Note that since the Haldane model only has one band, it is
not a good test case to see the collision of eigenvalues we are
concerned about in Section \ref{sec:logpb}, but rather illustrates how our
algorithm fails on a topological insulator with a non-zero Chern
number, as any algorithm must, since continuous frames cannot exist in
this case \cite{Brouder2007}.

The Haldane model is a time-reversal-symmetric insulator for
$\phi = 0$. When $\phi$ is increased and the time-reversal symmetry is
broken, the band gap decreases until it closes at\hfill \break
$\phi_{\rm c} = \arcsin(M/(3\sqrt3 t_{2})) \approx 0.019$. After
that point, it is a Chern insulator (with Chern number $1$).

\begin{figure}[h!]
  \centering
  \includegraphics[width=.3\textwidth]{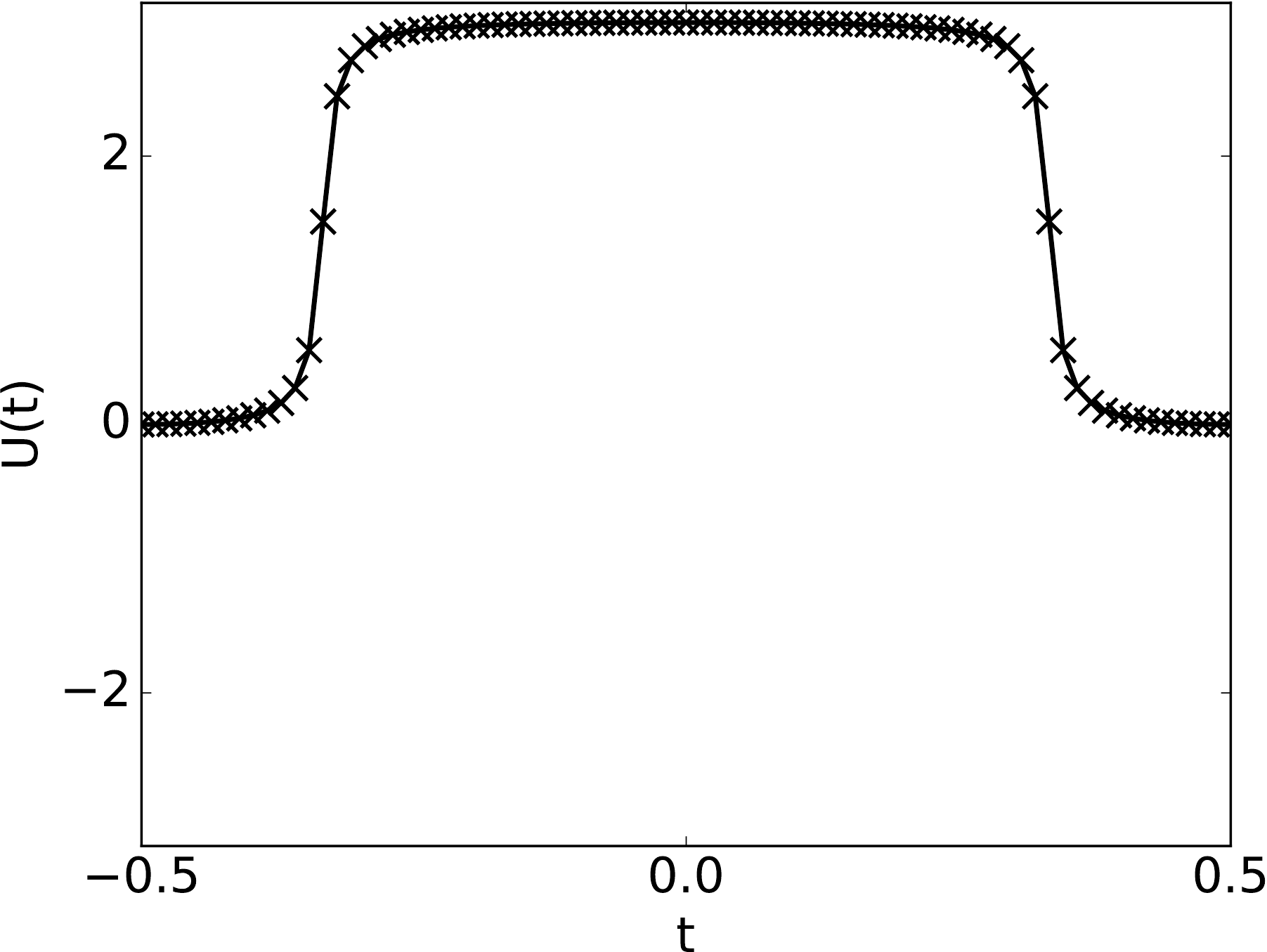}
  \includegraphics[width=.3\textwidth]{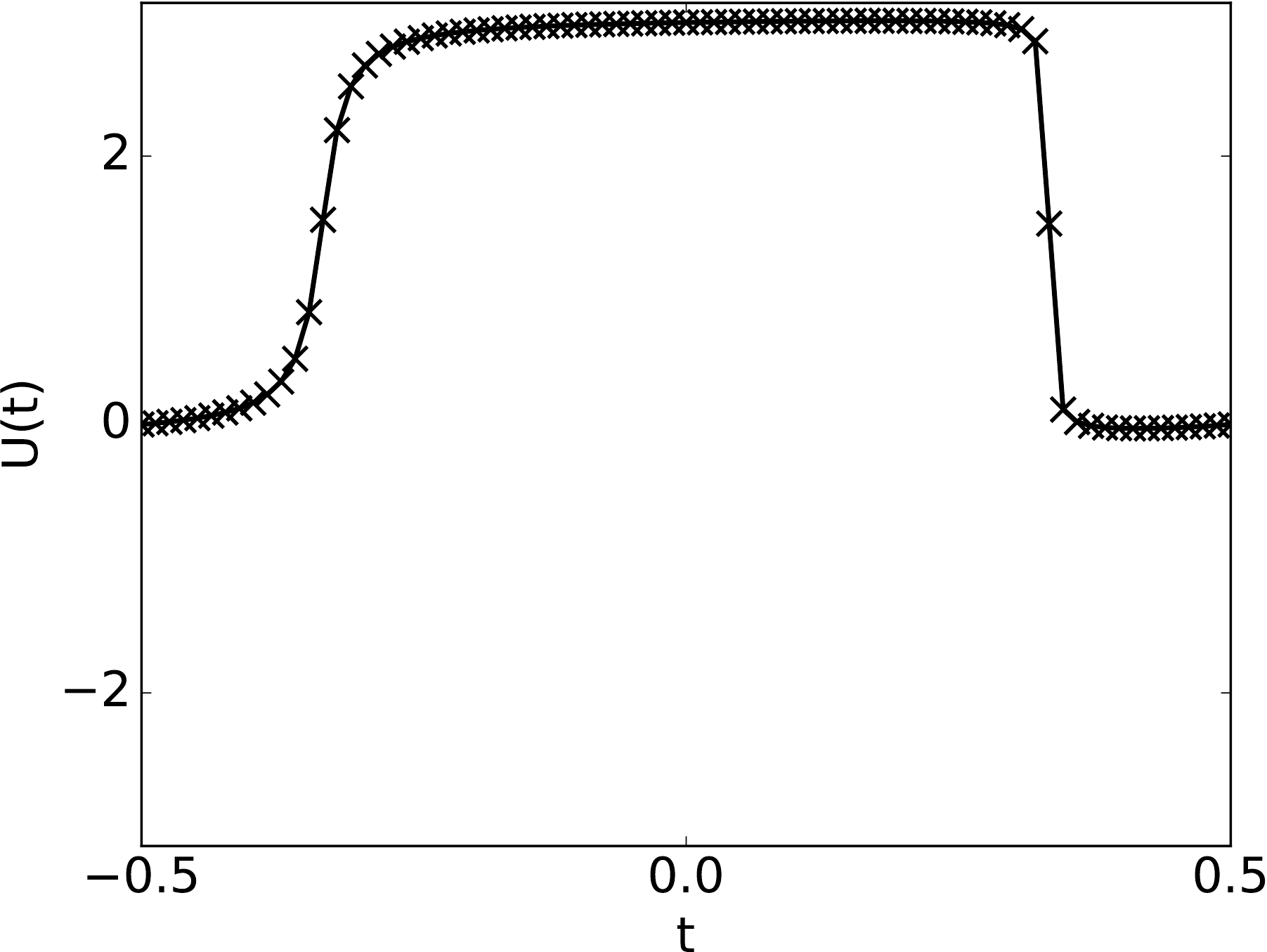}
  \includegraphics[width=.3\textwidth]{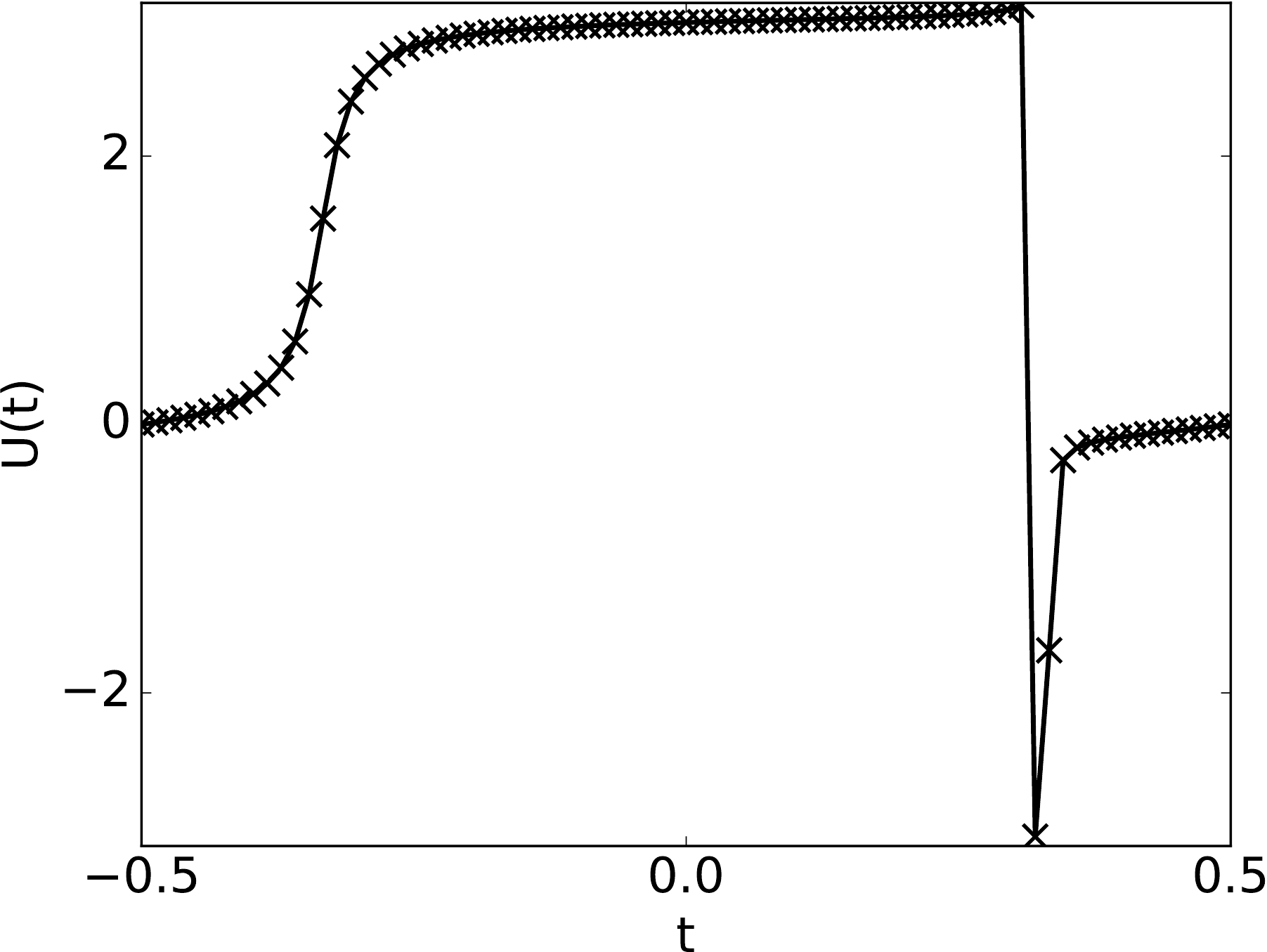}
  \caption{Phase of $\Uobs(k_{2})$ as a function of $k_{2}$ for the
    Haldane model with $\phi_{\rm c} \approx 0.019$, for $\phi = 0$
    (left), $\phi = \phi_{\rm c} - 0.005$ (middle) and
    $\phi = \phi_{\rm c} + 0.005$ (right). As $\phi$ increases, the
    time-reversal symmetry breaks down, and at $\phi_{\rm c}$, the gap
    closes and $U$ jumps abruptly from $-1$ to $1$. When
    $\phi_{\rm c}$ is increased into the Chern insulator regime, the
    map $U$ has degree 1, and no choice of branch cut can make its
    phase continuous and periodic.}
  \label{fig:haldane}
\end{figure}

As can be seen in Figure \ref{fig:haldane}, when $\phi$ is small, the
phase of the obstruction matrix (here, a single number) evolves
smoothly and symmetrically with respect to the origin: $\Uobs$ evolves
in the upper half-circle $|\Uobs| = 1, \Im \Uobs \geq 0$, and goes
from $1$ to $-1$ and back again, taking the same path in reverse. When
$\phi$ increases, the transition from $-1$ back to $1$ gets sharper,
until it is discontinuous at the critical threshold $\phi_{\rm
  c}$. When $\phi$ increases again, this discontinuity is resolved,
but this time $\Uobs$ goes through the lower half-circle on its way
back. The net result is that a logarithm of $\Uobs(k_{2})$ will pick
up a phase factor of $2\pi$ when going from $k_{2} = -1/2$ to
$k_{2} = 1/2$, and cannot therefore be continuous and periodic.

\subsection*{$\Z_{2}$ topological insulators: the Kane-Mele model}
\label{sec:kane-mele}
We now turn to the case of $\Z_{2}$ topological
insulators\cite{hasan2010colloquium}. $\Z_{2}$ topological insulators
are characterized by a \textit{fermionic} time-reversal symmetry,
squaring to $-1$ instead of $1$ as the one considered in
this paper, and possess a $\Z_{2}$ topological invariant. Systems with
an odd invariant have a topological obstruction to the construction of
frames respecting the time-reversal symmetry, but no obstruction to
the construction of non-symmetric frames \cite{FMP16}.

We test our algorithm on the Kane-Mele model \cite{kane2005z}. With
the same notation as in \cite{kane2005z}, we choose
$a = 1, t = 1, \lambda_{R} = 0, \lambda_{\rm SO} = 1$. For this choice
of parameters, the system is in a regular insulator phase for
$\lambda_{\nu} > \lambda_{\nu, \rm c} = 3 \sqrt 3$, and in a quantum
spin Hall phase for $\lambda_{\nu} < \lambda_{\nu,\rm c}$.

\begin{figure}[h!]
  \centering
  \includegraphics[width=.49\textwidth]{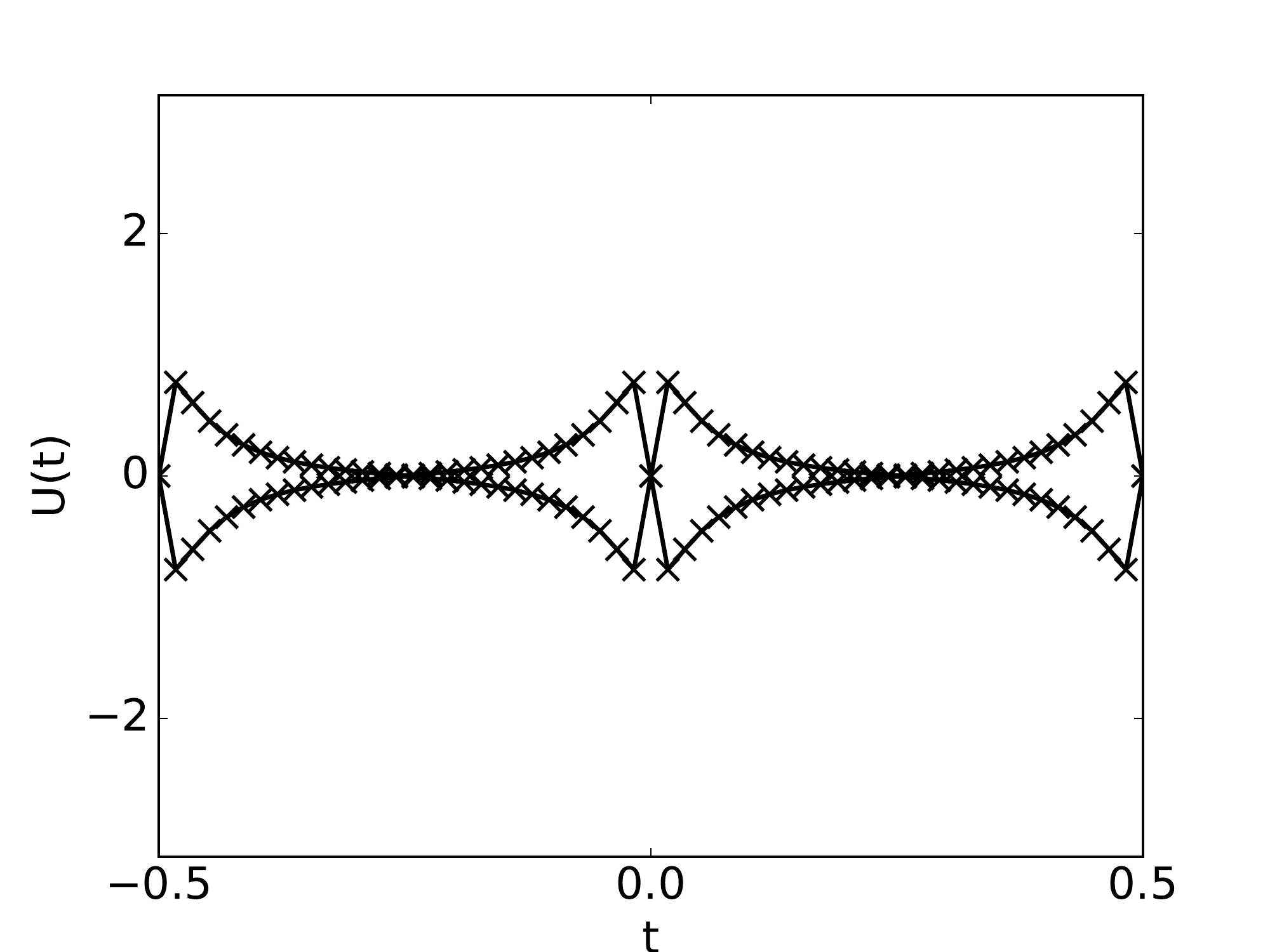}
  \includegraphics[width=.49\textwidth]{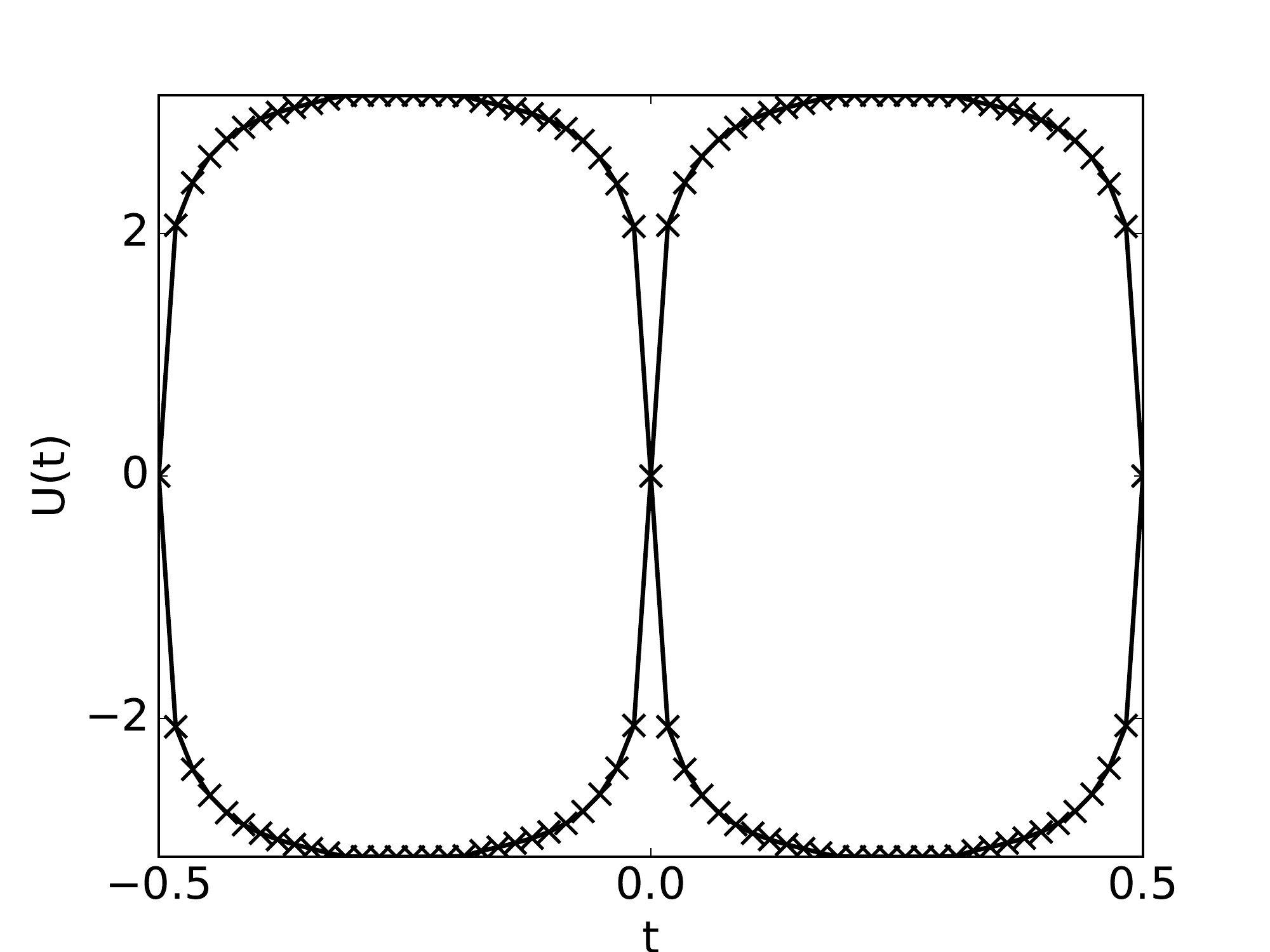}
  \caption{Phase of $\Uobs(k_{2})$ as a function of $k_{2}$ for the
    Kane-Mele model, for $\lambda_{\nu,\rm c} = 3 \sqrt 3$, with
    $\lambda_{\nu} = \lambda_{\nu,\rm c} + 0.02$ (left) and
    $\lambda_{\nu,\rm c} = \lambda_{\nu, \rm c} - 0.02$ (right). As
    $\lambda_{\nu, \rm c}$ decreases, the system transitions from an
    even to an odd $\Z_{2}$ invariant, and eigenvalues with different
    numbers of turns collide.}
  \label{fig:kane-mele}
\end{figure}

We show in Figure \ref{fig:kane-mele} that this transition introduces
a collision of eigenvalues with different numbers of turns. The
correspondence between the eigenvalue collisions and the $\Z_{2}$
invariant was recently proved in \cite[Proposition
5.7]{cornean2016wannier}. Let us stress that this collision occurs
because of the non-trivial topological states of systems with
\textit{fermionic} time-reversal symmetry. For systems with a
\textit{bosonic} form of time-reversal (the ones we consider in this
paper), we found no such crossings.


\begin{thebibliography}{10}

\bibitem{Blount1962}
E.I. Blount.
\newblock Formalism of band theory.
\newblock In F.~Seitz and D.~Turnbull, editors, {\em Solid State Physics},
  pages 305--373. Academic Press, 1962.

\bibitem{Brouder2007}
C.~Brouder, G.~Panati, M.~Calandra, C.~Mourougane, and N.~Marzari.
\newblock Exponential localization of {Wannier} functions in insulators.
\newblock {\em Physical Review Letters}, 98(4):046402, 2007.

\bibitem{cohen1966band}
M.L. Cohen and T.K. Bergstresser.
\newblock Band structures and pseudopotential form factors for fourteen
  semiconductors of the diamond and zinc-blende structures.
\newblock {\em Physical Review}, 141(2):789, 1966.

\bibitem{cornean2015construction}
H.~Cornean, I.~Herbst, and G.~Nenciu.
\newblock On the construction of composite {Wannier} functions.
\newblock {\em to appear in Ann. H. Poincar\'e}, 2015.

\bibitem{cornean2016wannier}
H.~Cornean, D.~Monaco, and S.~Teufel.
\newblock Wannier functions and $\mathbb{Z}_2$ invariants in time-reversal
  symmetric topological insulators.
\newblock {\em arXiv preprint arXiv:1603.06752}, 2016.

\bibitem{corsetti2012properties}
F.~Corsetti.
\newblock {\em On the properties of point defects in silicon nanostructures
  from ab initio calculations}.
\newblock PhD thesis, Imperial College London, 2012.

\bibitem{damle2015compressed}
A.~Damle, L.~Lin, and L.~Ying.
\newblock Compressed representation of {K}ohn--{S}ham orbitals via selected
  columns of the density matrix.
\newblock {\em Journal of Chemical Theory and Computation}, 11(4):1463--1469,
  2015.

\bibitem{damle2015scdm}
A.~Damle, L.~Lin, and L.~Ying.
\newblock {SCDM}-k: Localized orbitals for solids via selected columns of the
  density matrix.
\newblock {\em arXiv preprint arXiv:1507.03354}, 2015.

\bibitem{Cloizeaux1964b}
J.~des Cloizeaux.
\newblock Analytical properties of n-dimensional energy bands and {Wannier}
  functions.
\newblock {\em Physical Review}, 135(3A):A698--A707, 1964.

\bibitem{Cloizeaux1964a}
J.~des Cloizeaux.
\newblock Energy bands and projection operators in a crystal: Analytic and
  asymptotic properties.
\newblock {\em Physical Review}, 135(3A):A685--A697, 1964.

\bibitem{Dubrovin1980}
B.A. Dubrovin and S.P. Novikov.
\newblock Ground states of a two-dimensional electron in a periodic magnetic
  field.
\newblock {\em Zh. Eksp. Teor. Fiz.}, 79(3), 1980.

\bibitem{fiorenza2016construction}
D.~Fiorenza, D.~Monaco, and G.~Panati.
\newblock Construction of real-valued localized composite {Wannier} functions
  for insulators.
\newblock {\em Ann. Henri Poincar{\'e}}, 17(1):63--97, 2016.

\bibitem{FMP16}
D.~Fiorenza, D.~Monaco, and G.~Panati.
\newblock $\mathbb{Z}_2$ invariants of topological insulators as geometric
  obstructions.
\newblock {\em Commun. Math. Phys.}, 343(3):1115--1157, 2016.

\bibitem{giannozzi2009quantum}
P.~Giannozzi, S.~Baroni, N.~Bonini, M.~Calandra, R.~Car, C.~Cavazzoni,
  D.~Ceresoli, G.~Chiarotti, M.~Cococcioni, I.~Dabo, et~al.
\newblock {Q}uantum {E}spresso: a modular and open-source software project for
  quantum simulations of materials.
\newblock {\em Journal of physics: Condensed matter}, 21(39):395502, 2009.

\bibitem{guillemin2010differential}
V.~Guillemin and A.~Pollack.
\newblock {\em Differential topology}, volume 370.
\newblock American Mathematical Soc., 2010.

\bibitem{Haldane1988}
F.D.M. Haldane.
\newblock Model for a quantum {Hall} effect without {L}andau levels:
  Condensed-matter realization of the "parity anomaly".
\newblock {\em Physical Review Letters}, 61(18):2015--2018, 1988.

\bibitem{hasan2010colloquium}
M.~Hasan and C.~Kane.
\newblock Colloquium: topological insulators.
\newblock {\em Reviews of Modern Physics}, 82(4):3045, 2010.

\bibitem{HelfferSjostrand1989}
B.~Helffer and J.~Sj\"ostrand.
\newblock {\'E}quation de {S}chr\"odinger avec champ magn\'etique et \'equation
  de {H}arper.
\newblock {\em Lecture Notes in Physics}, 345:118--197, 1989.

\bibitem{kane2005z}
C.J. Kane and E.J. Mele.
\newblock $\mathbb{Z}_2$ topological order and the quantum spin hall effect.
\newblock {\em Physical Review Letters}, 95(14):146802, 2005.

\bibitem{Kohn1959}
W.~Kohn.
\newblock Analytic properties of {Bloch} waves and {Wannier} functions.
\newblock {\em Physical Review}, 115(4):809--821, 1959.

\bibitem{MMYSV12}
N.~Marzari, A.A. Mostofi, J.R. Yates, I.~Souza, and D.~Vanderbilt.
\newblock Maximally localized {W}annier functions: Theory and applications.
\newblock {\em Reviews of Modern Physics}, 84(4):1419--1475, 2012.

\bibitem{marzari1997maximally}
N.~Marzari and D.~Vanderbilt.
\newblock Maximally localized generalized {Wannier} functions for composite
  energy bands.
\newblock {\em Physical Review B}, 56(20):12847, 1997.

\bibitem{MonacoPanati2014}
D.~Monaco and G.~Panati.
\newblock Topological invariants of eigenvalue intersections and decrease of
  {Wannier} functions in graphene.
\newblock {\em Journal of Statistical Physics}, 155(6):1027--1071, 2014.

\bibitem{MPPT2016}
D.~Monaco, G.~Panati, A.~Pisante, and S.~Teufel.
\newblock Optimal decay of {Wannier} functions in {Chern} and {Quantum Hall}
  insulators.
\newblock in preparation, 2016.

\bibitem{mostofi2008wannier90}
A.~Mostofi, J.~Yates, Y.-S. Lee, I.~Souza, D.~Vanderbilt, and N.~Marzari.
\newblock wannier90: A tool for obtaining maximally-localised {Wannier}
  functions.
\newblock {\em Computer Physics Communications}, 178(9):685--699, 2008.

\bibitem{mostofi2014updated}
A.~Mostofi, J.~Yates, G.~Pizzi, Y.-S. Lee, I.~Souza, D.~Vanderbilt, and
  N.~Marzari.
\newblock An updated version of {W}annier90: A tool for obtaining
  maximally-localised {W}annier functions.
\newblock {\em Computer Physics Communications}, 185(8):2309--2310, 2014.

\bibitem{mustafa2015automated}
J.I. Mustafa, S.~Coh, M.L. Cohen, and S.G. Louie.
\newblock Automated construction of maximally localized {Wannier} functions:
  Optimized projection functions method.
\newblock {\em Physical Review B}, 92(16):165134, 2015.

\bibitem{Nenciu1982}
A.~Nenciu and G.~Nenciu.
\newblock Dynamics of {Bloch} electrons in external electric fields. ii. the
  existence of {Stark-Wannier} ladder resonances.
\newblock {\em Journal of Physics A: Mathematical and General},
  15(10):3313--3328, 1982.

\bibitem{Nenciu1983}
G.~Nenciu.
\newblock Existence of the exponentially localised {Wannier} functions.
\newblock {\em Communications in Mathematical Physics}, 91(1):81--85, 1983.

\bibitem{Nenciu1991}
G.~Nenciu.
\newblock Dynamics of band electrons in electric and magnetic fields: Rigorous
  justification of the effective hamiltonians.
\newblock {\em Reviews of Modern Physics}, 63(1):91--127, 1991.

\bibitem{Panati2007}
G.~Panati.
\newblock Triviality of {Bloch} and {Bloch-Dirac} bundles.
\newblock {\em Annales Henri Poincar\'e}, 8(5):995--1011, 2007.

\bibitem{panati2013bloch}
G.~Panati and A.~Pisante.
\newblock {Bloch} bundles, {Marzari-Vanderbilt} functional and maximally
  localized {W}annier functions.
\newblock {\em Communications in Mathematical Physics}, 322(3):835--875, 2013.

\bibitem{Park2011}
C.-H. Park and N.~Marzari.
\newblock Berry phase and pseudospin winding number in bilayer graphene.
\newblock {\em Physical Review B - Condensed Matter and Materials Physics},
  84(20):205440, 2011.

\bibitem{resta2007theory}
R.~Resta and D.~Vanderbilt.
\newblock Theory of polarization: a modern approach.
\newblock In {\em Physics of Ferroelectrics}, pages 31--68. Springer, 2007.

\bibitem{sakuma2013symmetry}
R.~Sakuma.
\newblock Symmetry-adapted {W}annier functions in the maximal localization
  procedure.
\newblock {\em Physical Review B}, 87(23):235109, 2013.

\bibitem{spaldin2012beginner}
N.A. Spaldin.
\newblock A beginner's guide to the modern theory of polarization.
\newblock {\em Journal of Solid State Chemistry}, 195:2--10, 2012.

\bibitem{yates2007spectral}
J.~Yates, X.~Wang, D.~Vanderbilt, and I.~Souza.
\newblock Spectral and {Fermi} surface properties from {Wannier} interpolation.
\newblock {\em Physical Review B}, 75(19):195121, 2007.

\end{thebibliography}
\end{document}